\RequirePackage[2020-02-02]{latexrelease}
\documentclass[a4paper]{aa}

\RequirePackage[utf8]{inputenc}
\RequirePackage[T1]{fontenc}
\usepackage[english]{babel}

\usepackage{csquotes}

\usepackage{amsmath}
\usepackage{amssymb}

\usepackage{multirow}

\usepackage{siunitx}

\usepackage[switch]{lineno}

\PassOptionsToPackage{dvipsnames}{xcolor} 
\usepackage{graphicx,pstricks}
\usepackage{pgfplots}
\usepackage{color}
\usepackage{float}

\usepackage{url}
\usepackage{xspace}

\usepackage{hyperref}
\hypersetup{colorlinks=true, linktocpage=true, pdfstartpage=3,
  pdfstartview=FitV, breaklinks=true, pdfpagemode=UseNone,
  pageanchor=true, pdfpagemode=UseOutlines, plainpages=false,
  bookmarksnumbered, bookmarksopen=true, bookmarksopenlevel=1,
  hypertexnames=true, pdfhighlight=/O, urlcolor=RoyalBlue,
  linkcolor=RoyalBlue, citecolor=RoyalBlue}

\usepackage{aas_macros}
\bibpunct{(}{)}{;}{a}{,}{,}

\usepackage[nameinlink]{cleveref}

\usepackage{catoptions}

\usepackage[symbol]{footmisc}

\makeatletter

\def\Autoref#1{%
  \begingroup
  \edef\reserved@a{\cpttrimspaces{#1}}%
  \ifcsndefTF{r@#1}{%
    \xaftercsname{\expandafter\testreftype\@fourthoffive}
      {r@\reserved@a}.\\{#1}%
  }{%
    \ref{#1}%
  }%
  \endgroup
}
\def\testreftype#1.#2\\#3{%
  \ifcsndefTF{#1autorefname}{%
    \def\reserved@a##1##2\@nil{%
      \uppercase{\def\ref@name{##1}}%
      \csn@edef{#1autorefname}{\ref@name##2}%
      \autoref{#3}%
    }%
    \reserved@a#1\@nil
  }{%
    \autoref{#3}%
  }%
}
\makeatother

\newcommand{\numb}[1]{\textcolor{black}{#1}} 
\newcommand{\skybot}{\texttt{SkyBoT}\xspace}
\newcommand{\sm}{\texttt{SkyMapper}\xspace}
\newcommand{\sdss}{\texttt{SDSS}\xspace}
\newcommand{\smss}{\texttt{SMSS}\xspace}
\newcommand{\gaia}{\texttt{Gaia}\xspace}
\newcommand{\classy}{\texttt{Classy}\xspace}
\newcommand{\neorocks}{\texttt{NEOROCKS}\xspace}
\newcommand{\rocks}{\texttt{rocks}\xspace}
\newcommand{\ssodnet}{\texttt{SsODNet}\xspace}
\newcommand{\source}[1]{\textsuperscript{\textcolor{blue}{[citation needed]}}\xspace}

\newcommand{\filtu}{\ensuremath{u}\xspace}
\newcommand{\filtg}{\ensuremath{g}\xspace}
\newcommand{\filtr}{\ensuremath{r}\xspace}
\newcommand{\filti}{\ensuremath{i}\xspace}
\newcommand{\filtz}{\ensuremath{z}\xspace}
\newcommand{\colorgr}{\ensuremath{g}-\ensuremath{r}\xspace}
\newcommand{\colorgi}{\ensuremath{g}-\ensuremath{i}\xspace}
\newcommand{\colorri}{\ensuremath{r}-\ensuremath{i}\xspace}
\newcommand{\coloriz}{\ensuremath{i}-\ensuremath{z}\xspace}

\newcommand\sdssTotObs{11,142\xspace}
\newcommand\sdssTotAst{5,425\xspace}
\newcommand\sdssFastAst{470\xspace}

\newcommand\skmTotObs{12,001\xspace}
\newcommand\skmTotAst{3,149\xspace}

\newcommand\gaiaNeoAst{838\xspace} 
\newcommand\gaiaTotAst{60,518\xspace} 

\newcommand\skmTotColors{9,212\xspace} 
\newcommand\skmAstColors{2,081\xspace} 

\newcommand\clsTotObs{4,548\xspace} 
\newcommand\clsTotAst{3,157\xspace} 
\newcommand\clsNeoObs{1,072\xspace} 
\newcommand\clsNeoAst{846\xspace} 

\newcommand\TotAst{7,401\xspace}

\begin{document}

\title{Compositional properties of planet-crossing asteroids from astronomical surveys}
\author{A. V. Sergeyev\inst{1,2}
      \thanks{Corresponding author: alexey.v.sergeyev@gmail.com}
      \and B. Carry\inst{1}
      \and M. Marsset\inst{3,4}
      \and P. Pravec\inst{5}
      \and D. Perna\inst{6}
      \and F. E. DeMeo\inst{7, 4}
      \and V. Petropoulou\inst{8}
      \and M. Lazzarin\inst{9}
      \and F. La Forgia\inst{9}
      \and I. Di Petro \inst{10}
      \and the NEOROCKS team \thanks{The NEOROCKS team: E. Dotto, M. Banaszkiewicz, 
      S. Banchi, M.A. Barucci, F. Bernardi, M. Birlan, A. Cellino, J. De Leon, M. 
      Lazzarin, E. Mazzotta Epifani, A. Mediavilla, J. Nomen Torres, E. Perozzi, 
      C. Snodgrass, C. Teodorescu, S. Anghel, A. Bertolucci, F. Calderini, F. Colas, 
      A. Del Vigna, A. Dell'Oro, A. Di Cecco, L. Dimare, P. Fatka, S. Fornasier, 
      E. Frattin, P. Frosini, M. Fulchignoni, R. Gabryszewski, M. Giardino, 
      A. Giunta, T. Hromakina, J. Huntingford, S. Ieva, J.P. Kotlarz, M. Popescu,
      J. Licandro, H. Medeiros, F. Merlin, F. Pinna, G. Polenta, A. Rozek, 
      P. Scheirich,  A. Sonka, G.B. Valsecchi, P. Wajer, A. Zinzi.}
\\}
\institute{
  Universit\'e C{\^o}te d'Azur, Observatoire de la C{\^o}te d'Azur, CNRS, Laboratoire Lagrange, France
  \label{i:oca}
  \and
  V. N. Karazin Kharkiv National University, 4 Svobody Sq., Kharkiv, 61022, Ukraine
  \label{i:kha}
  \and 
  European Southern Observatory (ESO), Alonso de Cordova 3107, 1900, Casilla Vitacura, Santiago, Chile
  \label{i:eso}
  \and
  Department of Earth, Atmospheric and Planetary Sciences, MIT, 77 Massachusetts Avenue, Cambridge, MA, 02139, USA
  \label{i:mit2}
  \and
  Astronomical Institute, Academy of Sciences of 
  the Czech Republic, CZ-25165 Ond\v{r}ejov, Czech Republic
  \label{i:ondrejov}
  \and
  INAF - Osservatorio Astronomico di Roma, Via Frascati 33, I-00078 Monte Porzio Catone, Italy
  \label{i:inaf_rome}
  \and
  Department of Earth, Atmospheric, and Planetary Sciences, Massachusetts Institute of Technology, 77 Massachusetts Avenue, Cambridge, MA 02139, USA
  \label{i:cambridge}
  \and
  INAF - Osservatorio Astronomico di Roma, Monte Porzio Catone (RM), Italy
  \label{i:inaf_rome2}
  \and
  INAF - Department of Physics and Astronomy, University of Padova, Vicolo dell'Osservatorio, 3, I-35122 Padova, Italy
  \label{i:inaf_padova}
  \and
  Agenzia Spaziale Italiana (ASI), Via del Politecnico 00133 Roma, Italy
  \label{i:asi}
}

\titlerunning{Properties of NEOs from surveys}
\authorrunning{Sergeyev et al.}

\abstract
{The study of planet-crossing asteroids is of both practical and fundamental importance.
As they are closer than asteroids in the Main Belt, we have access to a smaller size range, 
and this population frequently impacts planetary surfaces and can pose a threat
to life.}
{We aim to characterize the compositions of a large corpus of planet-crossing asteroids
and to study how these compositions are related to orbital and physical parameters.}
{We gathered publicly available visible colors 
of near-Earth objects (NEOs) 
from the Sloan Digital Sky Survey (\sdss) and \sm surveys. 
We also computed \sdss-compatible
colors  from reflectance spectra of the \gaia mission and a
compilation of ground-based observations.
We determined the taxonomy of each NEO from its colors and studied
the distribution of the taxonomic classes and spectral slope against
the orbital parameters and diameter.
}
{We provide updated photometry for \numb{\sdssFastAst} NEOs from the \sdss, 
and taxonomic classification of \numb{\TotAst} NEOs.
We classify \numb{42} NEOs that are mission-accessible, including
\numb{six} of the \numb{seven} flyby candidates of the ESA Hera mission.
We confirm the perihelion dependance of spectral slope among S-type
NEOs, likely related to a rejuvenation mechanism linked with
thermal fatigue. We also confirm the clustering of A-type NEOs
around 1.5--2\,AU, and predict the taxonomic distribution of 
small asteroids in the NEO source regions in the Main Belt.
}
{}


\maketitle

\section{Introduction}

Asteroids are the remnants of the building blocks that accreted
to form the terrestrial planets and the core of the giant planets
in the early Solar System \SI{4.6}{\giga y} ago.
Asteroids are also the origin of the meteorites that fell on the planets,
including the Earth. 
These meteorites represent the only possibility to study 
in detail the composition of asteroids in the laboratory 
\citep[e.g.,][]{2008-ChEG-68-Consolmagno,2015-Icarus-252-Cloutis}, 
with the exception of the tiny samples of rock, 
provided by return-sample missions:
JAXA Hayabusa \citep{2011-Science-333-Yurimoto}
and
Hayabusa-2 \citep{2022Sci...375.1011T},
as well as the soon due 
NASA OSIRIS-REx \citep{2017SSRv..212..925L}.

In contrast to targeted sample collection, we cannot choose the origin
of meteorites striking the Earth. Identifying their source regions 
is therefore crucial to determining the physical
conditions and abundances in elements that reigned in the
protoplanetary nebula around the young Sun 
\citep{2006-MESS2-McSween}.
From the analysis of a bolide trajectory, it is possible
to reconstruct a meteorite's heliocentric orbit
\citep{2006-MPS-41-Gounelle},
although such determinations have been limited
to only a few meteorites \citep{2018-Icarus-311-Granvik}.

\begin{figure*}[ht]
  \centering
  \includegraphics[width=1\hsize]{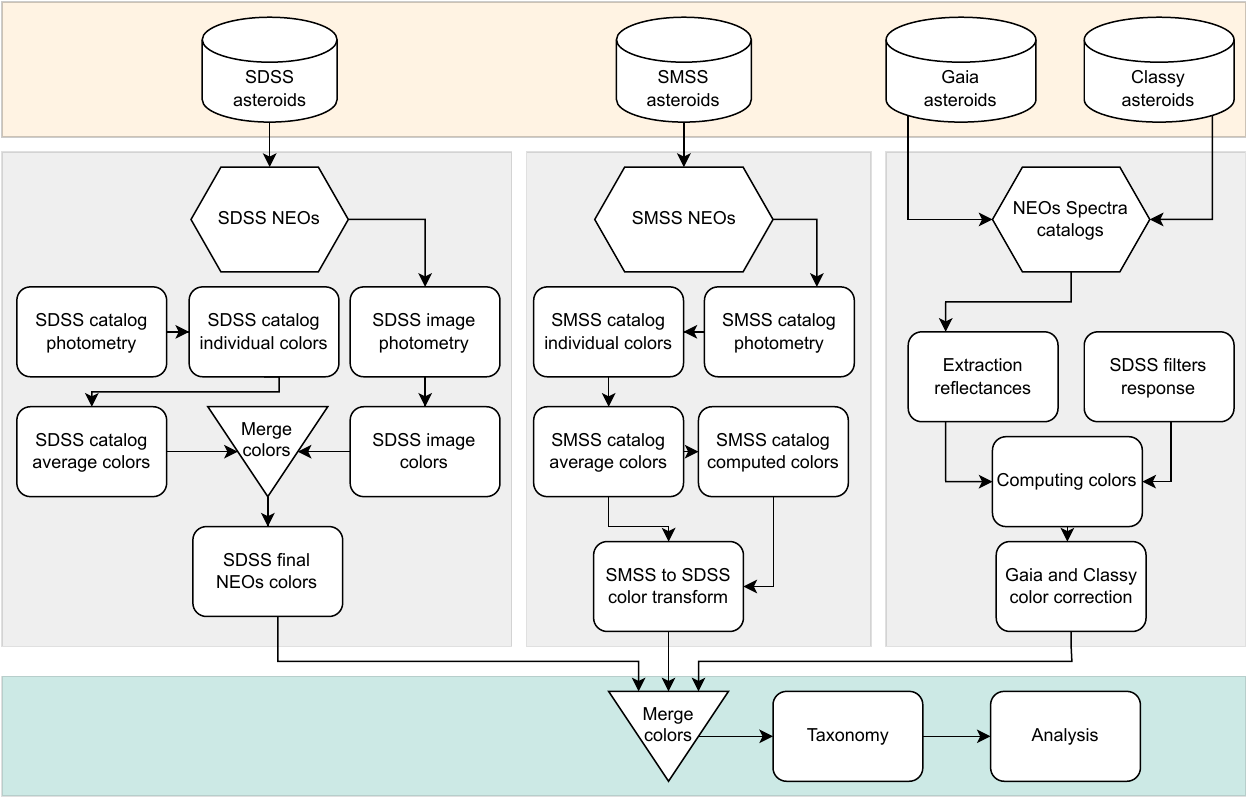}
  \caption{Schematic view of the extraction, convertion, and merging of
    NEOs from \sdss, \smss, \gaia, and \classy catalogs.
    }
  \label{fig:workflow}
\end{figure*}

Among the different dynamical classes of asteroids, the near-Earth
and Mars-crosser asteroids (NEAs and MCs), whose orbits
cross that of the telluric planets, form a transient population. Their
typical lifetime is of only a few million years 
before they are ejected from the Solar System,
fall into the Sun, or impact a planet
\citep{1997Sci...277..197G}. 
We refer here to near-Earth objects (NEOs) in a liberal
sense, encompassing both asteroid-like and comet-like objects
whose orbits cross that of a terrestrial planet (hence including
NEAs, MCs, and some Hungarias).

These populations are of both scientific and pragmatic interest.
As they are closer to the Earth than the asteroid belt, we have access to
smaller objects from ground-based telescopes. Their orbital proximity
implies a much smaller impulsion to reach them with a spacecraft
and make them favorable targets for space exploration \citep{2012-DPS-Abell}.
On the other hand, these objects could potentially pose a threat,
and studying their properties is a key aspect in planning 
risk mitigation \citep{2015hchp.book..763D}, 
of which the National Aeronautics and Space Administration (NASA)  Demonstration for Autonomous Rendezvous Technology (DART) and European Space Agency (ESA) Hera missions
are lively demonstrators \citep{2021PSJ.....2..173R, 2022PSJ.....3..160M}.

We focus here on the compositional properties of a large corpus
of NEOs as part of the \neorocks project
\citep{2021plde.confE.221D}, whose goal is the characterization of the NEO population.
The article is organized as follows:.
In \Autoref{sec:data} we present the data we have collected and 
the way in which we are building a large catalog of NEOs with visible colors
(including a refinement of the photometry of
the NEOs present in the \sdss catalog,
\Autoref{sec:photometry}).
We then present in \Autoref{sec:taxo} the way in which we determine the
taxonomic class of each NEO.
We focus on the taxonomy of the potential targets for 
space missions in \Autoref{sec:space}, and finally, we discuss
the distribution of taxonomic classes, 
the effect of space weathering and planetary encounters,
and NEO source regions 
in \Autoref{sec:disc}.

\section{Data sources\label{sec:data}}

In this section, we describe the data sets we collect, 
how they compare in terms of precision, and the way in which
we merge them into a single catalog of colors.
The entire process is summarized in \Autoref{fig:workflow}.

\subsection{Collecting data sets\label{ssec:collect}}

We gathered the colors of NEOs from four 
recently published sources:
the Sloan Digital Sky Survey \citep[\sdss,][]{SergeyevCarry2021},
the \sm Southern Survey \citep[\smss,][]{Sergeyev2022}, 
the \gaia DR3 visible spectra \citep[\gaia,][]{gaia3-spectra},
and a compilation of ground-based spectra \citep[\classy,][]{2022A&A...665A..26M}.
For the last two sources, we converted the reflectance to colors
in order to obtain the largest possible homogeneous data set
(\Autoref{app:gaia_color}).

Each \sdss observation sequence contains quasi-simultaneous 
measurements in five filters (\filtu, \filtg, \filtr, \filti, \filtz), 
providing colors of all combinations.
There is a constant time difference between two exposures in consecutive filters,
equal to \SI{57}{\second}. The largest time difference between two 
exposures occurs for the \filtg and \filtr filters, and is approximately \SI{230}{\second}.
The initial \sdss catalog contains \numb{\sdssTotObs} individual multi-filter 
observations of \numb{\sdssTotAst} unique NEOs.
For each NEO, we computed the weighted mean of each color from
multiple measurements. 
Owing to potential biases on the \sdss photometry for fast-moving NEOs
\citep{2014AN....335..142S,2016Icar..268..340C}, we remeasured \numb{\sdssFastAst} NEO colors on \sdss frames (see \Autoref{sec:photometry}).

The \sm includes several observing strategies. 
A shallow six-filter sequence with exposure times between 
\SI{5}{\second} and \SI{40}{\second}, 
a deep ten-image sequence of $uvgruvizuv$ with \SI{100}{\second} exposures, 
and pairs of deep exposures in (\filtg,\filtr) and (\filti,\filtz).
This observing strategy, in conjunction with the enhanced 
sensitivity in \filtg and \filtr, implies a predominance of $g-r$ colors 
in the results, but almost always leads to the measurement of at least 
one photometric color obtained within $\lesssim$ \SI{2}{\minute}
\citep[see][for more details]{Sergeyev2022}.
The initial \sm catalog contains \numb{\skmTotObs} individual observations 
of \numb{\skmTotAst} unique NEOs. 
We computed the asteroid color indexes by limiting the observation time
between two filters to 20 minutes and weighted the mean color 
of multiple asteroid measurements whenever possible. Through this method,
we retrieved \numb{\skmTotColors} colors of \numb{\skmAstColors} individual NEOs.
The \sm filters are slightly different from those of \sdss.
We thus converted the \sm colors into \sdss colors using color-transformation coefficients 
that were computed from a wide range of stellar classes 
\citep{Sergeyev2022}.

\gaia DR3 \citep{2016AA...595A...1G, gaia3-survey} contains \numb{\gaiaTotAst}
low-resolution reflectance spectra
of asteroids \citep{gaia3-spectra}. 
Among these, \numb{\gaiaNeoAst} are NEOs, of which \numb{199} 
have not been recorded in other catalogs. 
These optical spectra range from \num{374} to \SI{1034}{\nano\meter}, meaning that
they almost fully 
overlap with \sdss \filtg to \filtz filters (see \Autoref{fig:gaia_spectrum}). 
We thus converted the \gaia reflectance spectra to \sdss colors to homogenize
the data set.
We detail the procedure in \Autoref{app:gaia_color}.

The \gaia DR3 represents the largest catalog of asteroid reflectance spectra.
However, the spectra of NEOs have regularly been acquired with 
ground-based facilities for decades, often over a larger 
wavelength range and with a higher spectral resolution
\citep[e.g., NEOSHIELD2, MITHNEOS, and MANOS surveys, 
see][]{2018PSS..157...82P, 2019Icar..324...41B, 2019AJ....158..196D}. 
Therefore, we used the preprocessed and resampled ground-based spectra 
from \citet{2022A&A...665A..26M}, which comprises 
\numb{\clsTotObs} spectra of \numb{\clsTotAst} unique asteroids.
We extracted \numb{\clsNeoObs} spectra of \numb{\clsNeoAst} unique NEOs
and converted them to \sdss colors with the same procedure as for the \gaia data
(\Autoref{app:gaia_color}).


\subsection{Comparing data sets\label{ssec:compare}}

\begin{figure}[]
  \centering
  \includegraphics[width=\hsize]{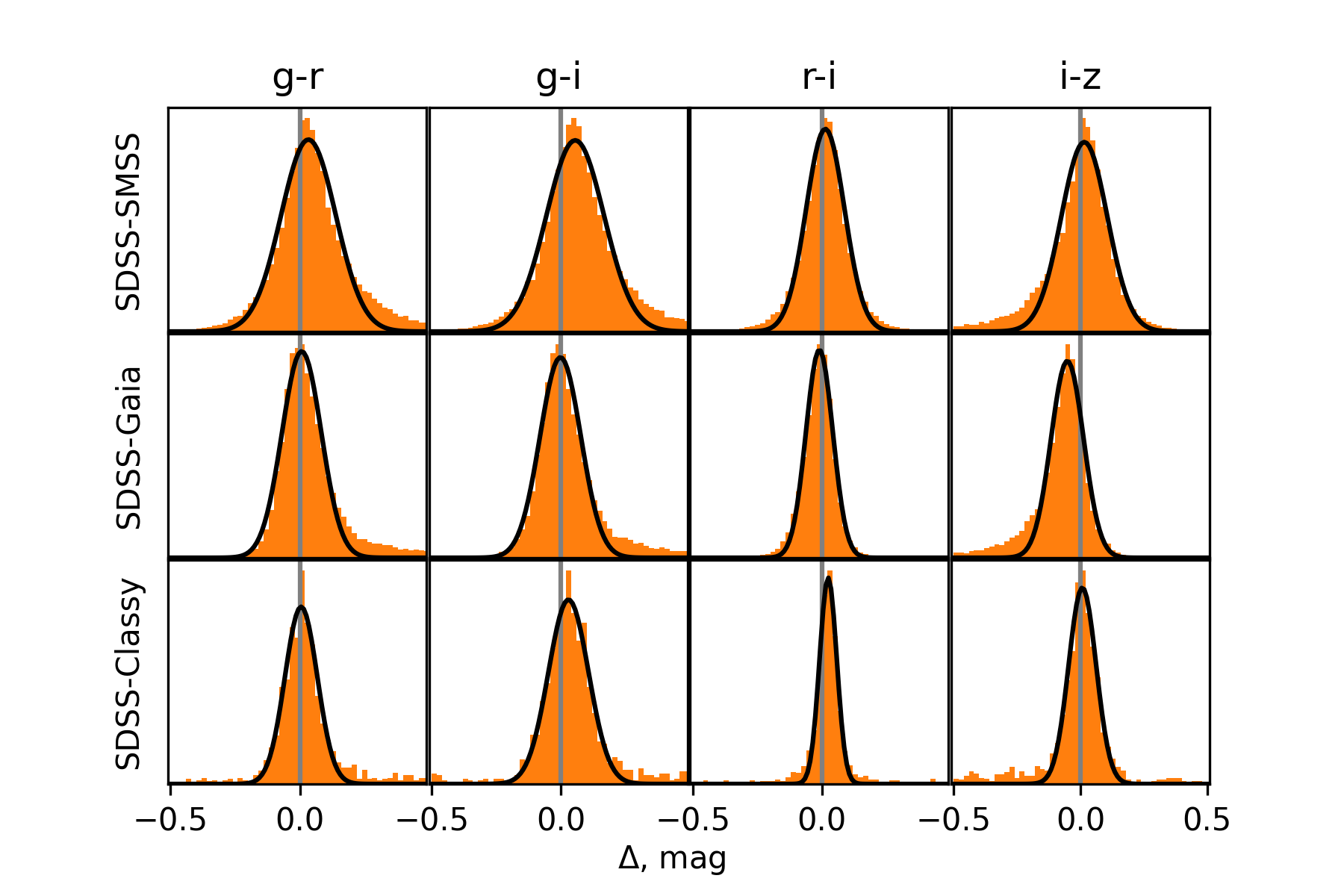}
  \caption{
    Distribution of color differences between the \smss, \gaia, and \classy with respect to the \sdss data set, using asteroids commonly found in these data sets.   
  The distribution was fitted with a Gaussian curve, represented by the black line. The central gray vertical line denotes the zero offset.
    }
  \label{fig:color_diff_distr}
\end{figure}

\begin{figure*}[t]
  \centering
  \includegraphics[width=1.0\hsize]{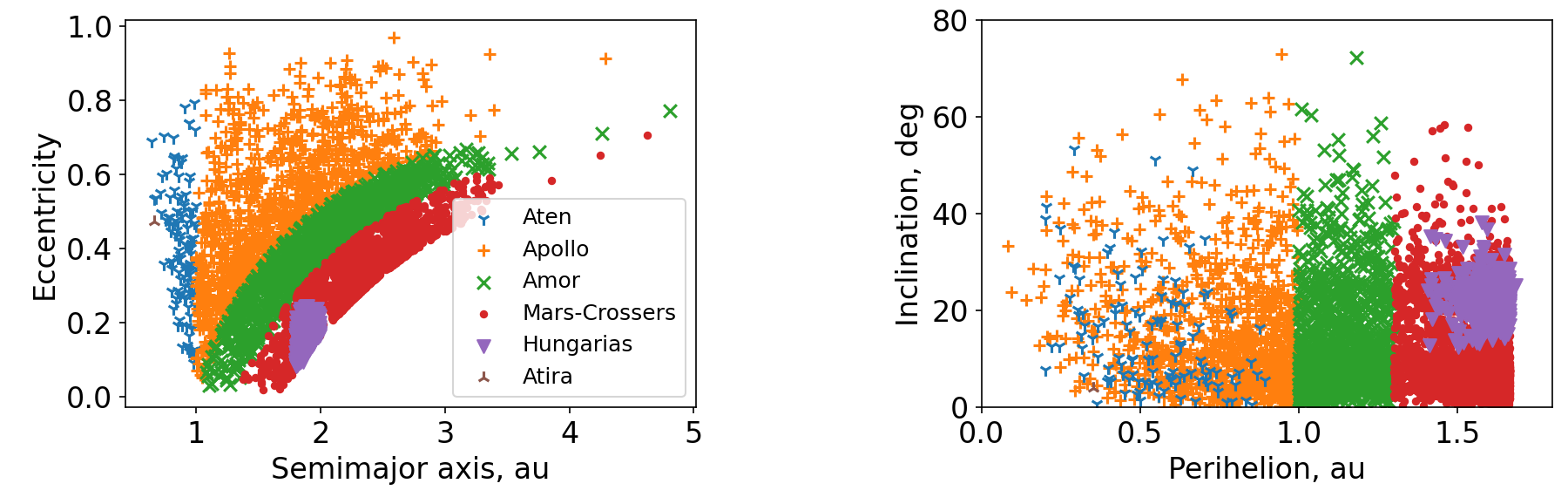}
  \caption{Distribution of the orbital elements of the NEOs, color-coded by dynamic class.
  }
  \label{fig:orbital}
\end{figure*}

\begin{table*}[t]
    \begin{tabular}{l|r|r|r|r|r|r|r|r}
        \hline
        \multirow{2}{*}{Sample} &
                \multicolumn{2}{c|}{\colorgr} &
                \multicolumn{2}{c|}{\colorgi} &
                \multicolumn{2}{c|}{\colorri} &
                \multicolumn{2}{c}{\coloriz} \\
                \cline{2-9}
                & \multicolumn{1}{c|}{Difference}
                & \multicolumn{1}{c|}{$n$}
                & \multicolumn{1}{c|}{Difference}
                & \multicolumn{1}{c|}{$n$}
                & \multicolumn{1}{c|}{Difference}
                & \multicolumn{1}{c|}{$n$}
                & \multicolumn{1}{c|}{Difference}
                & \multicolumn{1}{c}{$n$} \\
        \hline
        \sdss-\smss &$0.033\pm0.106$&54283&$0.056\pm0.112$&52546&$0.013\pm0.074$&59242&$0.016\pm0.091$&35252 \\
        \sdss-\gaia &$0.007\pm0.076$&27158&$-0.002\pm0.079$&27043&$-0.010\pm0.052$&28455&$-0.051\pm0.064$&24768 \\
        \sdss-\classy & $0.005\pm0.066$ & 1807 & $0.029\pm0.078$ & 1796 & $0.024\pm0.033$ & 1843 & $0.008\pm0.054$&1734\\
        \hline
        \\
        \end{tabular}

    \caption{The mean and standard deviation of the color difference between \sdss 
    and the other samples. The number of 
    asteroids in each of the samples is also reported. We limit the \sdss sample
    to asteroids with uncertainties below 0.1 mag.}
    \label{tab:color_diff_common}
  \end{table*}

Before merging the four catalogs of colors, we
checked for systematic differences in colors and uncertainties
among the four data sets.
To do this, we did not restrict the comparison to NEOs, but used 
all of the available asteroid colors from the entire four data sets:
\numb{400,894} for \sdss,
\numb{139,220} for \smss,
\numb{\gaiaTotAst} for \gaia, and
\numb{\clsTotAst} for ground-based (\classy).

We cross-matched the asteroid colors from the other sources
to the \sdss, which contains the largest number of asteroids and is used as a reference here.
We found
\numb{67,921},
\numb{28,948}, and
\numb{1,951} asteroids in common for the \smss,
\gaia, and \classy catalogs, respectively.
We then computed the color difference between the \sdss and the other catalogs.
The distribution of these differences were normal for all pairs of filters and catalogs,
with mean values close to zero (\Autoref{fig:color_diff_distr}).
The spread (standard deviation) of these differences reflects
a combination of several effects:
the measurement uncertainties of each catalog
(either magnitudes or spectra), the potential effect of asteroid rotation
\citep[due to the non-simultaneous acquisition 
of asteroid images in different filters; see, e.g.,][]{2018AA...609A.113C} 
and observations at different phase angles
\citep{2012Icar..220...36S, gaia3-spectra, 2020A&A...642A..80C}.

The detailed results of this comparison are presented in \Autoref{tab:color_diff_common}.
There are small systematic offsets between catalogs on average, 
much smaller than their standard deviation but larger than the standard 
error ($\sigma/\sqrt{n}$, where $n$ is the number of observations).
For instance, \smss matches \sdss with an average 
\colorgr color difference of \numb{0.033} mags and a standard deviation of 
\numb{0.106} magnitude. This was determined using \numb{44,005}
shared \colorgr color measurements that had an error of less than 0.1 magnitude.
Although the systematic offset is three times smaller than the standard deviation,
the standard error is approximately \numb{0.0005}. 
Therefore, these systematic biases were corrected by adding the precomputed offsets for each color before merging the data sets.

\begin{figure}[t]
  \centering
  \includegraphics[width=1.0\hsize]{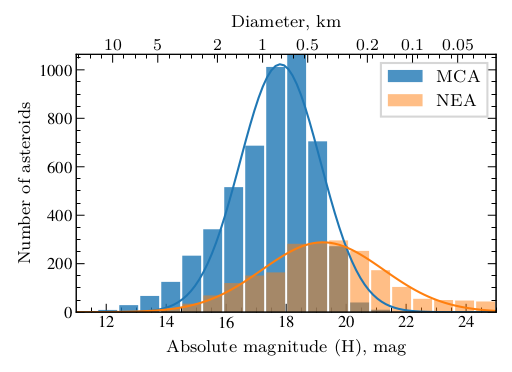}
  \caption{Distribution of absolute magnitude of MCs (blue) and 
  NEAs (orange). The diameter
  scale is a guideline, computed with an average albedo of \numb{0.24}.
  }
  \label{fig:abs_diameter}
\end{figure}

As visible in \Autoref{fig:color_diff_distr}, 
the width of the color difference distributions
is largest between \sdss and \smss, 
because both catalogs have the largest color uncertainties.
Once the color difference between the catalogs is corrected,
the standard deviation can be independently computed as 
\begin{equation}
\sigma_{\sdss-\smss} =\sqrt{\sigma_{\sdss}^2+\sigma_{\smss}^2}.
\end{equation}
We present a detailed comparison of the color differences and 
uncertainties in \Autoref{app:error}.
Based on this analysis, we note that some uncertainties 
are either over- or under-estimated (e.g., \gaia and \smss, respectively), 
and we applied multiplicatively correcting factors to select the best color value
 between identical asteroid color measurements in the catalogs 
 (see \Autoref{tab:individual_error}).

\subsection{Merging data sets\label{sec:full_catalog}}

\begin{figure*}[t]
  \centering
  \includegraphics[width=1.0\hsize]{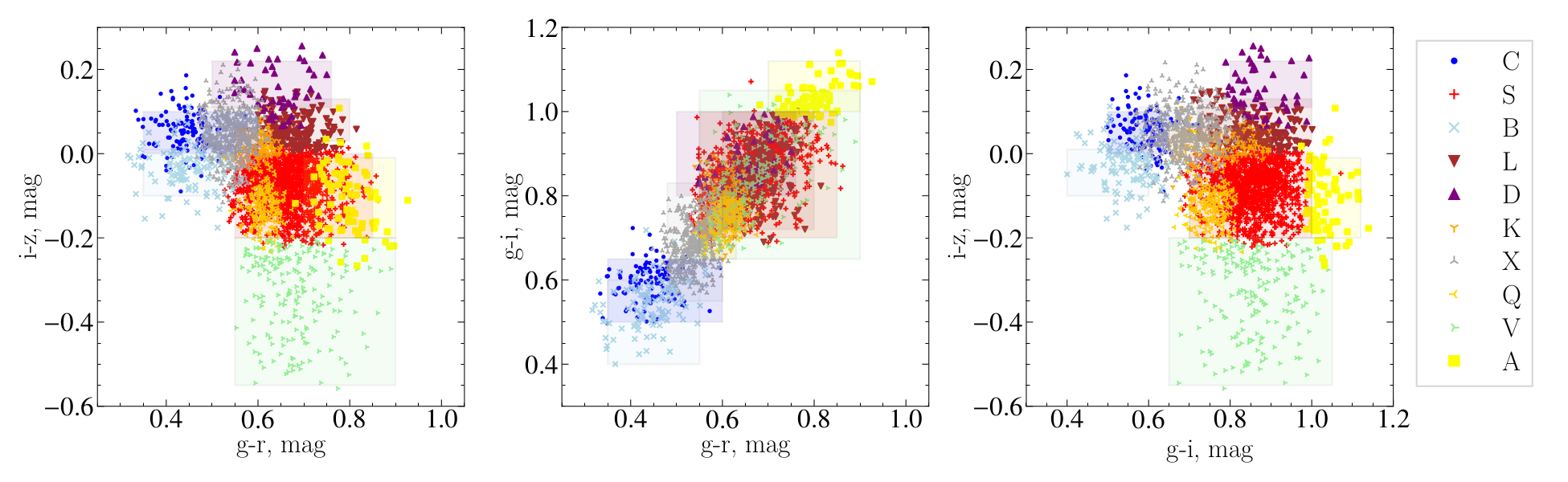}
  \caption{Color-color distribution NEOs with a taxonomy probability 
  above 0.2, color-coded by taxonomic classes.
  }
  \label{fig:taxo_grgiiz}
\end{figure*}

\begin{figure*}
  \centering
  \includegraphics[width=1.0\hsize]{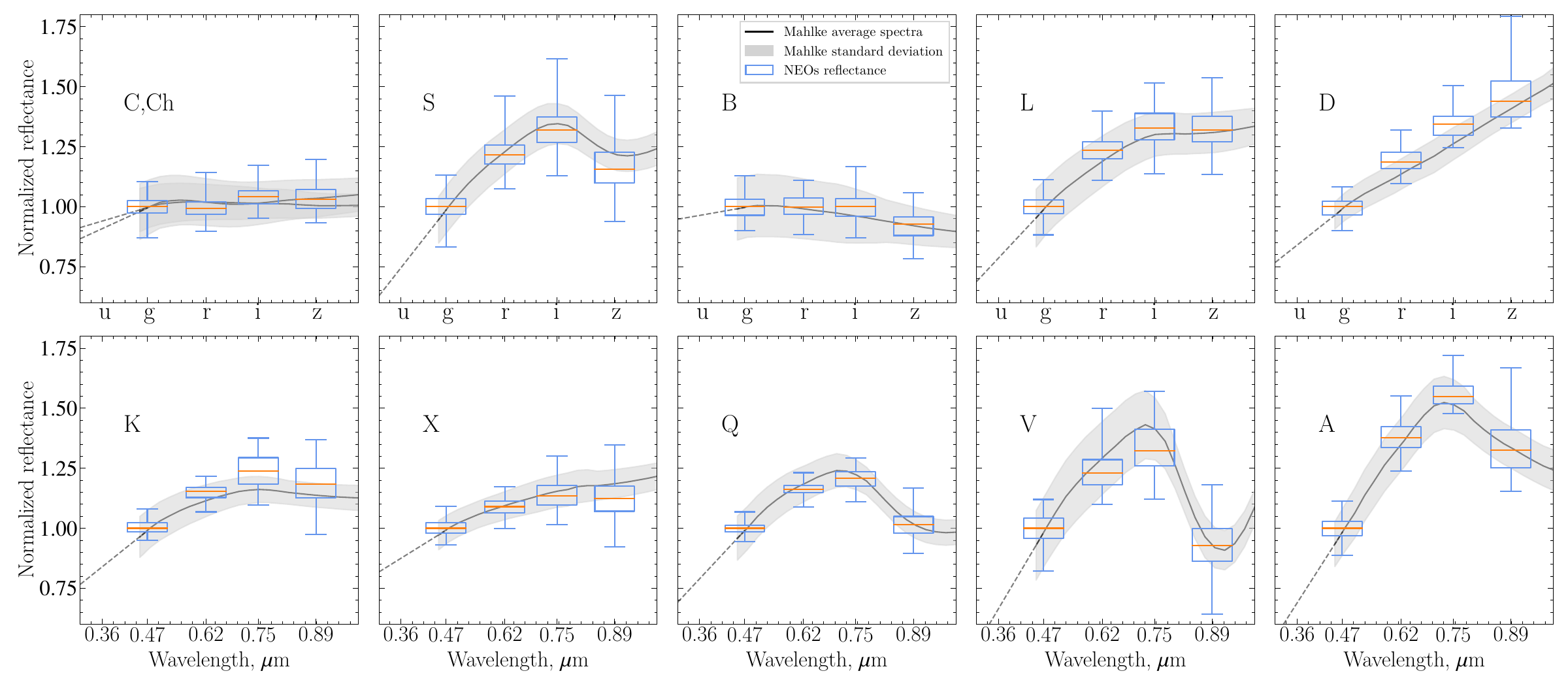}
  \caption{Pseudo-reflectance spectra of asteroids 
    based on their \colorgr, \colorgi, and \coloriz colors.
    The distribution of values for each band is
    represented by whiskers (95\% extrema, and the 25, 50, and 75\% quartiles).
    For each class, we also represent
    the associated template spectra of the \citet{2022A&A...665A..26M} taxonomy.        
  }
  \label{fig:grgiiz_reflectance}
\end{figure*}

We merged the four data sets based on asteroid designation
(we used the \rocks\footnote{\url{https://rocks.readthedocs.io}} interface
to the name resolver of 
\ssodnet\footnote{\url{https://ssp.imcce.fr/webservices/ssodnet/}}, 
see \citealt{ssodnet}).
Each catalog contains NEOs that have not been measured in the others.
The most prolific source is the \sdss, which contains \numb{4,398} unique NEOs,
followed by \sm, with \numb{964} unique NEOs.
The \classy and \gaia catalogs contain \numb{507} and \numb{199} unique NEOs, respectively.

For NEOs present in more than one catalog,
the color with the smallest uncertainty is selected.
This results in a catalog of \numb{\TotAst} NEOs
(i.e., NEAs and MCs)
with at least one color measurement,
which we call \neorocks.
We collected the ancilllary parameters of each asteroid in our
sample with \ssodnet, including orbital elements and albedo,
for instance. 
The description of the catalog is presented in \Autoref{sec:cat_desc}.

We present in \Autoref{fig:orbital} 
the orbital distribution of the \neorocks sample and
detail in \Autoref{tab:dynclass} the dynamical classes,
including \numb{2277} NEOs (Aten, Amor, Apollo, and Atira) and
\numb{5124} MCs. We also included the Hungarians that, owing to their eccentricity, have a perihelion within the orbit of Mars in the MC sample.

The absolute magnitudes in the \neorocks catalog
extracted from the virtual observatory
Solar System open database network (\ssodnet) \citep{2023A&A...671A.151B}
show a bimodal distribution (\Autoref{fig:abs_diameter}), resulting
from the typical larger distance of MCs compared with NEAs.
The average absolute magnitude of the NEAs is $19.2\pm2.0$,
while it is $17.8\pm 1.3$ for MCs.
Assuming an albedo of 0.24 for all NEOs results in 
an average diameter of 
$0.40^{+0.61}_{-0.24}$\,km for NEAs and
$0.76^{+1.34}_{-0.35}$\,km for MCs,
covering a complessive range from $\approx$ 10km down to 50m.
We chose this albedo as it is the mean albedo of S-type 
asteroids \citep{2022A&A...665A..26M}, the
most represented taxonomic class among NEOs
(\Autoref{sec:disc} and, e.g., \citet{2019Icar..324...41B}).

\begin{table}[]
  \centering
    \begin{tabular}{lr}
        \hline \hline
        Dynamical class & Number \\
        \hline
        Mars-Crossers &  4,380 \\
        Amor          &  1074 \\
        Apollo        &  1078 \\
        Hungarias     &   744 \\
        Aten          &   124 \\
        Atira         &     1 \\
        \hline
        Total         &  7,401 \\
        \hline
        \\
    \end{tabular}
  \caption{Distribution of NEOs among dynamic classes.
  }
  \label{tab:dynclass}
\end{table}

\section{Taxonomy\label{sec:taxo}}

Taxonomy is a convenient way to summarize observations into a simpler set
of labels that describe categories of objects that share the same properties.
Asteroid taxonomy is based on the spectral signatures of the light
reflected by the surface \citep[e.g.,][]{2015aste.book..151B,  2015aste.book...43R}.
Widely used asteroid taxonomy schemes include those of 
\citet{1984-PhD-Tholen}, using visible colors and albedo,
and \citet{2009Icar..202..160D}, using visible and near-infrared spectrum
\citep[itself an extension of][based on the visible spectrum]{2002-Icarus-158-BusI}.
These have recently been unified into a taxonomy using both
visible and near-infrared spectra and albedos
\citep{2022A&A...665A..26M}.

\begin{figure}[h]
  \centering
  \includegraphics[width=\hsize]{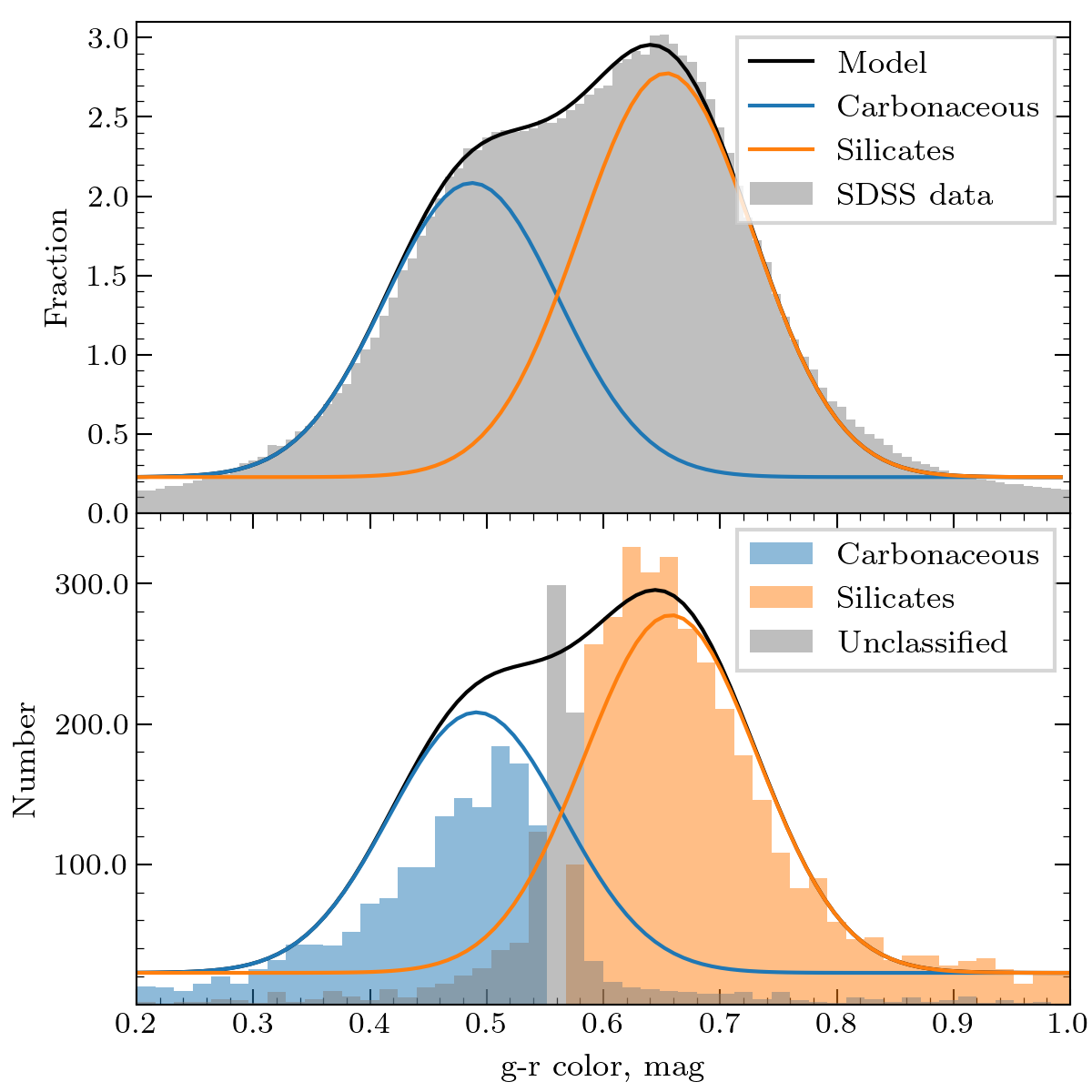}
  \caption{Distribution of g-r colors in SDSS asteroids and taxonomic categorization of NEOs.
   Top: Color distribution of one million asteroids obtained from
   the \sdss \citep{SergeyevCarry2021} data set modeled by 
   fitting a mixture of two Gaussians
   (represented by the black line). 
   The two main taxonomic classes, 
   silicate (depicted in orange) 
   and carbonaceous (depicted in blue), 
   were represented by the model.
  \newline
  Bottom: Distribution of \colorgr colors and the taxonomy of NEOs analyzed using the two-component mixture model of the two primary 
  classes in the \sdss data set (shown by lines). 
  The carbonaceous and silicate taxonomy complexes are represented 
  by blue and orange, respectively. 
  Unclassified asteroids, 
  where the probability of belonging to each complex is comparable, 
  are represented in gray.
  }
  \label{fig:gr_taxo_distr}
\end{figure}

\subsection{Classification of multi-color NEOs}

We used the same approach as earlier works on photometry, deriving consistent
classification with spectroscopy \citep[e.g.,][]{2013Icar..226..723D,
2018-AA-617-Popescu, SergeyevCarry2021}. We converted reference spectra into 
colors (\Autoref{app:gaia_color})
and used them to define the taxonomic class in the photometry space.
To determine the taxonomic class of each asteroid, 
we employed the probabilistic approach of \citet{Sergeyev2022}, 
which involves computing the intersection between the volume 
occupied by the color (with uncertainty) of an object and the 
regions of each taxonomic class.
We updated the regions to match the recent taxonomy by
\citet{2022A&A...665A..26M} instead of using the templates from Bus-DeMeo
\citep{2009Icar..202..160D} and computed the probability for each asteroid belonging to 
each of the ten broad taxonomy complexes: A, B, C, D, K, L, Q, S, V, and X. 

The final taxonomy for each asteroid was selected based on the most probable
 taxonomic complex. We also provided the second-highest probability 
 taxonomic complex. Asteroids with a likelihood of less than 10\% fitting into any taxonomy complex were labeled as U (unclassified). (\Autoref{tab:cat_color_ind}).

We present in \Autoref{fig:taxo_grgiiz} the color-color distribution of \numb{2341} NEOs
for which taxonomy is predicted with a probability higher than 20\%.
This constraint was selected to avoid the visual overloading of the figure.
The distribution follows the reported color distribution of asteroids
in the \sdss filter system \citep{2005Icar..173..132N, 2008-Icarus-198-Parker,
2016Icar..268..340C}.
We also present a comparison of pseudo-reflectance spectra based on the photometry of
our sample with the template spectra of the taxonomic class from 
\citet{2022A&A...665A..26M} in 
\Autoref{fig:grgiiz_reflectance}. 
The correspondence of the \sdss median spectra with the template spectra confirms the
chosen taxonomy boundaries. 
The method provides a reliable way to determine the taxonomic classification of NEOs using photometry data. 
With the increasing number of NEOs discovered every year, 
it is becoming increasingly important to be able to classify these objects accurately and efficiently. 
Spectroscopy is the most accurate method for determining asteroid taxonomy, but it is time-consuming and requires a significant amount of telescope time. 
On the other hand, photometry data can be obtained much more efficiently, 
making it a more practical choice for large-scale surveys.

\subsection{Classification based on a single color} \label{sec:gr_taxonomy}

\begin{figure}
  \centering
  \input{figs/confisuin_matrix_gr.pgf}
  \caption{Confusion matrix illustrating the correlation between predicted single-color (\colorgr) taxonomy outcomes and the results of a three-color taxonomy (\colorgr, \colorgi, \coloriz). This matrix displays the fractions of true positives, false positives, true negatives, and false negatives.}
  \label{fig:confusion_gr}
\end{figure}

\begin{figure}
\centering
\includegraphics[width=1\hsize]{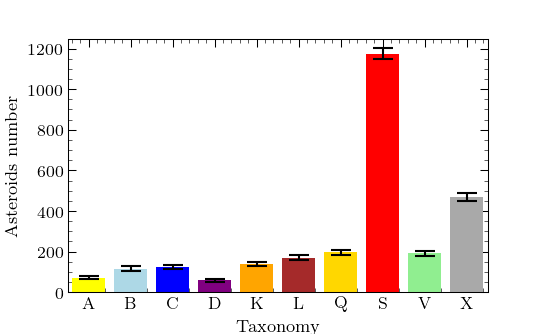}
\caption{NEO taxonomy distribution computed by 
(\colorgr, \colorgi, \coloriz) color indexes.
}
\label{fig:NEOs_complex_distr}
\end{figure}

\begin{figure*}
  \centering
  \includegraphics[width=1.0\hsize]{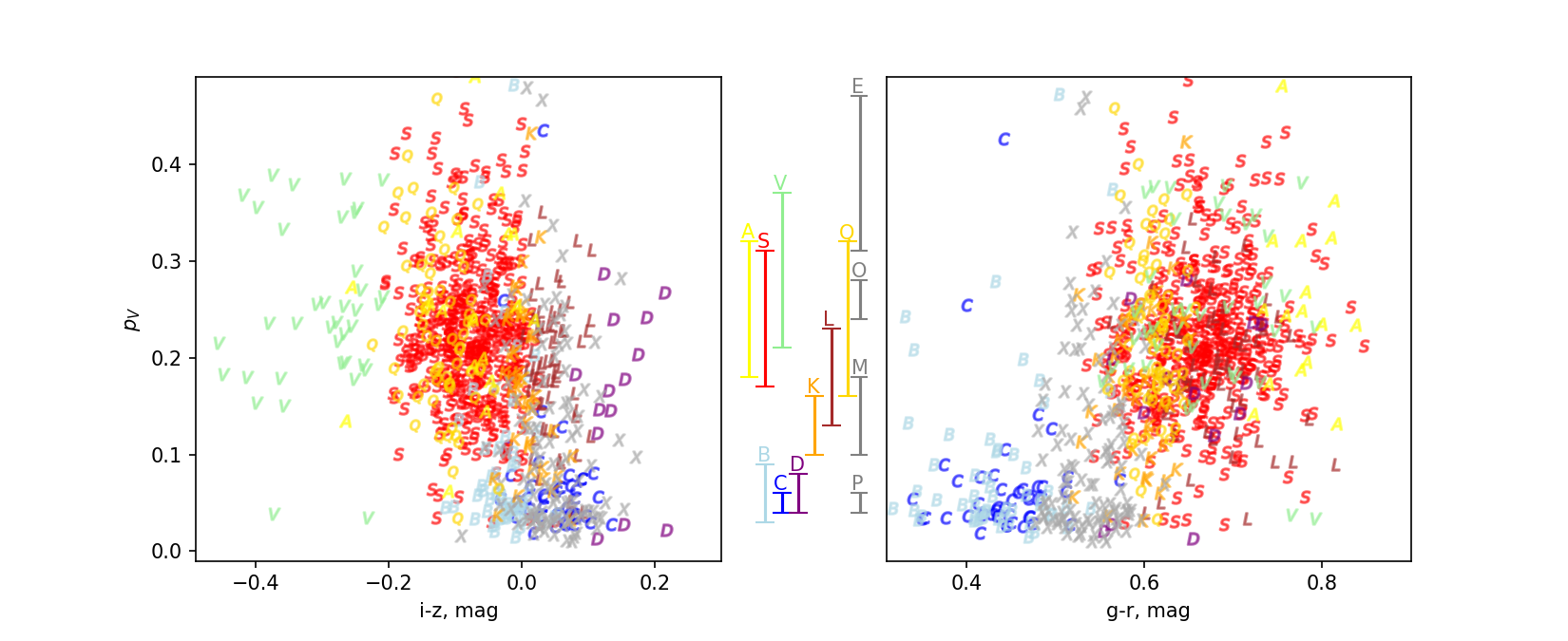}
  \caption{Colors and albedo of NEOs. 
    Taxonomy is marked by colored letters
    (same color-code as in Fig.~\ref{fig:taxo_grgiiz}).
    Vertical ranges between the panel indicate
    the one sigma range of albedo for each taxonomic class.
    \citep{2022A&A...665A..26M}.
    }
  \label{fig:albedo}
\end{figure*}

Many observations in the present data set have a significantly better 
signal-to-noise ratio in the \filtg and \filtr filters. Furthermore,
some of the asteroids from the \smss sample only have \colorgr color.
Thus we also classified asteroids from this single color.
We utilized the \colorgr color of one million asteroid observations from
\citet{SergeyevCarry2021} to build a reference distribution.
We fitted this distribution with two normal distributions,
corresponding to two wide complexes (carbonaceous, $C_1,$ and silicates, $S_1$).
We used these two distributions to compute the probability that a NEO
belongs to each wide complex, based on its \colorgr color.
Whenever the difference between the probabilities was smaller than 20 percent,
we marked these asteroids as unclassified.
We present the \colorgr color distribution of NEOs in \Autoref{fig:gr_taxo_distr}.
It is of course a cruder classification than the classification based
on three colors. However, it allows for discrimination between
``\textsl{red}'' (S, A, V, L, and D) and ``\textsl{blue}'' objects (C and B) in
a manner similar to \citet{2020ApJS..247...13E}.
A significant number of the unclassified asteroids belong to
the X complex, while the remainder are of the D- and K- asteroid types.
(\Autoref{fig:confusion_gr}).
Although a taxonomy based on a single color may appear limited, 
we present in \Autoref{fig:confusion_gr} the confusion matrix
between the one- and three-color classes. The $C_1$ and $S_1$
classes accurately separate asteroids belonging to the C complex
from those displaying an absorption band of around 1 micron (which are redder:
K types, L types, and S complex).

As a final step, we merged the taxonomy obtained with three colors
(\colorgr, \colorgi, and \coloriz) and that with a single color only (\colorgr).
The former is preferred over the latter (\Autoref{tab:cat_color_ind}).
If neither approach could classify an asteroid, 
we set the classification method to ``none.''

\subsection{Distribution of taxonomy and albedos}

The prevalence of S types is striking (\Autoref{fig:NEOs_complex_distr}). 
It is notable that the distribution presented here is influenced 
by the selection function of the observations, 
which introduces a bias, mainly due to the fact that the surveys used here
are magnitude limited, which will impact different taxonomic classes
of different albedos \citep{2013Icar..226..723D, 2022AJ....163..165M}.
The albedo is an important characteristic related to the composition
of asteroids \citep{1989aste.conf.1139T, 2022A&A...665A..26M}.
For instance, asteroids in the  B, C, and D classes have low albedos
(below 10\%) while mafic-silicate-rich asteroids
(e.g., A, Q, and S types) have albedos around 0.24.
The main advantage of taking the albedo into account is the 
possibility to split the degenerate X complex into
high albedo E-type asteroids (albedo above 0.30),
moderate albedo M (metallic) asteroids, and the
"dark" P asteroids (below 0.10).

We used \ssodnet \citep{ssodnet} to retrieve the albedo of the NEOs
in our data set for a consistency check.
In \Autoref{fig:albedo} we compare the \coloriz and \colorgr colors of
\numb{898} NEOs that have estimated albedo values.
There is an overall agreement between the range of albedos 
for the different taxonomic complexes, although outliers are visible.
These outliers are a consequence of either
misclassifications or biased albedos \citep{2021PSJ.....2...32M}, or both.
Mismatches occur mainly in classes with highly different
albedos but similar colors, such as D- and L-type asteroids
(here, some D types have albedos around 0.2, more consistent with L types).

The albedo distribution of X types reveals that approximately 45\% of them are actually
P types. The fraction of M types is approximately 45\% and the remaining 10\%
are high-albedo E-type asteroids \citep{2013-ApJ-762-Usui}.
However, P-type asteroids are very similar 
to C-type asteroids in both color and albedo, and can therefore be misclassified.

\subsection{Comparison with previous surveys}

\begin{figure}[t]
    \centering
    \includegraphics[width=1\hsize]{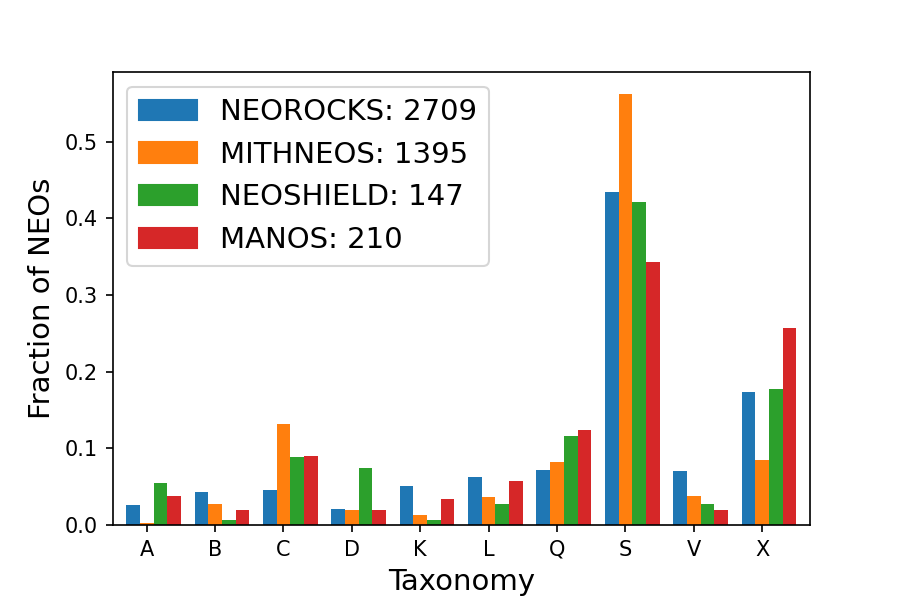}
    \caption{Comparison of the distribution of taxonomic classes of the NEOROCKS
    sample computed from (\colorgr, \colorgi, \coloriz) color indexes with the MITHNEOS, NEOSHIELD, and MANOS spectral surveys.
    }
    \label{fig:compare_taxo}
\end{figure}

  We compared the distribution of taxonomic 
  classes of the present \neorocks sample with the three previous main
  spectral surveys of NEOs:
  MITHNEOS \citep{2019Icar..324...41B}, 
  NEOSHIELD \citep{2018PSS..157...82P}, and
  MANOS \citep{2019AJ....158..196D} (see \Autoref{fig:compare_taxo}).
  The \neorocks sample overlaps almost completely with 
  the NEOSHIELD-2 and MANOS catalogs because 
  the \classy data include all available ground-based spectral observations. 
  The overlap with MITHNEOS is limited to approximately half of this catalog, for which
  a majority of spectra only cover the near-infrared range.
  While differences are visible \citep[and partly expected owing to the
  size dependence of taxonomic distribution; e.g.,][]{2019AJ....158..196D}, 
  we note an overall agreement with the different data sets. 

\begin{figure}[t]
    \centering
    \includegraphics[width=\hsize]{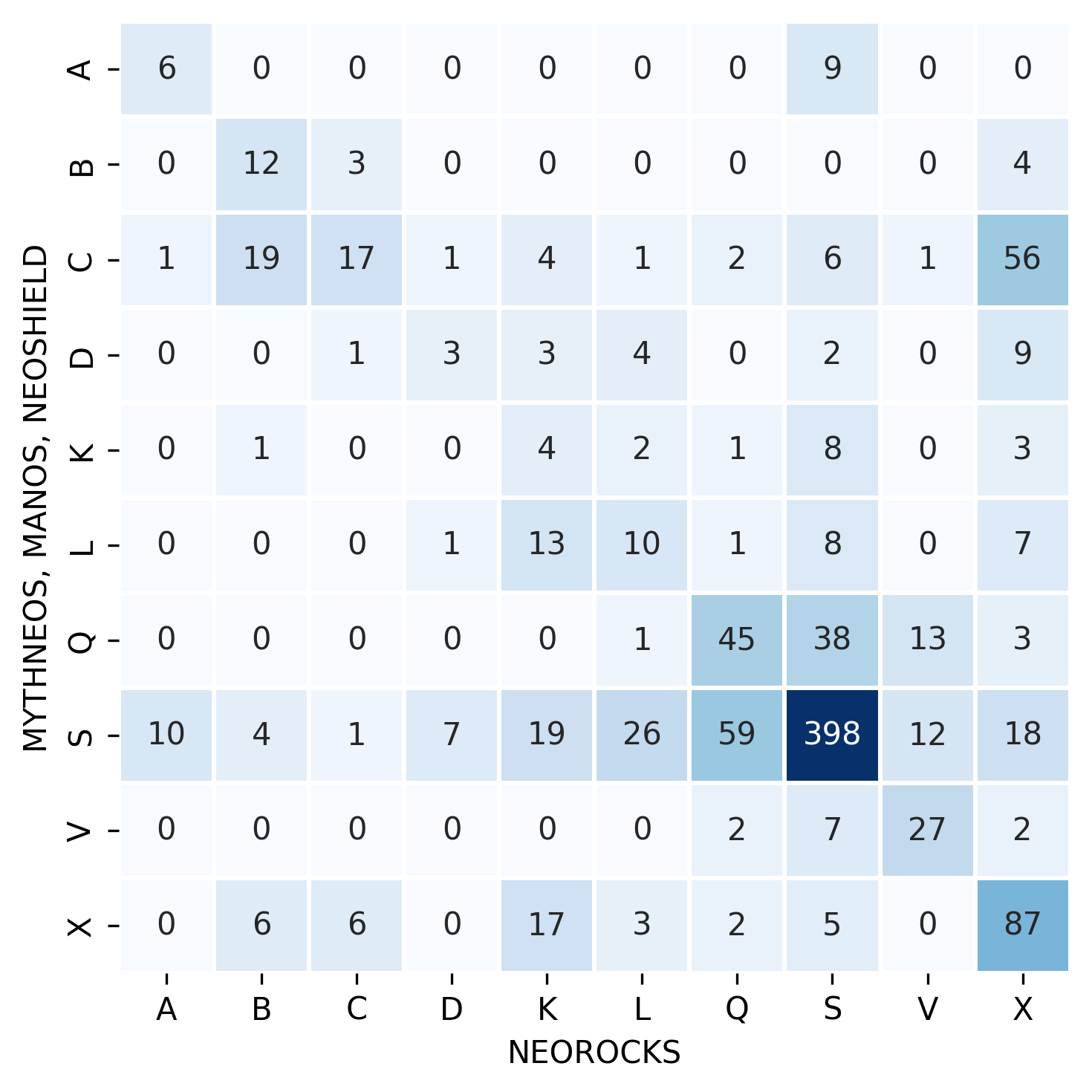}
    \caption{Comparison of the \neorocks
     NEO taxonomy with the MITHNEOS, MANOS, and NEOSHIELD-2
     catalogs.}
    \label{fig:confusion_griz}
\end{figure}

  The confusion matrix presented in \Autoref{fig:confusion_griz}   
  indicates that there is a high level of agreement in the taxonomic
   classification of S-, V-, and X-type asteroids. 
  However, some confusion is observed among the less common classes
  in the NEOs population, particularly K versus L and (A, L, Q) versus S. 
  Additionally, a significant number of C-type asteroids were classified
  as part of the wide X asteroid complexes, which also include
  P-type asteroids that share similar photometry and albedo properties
  with C-type asteroids. This highlights both the strengths and limitations 
  of using broadband colors as the basis for taxonomic classification.

\section{Targets accessible to space missions\label{sec:space}}

As opposed to other domains in astrophysics, the Solar System can almost
be considered as a close neighborhood. Distances are small enough that
we have sent space probes (some of which returned), providing ground-truths
for Earth-based studies and leading to great discoveries, such as
satellites of asteroids \citep{1995-Nature-374-Chapman, 1995Natur.374..785B},
the asteroid-meteorite link \citep{2006Sci...312.1330F, 2011-Science-333-Yurimoto},
and cryo-volcanism \citep[on Ceres,][]{2014Natur.505..525K, 2016Sci...353.4286R},
for instance.

Since the 1990s, opportunities to encounter an asteroid during 
an interplanetary mission have been considered, and 
dynamical studies have been conducted to find candidates for 
potential flyby missions
\citep[e.g.,][]{1990Icar...87..372D, 2022P&SS..21605476A}.
These candidates are often at the origin of characterization
efforts to select the actual target of the flyby and
prepare the spacecraft operations during the short encounter
\citep[e.g.,][]{1999A&A...352..697D, 2010-AA-523-Carry}.
As a result, there have been almost as many encounters (seven)
during opportunity flybys\footnote{%
  (21) Lutetia (Rosetta),
  (243) Ida (Galileo),
  (253) Mathilde (NEAR Shoemaker),
  (951) Gaspra (Galileo), 
  (2867) {\v S}teins (Rosetta), 
  (9969) Braille (Deep Space 1),
  (5525) Annefrank (Stardust).}
as targeted encounters with asteroids\footnote{%
  (1) Ceres (Dawn),
  (4) Vesta (Dawn), 
  (433) Eros (NEAR Shoemaker),
  (4179) Toutatis (Chang'e),
  (25143) Itokawa (Hayabusa),
  (65803) Didymos (DART), 
  (134340) Pluto (New Horizons),
  (162173) Ryugu (Hayabusa2),
  (486958) Arrokoth (New Horizons),
  (101955) Bennu (OSIRIS-REx).
  } (ten).

We searched in the present \neorocks data set for any candidate of upcoming
space missions \citep[e.g., NASA JANUS, JAXA Hayabusa-2 extension,][]{2020LPI....51.1965S,
2022cosp...44.3264Y} and found many
objects (\Autoref{tab:hera}) listed as flyby candidates for the ESA Hera mission
\citep[approximately one hundred, see][]{2020EPSC...14.1064F}.

\begin{table}[h]
  \centering
    \begin{tabular}{lrlcr}
        \hline
        Designation &  Number & Dyn.class & Taxo &  Prob \\
        \hline
        1995 OR     &   42532 & MB>Inner   & D &  -- \\
        2000 HJ89   &   54212 & MB>Inner   & V & 0.72\\
        2001 TJ72   &   88992 & MB>Inner   & S & 0.17\\
        Francismuir &   95802 & MB>Inne    & K & 0.84\\
        \hline
        Etiennemarey &   3456 & MB>Inner   & M & 0.95\\
        Gorlitsa     &   3818 & MB>Inner   & C & 0.98\\
        1981 EW30    &  10278 & MB>Inner   & S & 0.62\\
        2000 CC33    &  14710 & MB>Inner   & S & 0.24\\
        1998 WS9     &  49352 & MB>Inner   & M & 0.88\\
        1996 HL21    &  79317 & MB>Middle  & V & 0.03\\
        2000 EP110   &  86616 & MB>Inner   & S & 0.84\\
        2000 NF22    & 118687 & MB>Inner   & X & 0.93\\
        2001 UH40    & 125107 & MB>Inner   & K & 0.28\\
        2004 FQ111   & 128338 & MB>Inner   & S & 0.73\\
        2003 AQ28    & 151682 & MB>Inner   & S & 0.54\\
        2003 CB7     & 151738 & MB>Inner   & S & 0.90\\
        2001 QU65    & 189092 & MB>Inner   & V & 0.35\\
        2006 DR115   & 245739 & MB>Inner   & S & 0.57\\
        2008 EZ75    & 263476 & MB>Inner   & B & 0.21\\
        2008 FK125   & 274163 & MB>Inner   & X & 0.27\\
        2008 UE268   & 309745 & MB>Inner   & S & 0.51\\
        2007 UF127   & 355419 & MB>Middle  & S & 0.16\\
        2005 YX13    & 388155 & MB>Inner   & S & 0.08\\
        2012 AE1     & 392704 & NEA>Apollo & V & 0.17\\
        2008 FL108   & 431739 & MB>Inner   & C & 0.22\\
        2013 YQ49    & 479408 & MB>Inner   & V & 0.18\\
        2015 PT9     & 515878 & MB>Inner   & S & 0.28\\
        2011 HF9     &     -- & MB>Inner   & C & 0.36\\
        2013 LG2     &     -- & MB>Inner   & V & 0.06\\
        2014 JE85    &     -- & MB>Inner   & S & 0.09\\
        \hline
    \\
    \end{tabular}
  \caption{Flyby candidates of the ESA Hera mission. At the top of the table are candidates from the shortlist targets, and at the bottom, the candidates from the longlist targets.}
  \label{tab:hera}
\end{table}

  \begin{figure}
    \centering
    \includegraphics[width=1.0\hsize]{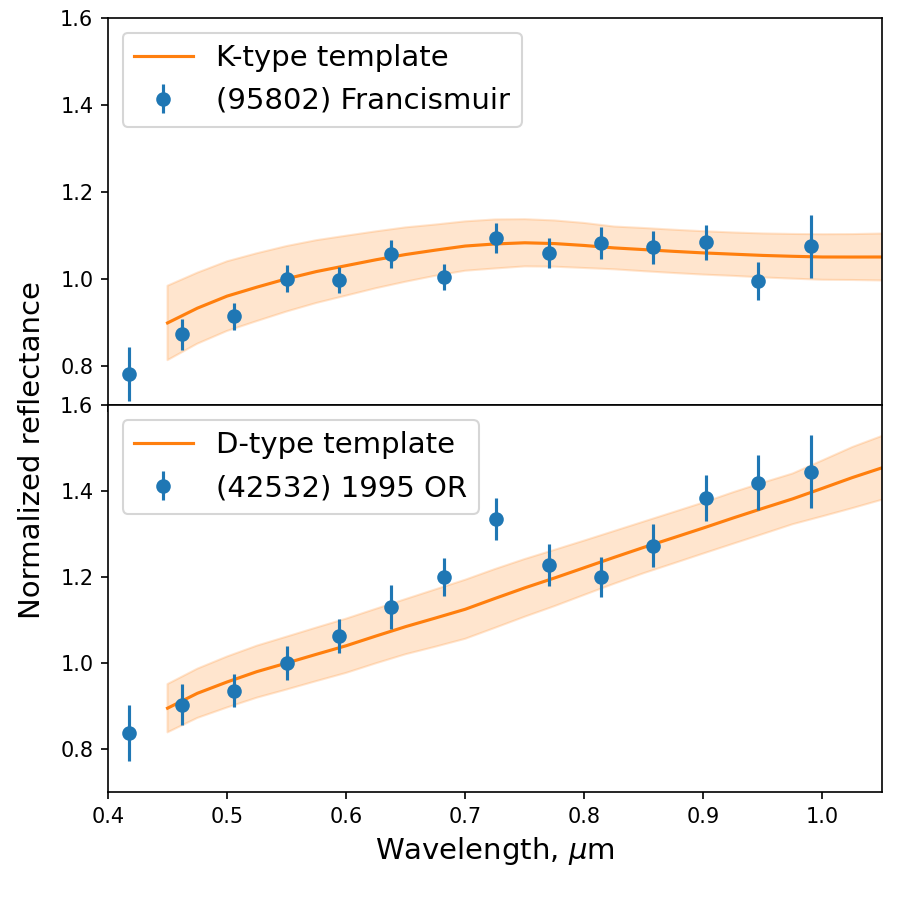}
    \caption{\gaia reflectance spectra of asteroid 
    (95802) Francismuir and (42532) 1995 OR, flyby candidates of 
    the ESA Hera mission. The orange line shows pre-computed reflectance 
    templates and their uncertainties \citep{2022A&A...665A..26M} for 
    P-type asteroids (top) and D-type asteroids (bottom).
    }
    \label{fig:hera}
  \end{figure}

  A critical parameter in selecting a space mission target is
  the amount of energy required to reach it. This quantity is
  often expressed as the total change of velocity, $\Delta v$.
  We collected $\Delta v$ computed and provided by
  L. Benner\footnote{\url{https://echo.jpl.nasa.gov/lance/delta_v.rendezvous.html}} and for NEOs in our \neorocks catalog with a
  $\Delta v$ $<$ \SI{6.5}{\kilo\meter}, the typical 
  $\Delta v$ required for a mission to Mars.
  We present in \Autoref{tab:deltav} the taxonomy of these
  \numb{42} mission-accessible NEOs.
  We also provide an analysis of the spectrum for the flyby candidate (10278) Virkki in \Autoref{sec:virkki}.

\begin{table}[t]
    \centering
    \begin{tabular}{lcccl}
        \hline
        Designation &  $\Delta V$ & Complex &  Prob & Dyn. Class \\
        {} &  km/sec  &  &   &  \\
        \hline
        2004 EU22   &      4.4 &       D &      0.55 &   Apollo \\
        1998 SF36   &      4.6 &       S &      0.96 &   Apollo \\
        2015 DP155  &      4.7 &       V &      0.81 &     Amor \\
        2008 DG5    &      4.8 &       S &      0.81 &   Apollo \\
        1996 GT     &      5.2 &       S &      1.00 &     Amor \\
        1994 CN2    &      5.2 &       S &      0.69 &   Apollo \\
        2001 SW169  &      5.3 &       S &      0.64 &     Amor \\
        1997 WT22   &      5.3 &       S &      0.96 &     Amor \\
        2002 LJ3    &      5.3 &       S &      0.96 &     Amor \\
        1973 EC     &      5.4 &       L &      0.96 &     Amor \\
        2006 UP     &      5.4 &       S &      0.67 &     Amor \\
        1982 HR     &      5.5 &       V &      1.00 &   Apollo \\
        1999 VG22   &      5.5 &       S &      0.75 &     Amor \\
        1980 PA     &      5.7 &       V &      1.00 &     Amor \\
        2010 WY8    &      5.7 &       S &      0.62 &     Amor \\
        2003 RB     &      5.7 &       S &      0.99 &     Amor \\
        2002 XP40   &      5.7 &       S &      1.00 &     Amor \\
        2001 FC7    &      5.8 &       X &      0.58 &     Amor \\
        1993 QA     &      5.9 &       S &      0.69 &     Amor \\
        2008 KZ5    &      6.0 &       S &      0.98 &     Amor \\
        1977 VA     &      6.0 &       X &      1.00 &     Amor \\
        2005 YY36   &      6.1 &       X &      0.56 &     Amor \\
        2001 WL15   &      6.1 &       S &      0.76 &     Amor \\
        2001 UA5    &      6.1 &       S &      0.54 &   Apollo \\
        A898 PA     &      6.1 &       S &      1.00 &     Amor \\
        2010 LJ14   &      6.2 &       Q &      0.55 &     Amor \\
        2007 VY7    &      6.2 &       V &      0.68 &   Apollo \\
        1998 KU2    &      6.3 &       B &      0.94 &     Amor \\
        1982 DV     &      6.3 &       S &      0.75 &     Amor \\
        2002 KL6    &      6.3 &       V &      0.97 &     Amor \\
        2000 JS66   &      6.3 &       S &      0.52 &   Apollo \\
        1929 SH     &      6.3 &       S &      1.00 &     Amor \\
        2005 RO33   &      6.3 &       S &      0.52 &     Amor \\
        2002 PG80   &      6.3 &       S &      0.62 &     Amor \\
        2001 FD90   &      6.3 &       V &      0.59 &     Amor \\
        1993 VW     &      6.3 &       V &      0.80 &   Apollo \\
        1981 CW     &      6.3 &       S &      0.64 &     Amor \\
        2004 VB     &      6.3 &       S &      0.97 &   Apollo \\
        2006 SV19   &      6.4 &       Q &      0.93 &     Amor \\
        2018 NB     &      6.4 &       S &      0.81 &     Amor \\
        2015 DV215  &      6.4 &       V &      0.58 &   Apollo \\
        2007 SJ     &      6.4 &       S &      0.57 &   Apollo \\
        \hline
        \\
        \end{tabular}
        
    \caption{Mission-accessible NEOs ($\Delta v$ $<$ \SI{6.5}{\kilo\meter})
      with a taxonomy probability above 0.5.
      }
    \label{tab:deltav}
\end{table}

\section{Discussion\label{sec:disc}}

We used the derived colors and taxonomic classes to address several topics.
In Section~\ref{ssec:sw}, we discuss the space weathering for the NEOs in the
S complex.
We then present the distribution of 
A types in 
Section~\ref{ssec:AX}.
We finally discuss the taxonomic distribution of small asteroids in the 
source regions of NEOs in Section~\ref{ssec:src}.

\subsection{Space weathering\label{ssec:sw}}

The surface of atmosphereless bodies in the Solar System is
aging from micro-meteorite impacts and ions of the solar
wind, commonly referred to as space weathering \citep{2004AREPS..32..539C}.
Space weathering changes the properties of the top-most
surface layer \citep[nanometer thick,][]{2011Sci...333.1121N},
as function of exposure (age and heliocentric distance) and composition.
Thanks to laboratory experiments \citep[e.g.,][]{2001Natur.410..555S,
2005Icar..174...31S,2006Icar..184..327B},
the effect of space weathering on mixtures of olivines and pyroxenes
(such as A, S, and V types) is well understood \citep{2015-AsteroidsIV-Brunetto}:
it reddens and darkens surfaces.
Its effects on the reflectance of more primitive material linked with
carbonaceous chondrites (such as B- and C-types)
is less straightforward, with both blueing and reddening as possible
outputs \citep{2017Icar..285...43L, 2018-Icarus-302-Lantz}.

In the case of S types, the effect is expected to be very fast, changing
ordinary chondrite-like material (the Q types) into S types in less than 
a million years \citep{2009-Nature-458-Vernazza}.
The presence of Q types among asteroids implies that their surfaces
are young. Considering the short timescale for space weathering
\citep[longer than the timescale to be injected from the Main Belt,][]{1997Sci...277..197G},
some rejuvenating mechanisms must be present \citep{2012-MNRAS-421-Marchi}.

Q-type asteroids were originally found among NEOs only, 
so planetary encounters were proposed as a rejuvenation mechanism
 \citep{2005Icar..173..132N, 2010Icar..209..510N, 2010Natur.463..331B}.
However, this early observation was due to an observing bias:
the fraction of Q increases toward smaller diameters, which
are harder to observe at larger distances
\citep{2012-Icarus-219-Thomas, 2016Icar..268..340C}.
As space weathering is a continuous process (ultimately resulting
 in asteroids being classified into two groups: S and Q),
the observed trend of shallower slopes among S/Q asteroids
with smaller diameters explains this bias, and
can be explained by a resurfacing due to landslides or failure
linked with Yarkovsky–O'Keefe–Radzievskii–Paddack (YORP) spin-up \citep{2018-Icarus-304-Graves}.

Recently, \citet{2019Icar..322....1G} tested another mechanism for rejuvenation
among NEOs:
a cracking mechanism due to thermal fatigue \citep{2010GeoRL..3718201V, 2014Natur.508..233D}.
Based on almost 300 NEOs \citep[from][]{2004Icar..169..373L, 
2005MNRAS.359.1575L, 2004Icar..170..259B},
this model explains the overall behavior of spectral slope against
perihelion, which was apparently misinterpreted as being linked to
planetary encounters.

\begin{figure}[t]
    \centering
    \includegraphics[width=\hsize]{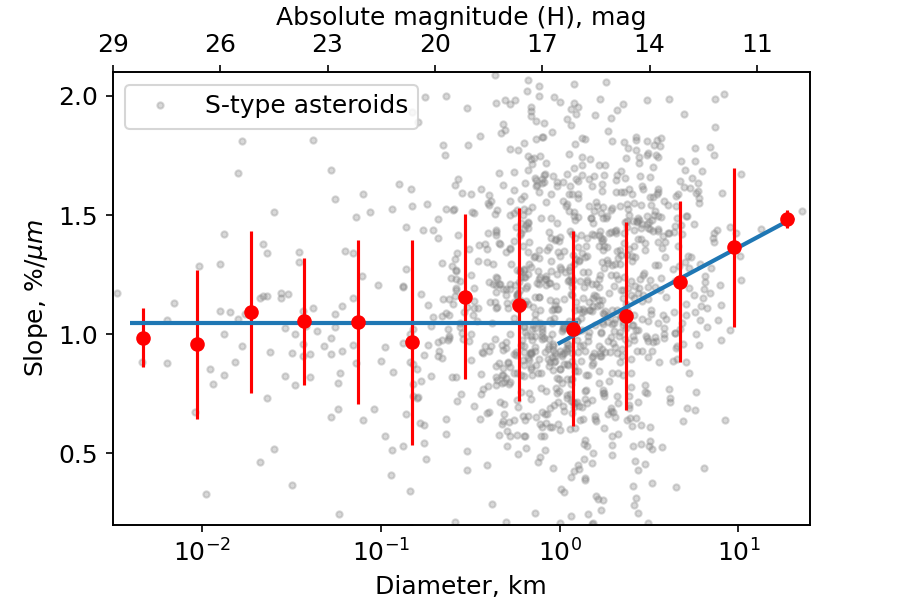}
    \caption{Spectral slope of S types as a function of asteroid diameters 
    (gray points),
    the weighted average in logarithmic size bins shown by red points.
    Weights were estimated by color uncertainty.
    }
    \label{fig:s_slope_diameter}
\end{figure}

\begin{figure}[]
  \centering
  \includegraphics[width=1.0\hsize]{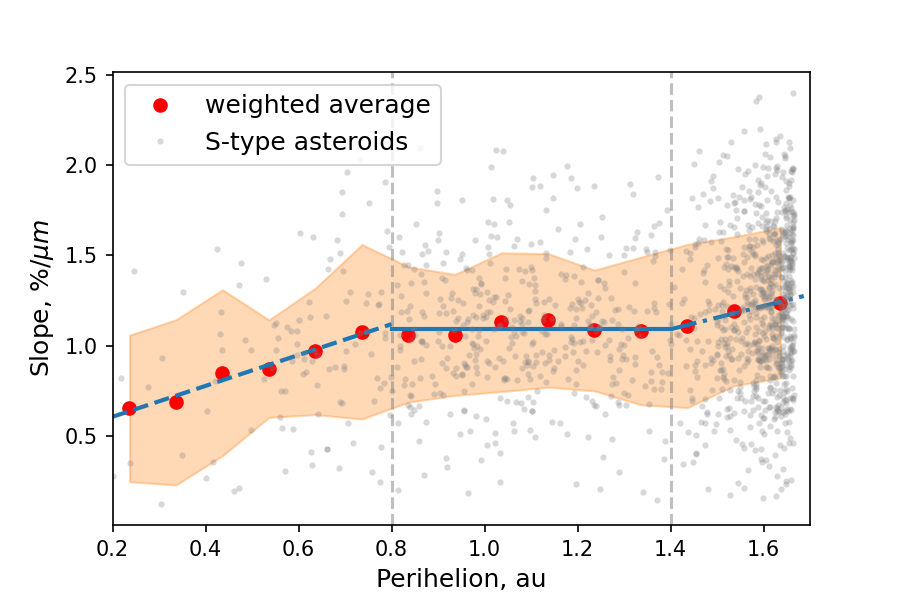}
  \caption{Spectral slope against perihelion for S types.
  Red dots and the shaded area are the running average and deviation, and
  blue lines are linear regressions on the running average.
  Although the entire sample presents a large spread, the running
  average shows two kinks.
  }
  \label{fig:s_slope_perihelion}
  \end{figure}

\begin{figure*}[!]
  \centering
    \input{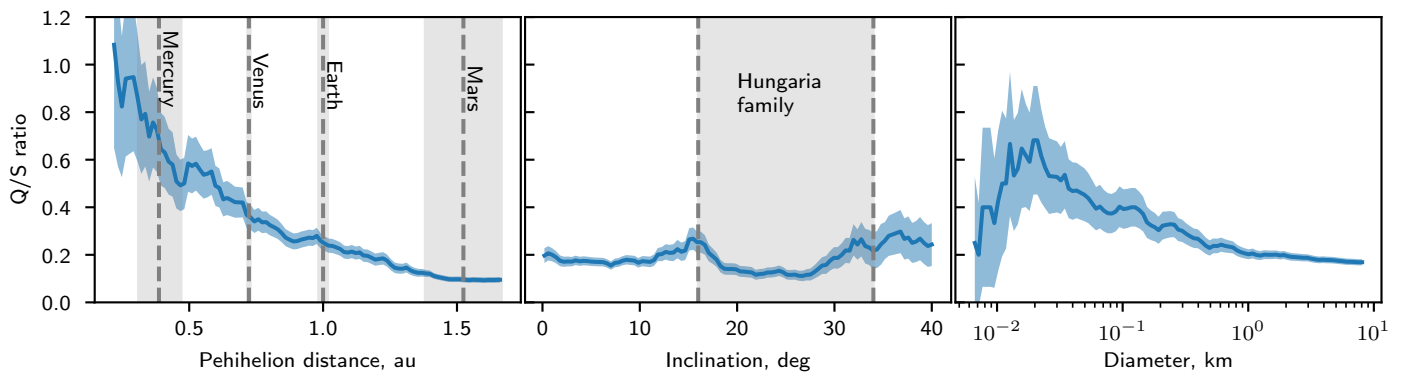}
    \caption{Running mean of the ratio between the number of Q and S asteroids as a
  function of perihelion, inclination, and diameter.
  Shaded areas correspond to the uncertainties considering Poisson statistic for the Q/S ratio.
  }
  \label{fig:qs_ratio}
\end{figure*}

\begin{figure*}[]
  \centering
    \input{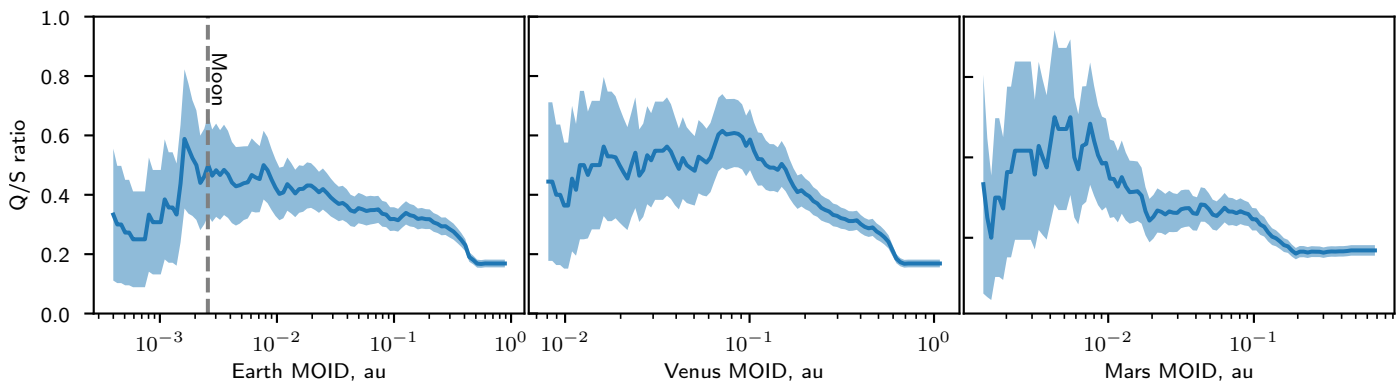}
    \caption{Running mean of the ratio between the number of Q and S asteroids as 
  a function of their MOID with the Earth, Venus, and Mars.
  }
  \label{fig:moid}
\end{figure*}

In the present section, we use the large \neorocks catalog
to address the question of space weathering.
Our sample contains \numb{1,175} S-type and \numb{196} Q-type
asteroids whose taxonomy is based on 
three colors with a probability higher than 0.2. 
We chose to use both the taxonomic types (i.e., the Q/S ratio) and the spectral
slope as indicators of space weathering. The former highlights the fraction
of very fresh surfaces in the sample, while the latter is more nuanced, with the 
weathering creating a continuous trend from blue to red surfaces.

We first studied the size dependence of space weathering of S-type asteroids. 
We present in \Autoref{fig:s_slope_diameter} their spectral slope 
(computed over the \filtg and \filti filters, expressed in \%/$\mu m$ 
consistently with reflectance spectroscopy) against their diameter.
A similar plot for the Q/S ratio is provided in \Autoref{fig:qs_ratio}.
The diameter of the asteroids ($D$) was estimated using their known
absolute magnitude ($H$) via the equation 
$D=1329\cdot p_V^{-0.5}\cdot 10^{-0.2H}$ 
\citep{2002aste.book..205H} and assuming an albedo of S-type asteroids $p_V=0.24$. 
The slope of S-type asteroids is constant for asteroids smaller than approximately 1--5\,km, 
and increases for larger asteroids.
This is consistent with the previous report by \citet{2004Icar..170..259B}.
Such behavior was indeed already reported \citep[e.g.,][]{2012-Icarus-219-Thomas, 
2023Icar..38915264D} and 
explained by resurfacing through YORP spin-up and failure \citep{2018-Icarus-304-Graves}.
The decrease in the Q/S ratio for the smallest NEOs may be attributed
to the increasing number of monoliths, for which resurfacing may be difficult.

In \Autoref{fig:s_slope_perihelion}, we present the relationship between the spectral 
slope of S-type asteroids and their perihelion. Our analysis shows a probable trend of 
increasing spectral slope with a more distant perihelion, which is consistent with 
the findings of previous studies \citet{2019Icar..322....1G}. 
The spectral slope 
remains constant until approximately 1.3--1.4 AU, beyond which it again increases.
As noted by \citet{2019Icar..322....1G}, this last behavior is likely 
an observing bias: the farther away the asteroids, the less we observe 
small diameters, and the fraction of fresh surfaces is not constant 
with diameters \citep{2016Icar..268..340C, 
2018-Icarus-304-Graves}.
A spectral slope value variation is 0.86$\pm$0.07(\%/$\mu$m)/AU 
from 0.2 to 0.8 AU and is 0.64 $\pm$ 0.07(\%/$\mu$m)/AU 
beyond 1.4 AU. Our analysis shows that within the orbit of Venus, 
the spectral slope is higher than previously estimated 
by \citet{2019Icar..322....1G}, who reported a value 
of 0.52 $\pm$ 0.21\%/$\mu$m/AU.

This behavior is also visible in the fraction of Q and S types
(\Autoref{fig:qs_ratio}). There is a strong correlation between
the Q/S ratio and the perihelion distance, with the fraction of Q types
increasing across a wide range of distances from 0.2 to 1.6 AU.
A similar trend was observed by \citep{2019AJ....158..196D}, 
who compared the perihelion distribution of \numb{138} S-type NEOs 
to that of \numb{178} NEOs, including \numb{91} Sq and \numb{87} Q subtypes 
for perihelions ranging from 0.7 to 1.0. Outside this range, however,
their data showed a flat behavior.
The recent study by \citep{2023Icar..38915264D} presents an almost linear 
trend of increasing Q-type asteroid fraction with decreasing 
perihelion in an interval from 0.5 to 1.3 AU, 
very similar to our result presented here.

We then tested the level of space weathering against planetary encounters, 
using the minimum orbit insertion
distance\footnote{Retrieved from the Minor Planet Center (MPC)}
(MOID) as an indicator of the proximity to the planets 
\citep[following, e.g.,][]{2010Natur.463..331B}.
The Q/S ratio is shown as a function of MOID for the
Earth, Venus, and Mars in \Autoref{fig:moid}.
While there is a trend of increasing fractions of Q-type asteroids toward smaller MOIDs,
it happens at distances too far to be due to the planetary encounter
and apparently is the result of the correlation with the perihelion distance.
\citep{2016Icar..268..340C, 2019Icar..322....1G}. For the Earth, it
even drops for MOIDs below the lunar distance, counterintuitively
(a similar situation occurs for Mars).
We note that here we use the current MOID of each NEO, while \citet{2010Natur.463..331B}
argued in favor of probing the dynamical history of individual objects
(which is beyond the scope of the present analysis).

We finally tested the ratio of Q to S types with the orbital
inclination. The ratio is overall flat, with a shallow peak around
15\degr and an increase above 30\degr.
The slightly decreasing fraction of Q asteroids 
in the inclination range of 15-35\degr~corresponds to the 
inclination range of the Hungarias and Phocaeas.
The maximum Q/S ratio at 5\degr~reported by \citep{2023Icar..38915264D} on \numb{477} S types is 
\numb{three} times larger than that of our sample. 
This disparity may be attributed to differences in the asteroid samples
and variations in the techniques employed to distinguish between Q- and S-type asteroids.

The present sample contains \numb{1,371} S- and Q-type NEOs,
a factor of 2-3 larger than the previous
studies. We confirm the trend of decreasing spectral slope
(increasing fraction of Q-types) toward smaller diameters.
We did not detect a clear signature against orbital inclination.
There is a clear increase in the fraction of 
Q types with smaller perihelion (also visible in the decrease in 
spectral slope), pointing to a strong effect of thermal
fatigue in refreshing asteroid surfaces.

\subsection{Distribution of A types\label{ssec:AX}}

\begin{figure}[t]
    \centering
    \includegraphics[width=0.5\textwidth]{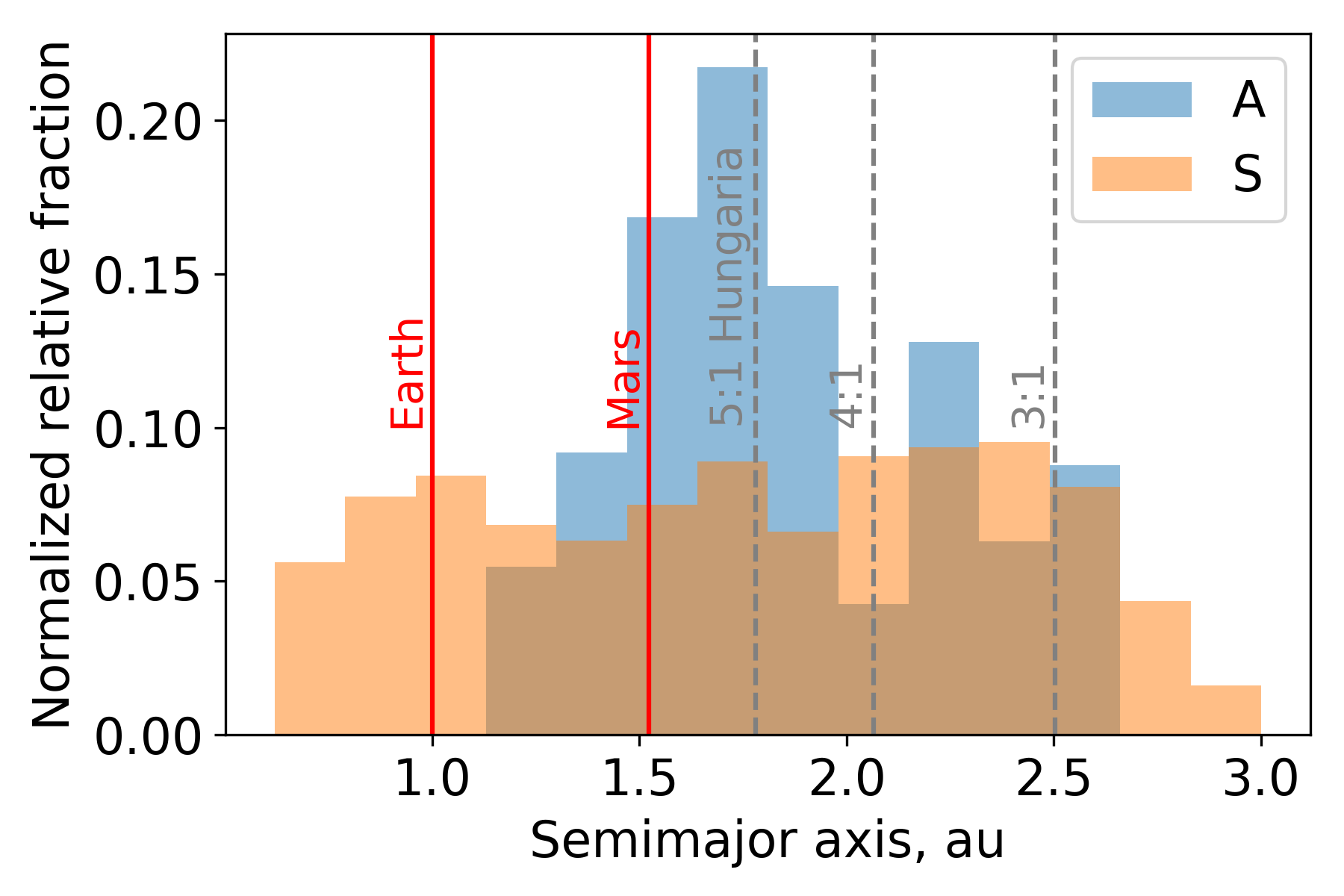}
    \caption{Relative distribution of A-types
    along the semi-major axis.
    }
    \label{fig:ax_dist}
\end{figure}

\begin{figure*}[!]
  \centering
  \includegraphics[width=1.0\textwidth]{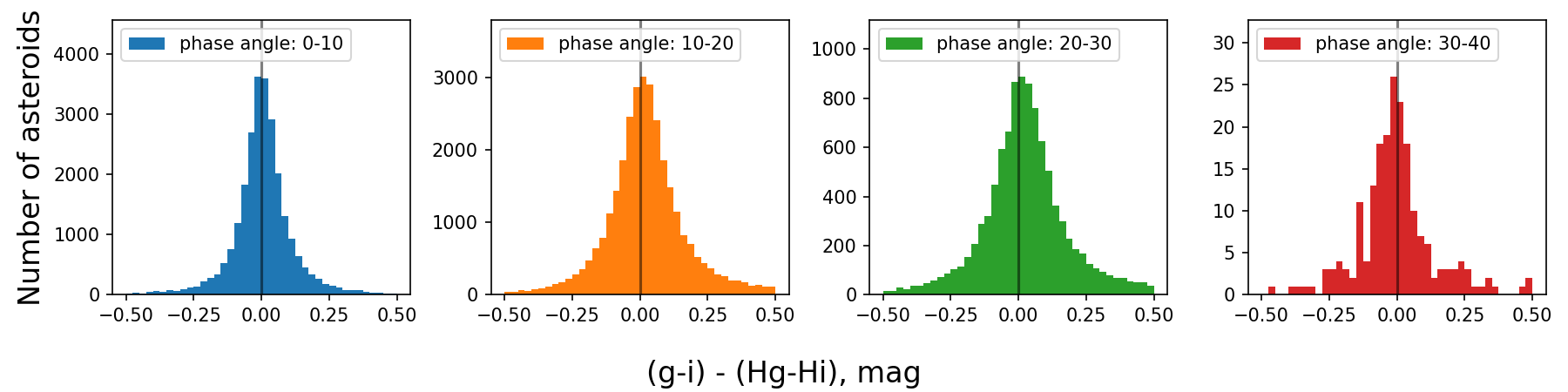}
  \caption{Distribution of the difference between the \sdss \colorgi asteroid colors and absolute magnitude (H) colors from \citep{2022A&A...657A..80A} as a function of phase angle.}
  \label{fig:phase_effect_hist}
\end{figure*}

A types are a rare type of asteroids in the Main Belt.
Their spectra exhibit a broad and deep absorption band around
1\,$\mu m$, indicating an olivine-rich composition \citep[e.g.,][]{2007Icar..192..434R}.
They have been thought to originate from the mantle of differentiated
planetesimals \citep{1984Sci...223..281C}, leading to the 
``missing mantle issue'' \citep{1996MPS...31..607B}.
The origin of A types is still debated \citep{2014Icar..228..288S, 2019Icar..322...13D}, although, the study of Mars Trojans indicates that certain A-type asteroids could be fragments that were ejected from Mars \citep{2017NatAs...1E.179P, 2021Icar..35413994C}.

The fraction of A-type asteroids by number in the entire Main Belt is estimated
at about 0.16\% and are believed to be homogeneously distributed \citep{2019Icar..322...13D}. 
We report here a fraction of \numb{2.5\,$\pm$\,0.2\%} A types among NEOs
(focusing on classifications with a probability higher than 0.5).
This much higher fraction of A types has already been reported by
both the MANOS and NEOSHIELD-2 surveys
\citep[Fig.~\ref{fig:compare_taxo}, from 1.7 to 5.5\%, ][]{2018MNRAS.477.2786P,
2019AJ....158..196D}.
While a classification based on visible wavelengths only may overestimate the
fraction of A \citep[misclassified from red S types owing to 
space weathering or observations at high phase angles,][]{2012Icar..220...36S},
the fraction of A types among NEOs appears
to be an order of magnitude higher than in the Main Belt.
Finally, \citet{2019AJ....158..196D} reported a concentration of A types
with a semi major axis close to that of Mars (1.5\,AU).

We present in \Autoref{fig:ax_dist} the fraction of A- and S-type NEOs
as a function of semi major axis.
While S types are evenly distributed, A types are concentrated between the orbit of Mars
and the 4:1 resonance with Jupiter, similar to the report by \citet{2019AJ....158..196D}.
Most A-type NEOs seem to be related to the Hungarias. In this region, the fraction of A-type asteroids 
increase by up to 4\%.
So, while the majority of the Hungarias are C and E types \citep{2014Natur.505..629D, 2019Icar..322..227L},
approximately 3\% of asteroids in this region are A types.

\subsection{The dependence of asteroid colors on phase angle}

The color of an asteroid is determined by the light it reflects, which is influenced by the composition of its surface material. However, the observed color of an asteroid can also change with the phase angle, which is the angle between the observer (usually Earth), the asteroid, and the Sun \citep{2000Icar..147...94B, 2015-AJ-150-Waszczak}. 
This change in color with phase angle is likely due to the way light scatters off the asteroid's surface. At higher phase angles, the light we see is more likely to have been scattered multiple times within the asteroid's surface before being reflected back to us. This multiple scattering can cause a redder object to appear bluer and vice versa, although this effect is only noticeable for phase angles of less than 7.5 degrees \citep{2022A&A...667A..81A}.
Considering the change in asteroid color with phase angle can be important for accurate taxonomy classification using color analysis techniques \citep{2022A&A...666A..77C}.

However, the exact mechanisms behind this color change with phase angle are still not fully understood and are an active area of research. The shape of the asteroid, its rotational state, and the macroscopic roughness of its surface can also influence the observed color and its change with phase angle \citep{2015-AA-580-Carvano}.

To investigate the impact of the phase effect on asteroid colors, we compared both the \sdss and \smss data sets with the absolute magnitude colors from the study by \citep{2022A&A...657A..80A}. The histograms of the \colorgi difference between the two data sets are shown in \Autoref{fig:phase_effect_hist}. 

We also selected asteroids that were observed at phase angles of greater than 20 degrees and had a phase difference of more than 5 degrees between observations. Subsequently, we determined the slope of the asteroid's \colorgi color as a function of phase angle. 
We found that color slope changes randomly and is comparable to the uncertainties in color. 

To investigate the trends among \textquote{red} and \textquote{blue} asteroids, we subdivided the asteroid data set into two groups based on their \colorgi colors. 
A red group is indicative of silicate asteroids, and a blue group is representative of carbonaceous asteroids. With this analysis, we did not catch any trends toward reddening or bluing within these subsets.
The random behavior of asteroid color slope indicates that more significant factors, such as the shape of the asteroid and uncertainties in photometry, may have a greater influence on the observed color and consequently on asteroid taxonomy.

Given that the phase effect could significantly alter the colors of asteroids only at large phase angles, and considering that our sample does not include NEOs observed at phase angles exceeding 40 degrees, we conclude that we cannot precisely predict and then correct the phase effect. Therefore, we did not take the phase effect into account in the color analysis of the NEOs data set.

\subsection{Source regions\label{ssec:src}}

\begin{figure*}[t]
    \centering
    \input{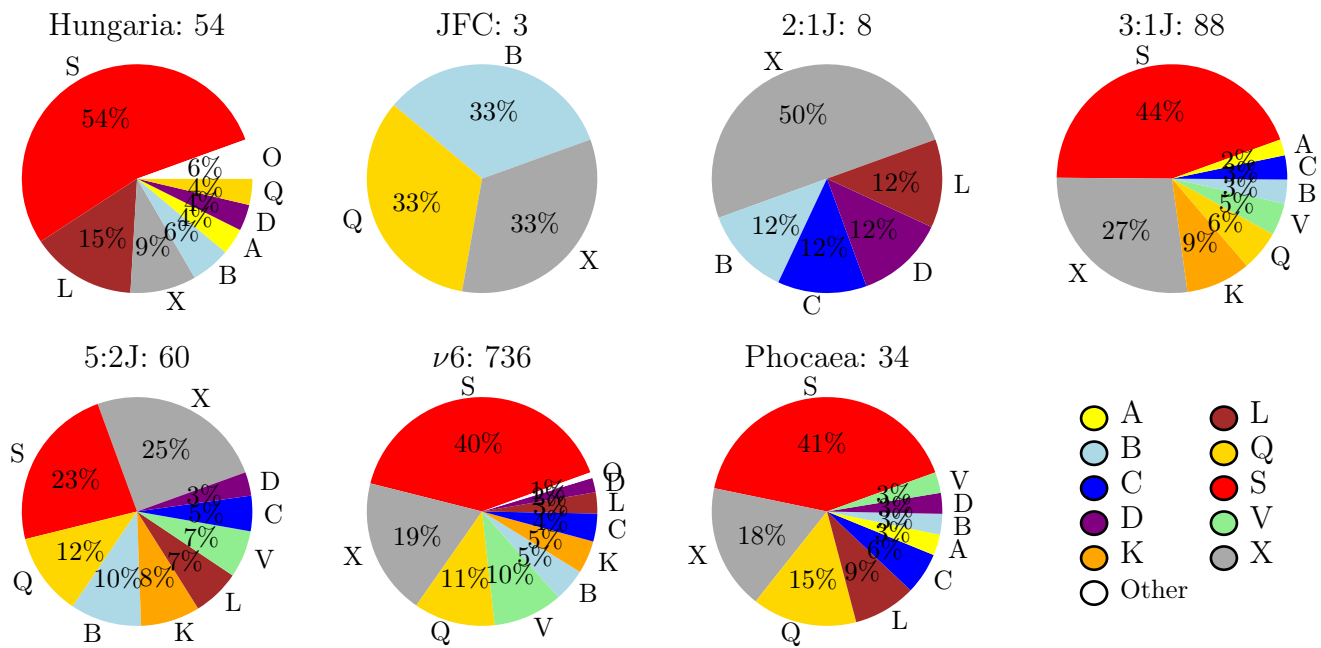}
    \caption{Taxonomic distribution of NEAs as per the seven-region model, previously calculated by 
    \citep{2018-Icarus-312-Granvik}
    }
    \label{fig:source_region}
  \end{figure*}

Investigating the orbital and size characteristics, as well as the origin 
of NEOs, is a crucial area of research in planetary sciences
\citep{2015-AsteroidsIV-Binzel, 2015aste.book..855A}. 
The dynamical pathway from the source regions to the planet-crossing space
is a crucial foundation for studying both individual NEAs and broader population-level 
questions. 
Understanding these distributions gives a holistic understanding of the dynamics,
origins, and potential risks associated with NEAs.

To deduce the probable origins of NEAs, we relied on what is known of their orbital properties in conjunction with previously simulated probabilities of seven-region\footnote{$\nu_6$ secular resonance,
2:1, 3:1, 5:2, mean-motion resonances (MMR) with Jupiter,
high inclination Phocaeas and Hungarias, and Jupiter family comets (JFC)} escape regions by \citet{2018-Icarus-312-Granvik}. 
We assigned each asteroid to its most probable origin area by employing a three-dimensional grid of orbital elements and a value of absolute magnitude as the fourth parameter. The grid includes semi major axis, a, eccentricity, e, and inclination, i, which was predicated on the calculations previously detailed in the research of \citet{2018-Icarus-312-Granvik}.
The orbital elements of these celestial bodies were obtained from the Minor Planet Center (MPC) database.

The most abundant source of NEOs is the $\nu_6$, which limits the 
inner border of the Main Belt. We predict it to be dominated by mafic-silicate-rich asteroids
(S, Q, V, see \Autoref{fig:source_region}).
The distribution of taxonomic classes is almost similar for the other source regions in the 
inner belt: the 3:1 MMR limiting the inner and middle belt, and the Phocaea and Hungaria
regions.
The fraction of mafic-silicate-rich asteroids decreases for source regions located further
from the Sun (5:2 and 2:1 MMR, JFC). These are dominated by opaque-rich asteroids
(B, C, D, see \Autoref{fig:source_region}).
Despite the observation biases (mainly related to albedo) and the relative low number
of NEOs predicted to originate from the outer regions, 
our results are in close agreement with \citet{2022AJ....163..165M}, in line
with the current understanding of taxonomic distribution \citep{2014Natur.505..629D}, but in a smaller size range.

\section{Conclusions\label{sec:conclusions}}

We combined a large sample of colors of planet-crossing asteroids, 
combining broadband photometry from the \sdss and \smss surveys
and reflectance spectroscopy from the ESA \gaia mission and
ground-based observations.
We determined the taxonomy of \numb{\TotAst} NEOs, with diameters
from approximately 10\,km to 50\,m. The sample is dominated by 
S-type asteroids (approximately 45\%), as occurs for other NEOs surveys. 
However, it is notable that the proportion of S types is 
overestimated due to observational bias.
We also report a much higher (up to 4\%) fraction of A types among NEOs
as compared to the Main Belt. These A types are concentrated
on a semi major axis between 1.5 and 2\,AU.
We confirm a strong
dependence of the spectral slope of S types with perihelion, based
on a sample of over one thousand objects. The distribution of slope
is consistent with the recently proposed rejuvenation model through thermal fatigue.

\begin{acknowledgements}
  This research has been conducted within the NEOROCKS
  project, which has received funding from the European Union's Horizon 2020
  research and innovation programme under grant agreement No 870403.
  The NEOROCKS team is composed by E. Dotto, M. Banaszkiewicz, S. Banchi, M.A. Barucci, F. Bernardi,
  M. Birlan, A. Cellino, J. De Leon, M. Lazzarin, E. Mazzotta Epifani, A. Mediavilla, D. Perna,
  E. Perozzi, P. Pravec, C. Snodgrass, C. Teodorescu, S. Anghel, A. Bertolucci, F. Calderini,
  F. Colas, A. Del Vigna, A. Dell’Oro, A. Di Cecco, L. Dimare, I. Di Pietro, P. Fatka, S.
  Fornasier, E. Frattin, P. Frosini, M. Fulchignoni, R. Gabryszewski, M. Giardino, A. Giunta,
  T. Hromakina, J. Huntingford, S. Ieva, J.P. Kotlarz, F. La Forgia, J. Licandro, H. Medeiros,
  F. Merlin, J. Nomen Torres, V. Petropoulou, F. Pina, G. Polenta, M. Popescu, A. Rozek, P.
  Scheirich, A. Sonka, G.B. Valsecchi, P. Wajer, A. Zinzi.\\

  This research has made use of the SVO Filter Profile
  Service supported from the Spanish MINECO through grant AYA2017-84089
  \citep{2012ivoa.rept.1015R, 2020sea..confE.182R}.
  We did an extensive use of the
  Virtual Observatory (VO) TOPCAT software
  \citep{2005ASPC..347...29T}, and
  IMCCE's VO tools
  SkyBot \citep{2006-ASPC-351-Berthier}
  and
  \ssodnet \citep{ssodnet}.
  This work made use of Astropy:\footnote{\url{http://www.astropy.org}}
  a community-developed core Python package and an ecosystem of tools 
  and resources for astronomy \citep{astropy:2013, astropy:2018, astropy:2022}. 
  This research made use of Photutils, an Astropy package for
  detection and photometry of astronomical sources
  \citep{photutils}.
  Thanks to all the developers and maintainers.\\

  This work has made use of data from the European Space Agency (ESA) mission
  {\it Gaia} \footnote{\url{https://www.cosmos.esa.int/gaia}}, processed by the {\it Gaia}
  Data Processing and Analysis Consortium (DPAC,
  \footnote{\url{https://www.cosmos.esa.int/web/gaia/dpac/consortium}}). Funding for the DPAC
  has been provided by national institutions, in particular the institutions
  participating in the {\it Gaia} Multilateral Agreement.\\
  
  Funding for SDSS-III has been provided by the Alfred P. Sloan Foundation, 
  the Participating Institutions, the National Science Foundation, and the U.S. Department 
  of Energy Office of Science. The SDSS-III web site is \footnote{\url{http://www.sdss3.org}}.
  SDSS-III is managed by the Astrophysical Research Consortium for the 
  Participating Institutions of the SDSS-III Collaboration including the 
  University of Arizona, the Brazilian Participation Group, Brookhaven 
  National Laboratory, Carnegie Mellon University, University of Florida, 
  the French Participation Group, the German Participation Group, Harvard University,
  the Instituto de Astrofisica de Canarias, the Michigan State/Notre 
  Dame/JINA Participation Group, Johns Hopkins University, Lawrence 
  Berkeley National Laboratory, Max Planck Institute for Astrophysics, 
  Max Planck Institute for Extraterrestrial Physics, New Mexico State University, 
  New York University, Ohio State University, Pennsylvania State University, 
  University of Portsmouth, Princeton University, the Spanish Participation Group, 
  University of Tokyo, University of Utah, Vanderbilt University, 
  University of Virginia, University of Washington, and Yale University.
\end{acknowledgements}

\appendix

\section{Conversion of reflectance to color\label{app:gaia_color}}

\begin{figure}[h]
    \centering
    \includegraphics[width=0.8\hsize]{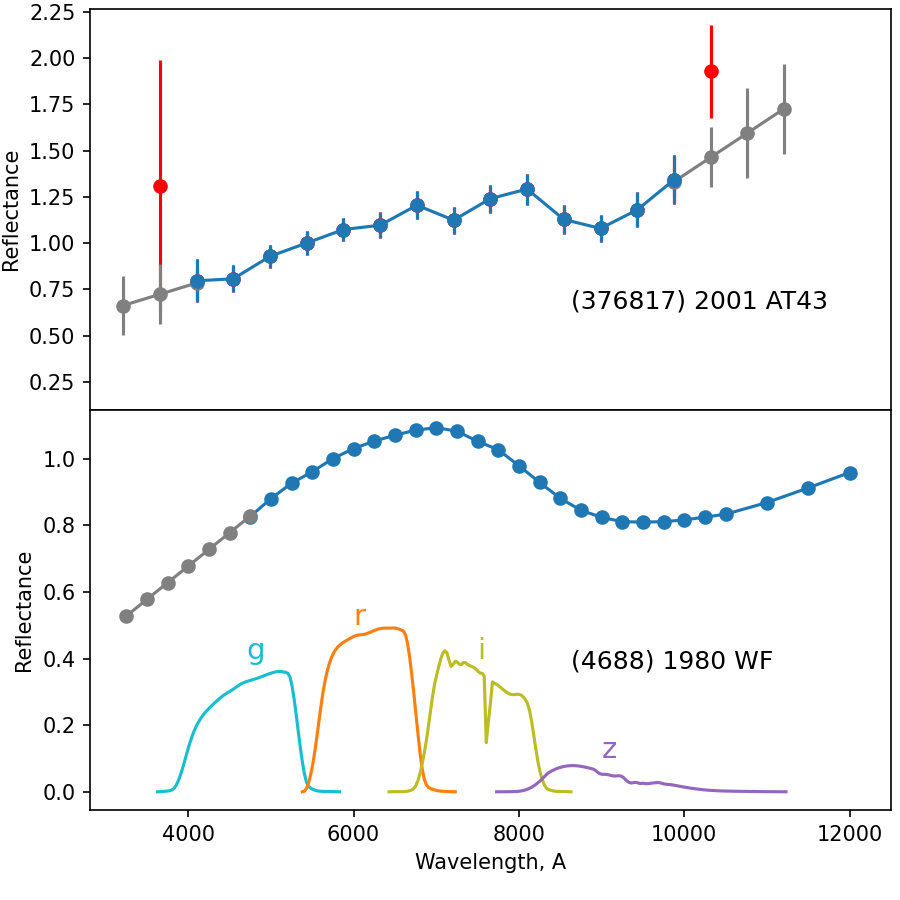}
    \caption{Examples of \gaia reflectance spectra of the asteroid (376817) 2001 AT43 (top) and \classy reflectance of the asteroid (4688) 1980 WF (bottom). 
    Red points indicate outliers \citep[see][]{gaia3-spectra}, while gray points indicate extrapolated values (see text).
    We overplot the transmission curves of the \sdss 
    \filtg, \filtr, \filti, and \filtz filters to show the 
    wavelength range covered by the reflectance spectra.
    }
    \label{fig:gaia_spectrum}
\end{figure}

A common strategy used to improve the detection of compositionally similar dependences in data analysis is to reduce the data's dimensionality. This process involves simplifying the data without losing critical information. When working with reflectance spectra that cover the same wavelength range, they can be transformed into colors, under the condition that the data encompass similarly broad wavelength ranges. This conversion facilitates better visualization and comparison of our data.

The transformation procedure involves several steps. Initially, each reflectance value was multiplied by a solar spectrum, which was taken from \citet{2014PASP..126..711B} and the filter transmission curves. The solar spectrum was used because it is the light from the Sun that is being reflected off the surfaces of the asteroids. The filter transmission curves were used to mimic the response of the instrument that would be observing this reflected light. Once these multiplications were completed, we calculated the integrals by summing up all of the individual product values across the wavelength range to produce a single, complete value that characterizes the source photometry value in the filter. This process is detailed in \citep[Chap. 7 in ][]{2021-riea}. The color index value is the logarithm relation of photometry obtained in two filters, which provides a measure of the object's color.

We have two types of reflectance spectra under consideration. The Gaia reflectance spectra consist of 16 data points, each a measurement of how much light an asteroid reflects at specific wavelengths. These values span from 374 nm to 1034 nm, increasing by 44 nm increments. The Classy reflectance spectra are more extensive, comprising 53 tabulated values ranging from 0.45 $\mu m$ to 2.45 $\mu m$. The intervals between these values are 0.025 $\mu m$ up to 1.025 $\mu m$, and increase to 0.05 $\mu m$ beyond this point.

We note that not all data from the Gaia spectra are reliable. Specifically, the first two values (which represent blue light) and the last two values (representing red light) can sometimes be inaccurate or spurious. Although these suspect values are often flagged in the Gaia DR3 catalog \citep{gaia3-spectra}, this is not always the case. To address this problem, we discarded these unreliable values and replaced them with extrapolated values. This extrapolation was based on the trend observed in the three nearest (and more reliable) reflectance values, as illustrated in (\Autoref{fig:gaia_spectrum}).

Having carried out these preliminary steps, we proceeded to convert the refined reflectance spectra into standard color indexes used in astronomy, namely \colorgr, \colorgi, \colorri, and \coloriz colors. This transformation is undertaken within the photometric system of the SDSS, a major astronomical survey that has provided extensive data on the night sky. To ensure accuracy in this conversion, we retrieved the transmission curves of the SDSS filters from the SVO filter profile service\footnote{\url{http://svo2.cab.inta-csic.es/theory/fps/index.php?mode=voservice}}. This service contains a variety of transmission curves from a multitude of observatories and astronomical instruments \citep{2012-IVOA-Rodrigo, 2020sea..confE.182R}.

Finally, we calculated the uncertainties associated with these color values. These uncertainties provide a measure of the potential error or variability in our color measurements. They are calculated as half the difference of color computed using the reflectance, plus and minus uncertainties. This gives us a measure of the range within which the true color value is likely to lie, thereby providing us with a more comprehensive understanding of the asteroids' color data.

\section{Photometry of fast-moving targets\label{sec:photometry}}

\begin{figure}[t]
    \centering
    \includegraphics[width=1\hsize]{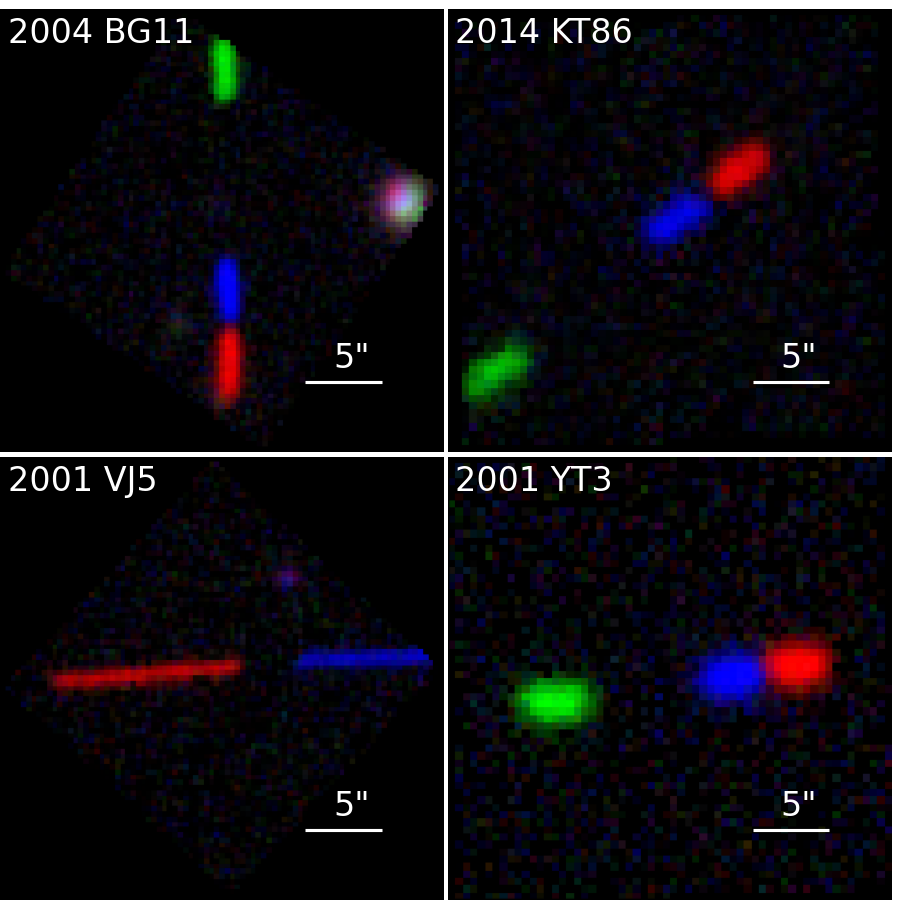}
    \caption{Examples of fast-moving NEOs in \sdss images.
      The color images are a combination of FITS images in
      \filtg (green), \filtr (red), and \filti (blue) filters.
      }
    \label{fig:fast_neas}
\end{figure}

The apparent motion of Main Belt asteroids is typically
about \SI{40}{\arcsec\per\hour}.
The length of the streak during the exposure (54\,s) is thus
comparable with the typical seeing of \sdss images.
However, NEOs have significantly faster apparent motion, up to hundreds of
arcseconds per hour, leading to trailed signatures
(\Autoref{fig:fast_neas}, \citet{2014AN....335..142S}).

\begin{figure}[t]
    \centering
    \includegraphics[width=1.0\hsize]{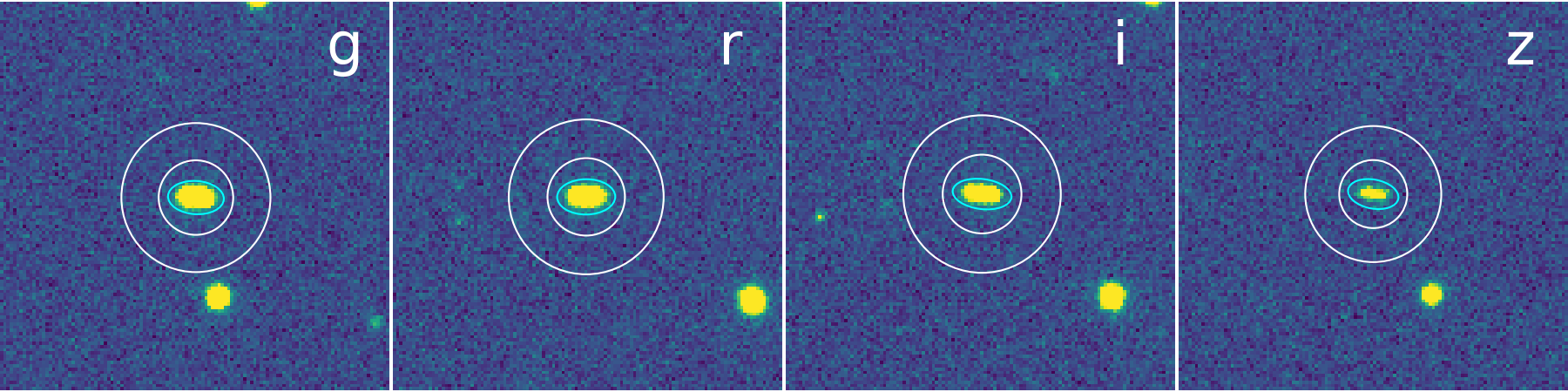}
    \caption{Photometry of 2006 UA on \sdss images, illustrating the elliptical aperture.
    The inner ellipse shows the region in which photons are counted. The two outer circles show the annulus used to estimate the sky background.
    }
    \label{fig:photometry}
\end{figure}

\begin{figure*}[t]
  \centering
  \includegraphics[width=1.0\hsize]{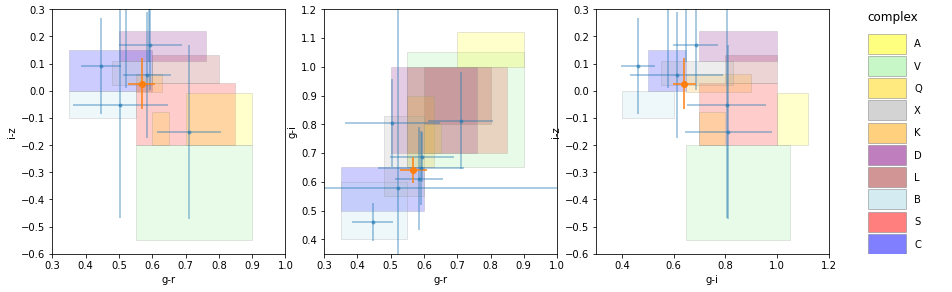}
  \caption{Colors of (277958) 2006 SP134 from individual \sdss catalog values  
  (in blue)
  and from our elliptical photometry (orange).
  The color boxes represent the limits of taxonomic classes.
  }
  \label{fig:2008_QC}
\end{figure*}

Through visual inspection of random NEOs images from the \sdss database,
and checking their photometry from the \sdss pipeline
\citep[as reported by][]{SergeyevCarry2021},
we found that fast-moving NEOs sometimes have incorrect photometry.
This is likely because they were not recognized as a single object
in the different filters by the \sdss pipeline.
Furthermore, the \sdss PSF photometry of elongated NEO tracks is biased.
We finally identified a few cases of erroneous estimate of the zero-point
in individual \sdss Flexible Image Transport System (FITS) frames.
We note that \sdss magnitudes are expressed as inverse hyperbolic 
sines \citep[``asinh'' magnitudes,][]{1999AJ....118.1406L}.
They are virtually identical to the usual Pogson astronomical magnitude in the
high signal-to-noise ratio (S/N) regime, but can diverge for faint objects such
as NEOs.

We overcame these issues by remeasuring the photometry of NEOs 
moving faster than \SI{80}{\arcsec\per\hour}
on \sdss images. We selected \numb{\sdssFastAst} NEOs 
with either an expected S/N above 10 in the $z$ filter
or multiple measurements. 
We used these criteria to ensure meaningful colors for taxonomy:
typical color differences between classes are on the order of 0.1 mag
\citep{2013Icar..226..723D}, and the
$z$ filter is crucial for probing the presence of an
absorption band around 1\,$\mu$m
\citep{2016Icar..268..340C}, which has been one of the major discriminants
in all taxonomies for the past half a century \citep{1975-Icarus-25-Chapman}.

For this task, we developed a python software using the
\texttt{astropy} \citep{astropy:2013, astropy:2018, astropy:2022}
\texttt{photoutils} \citep{photutils},
\texttt{astroquery} \citep{astroquery}, and
\texttt{sep} \citep[the core algorithms of \texttt{SExtractor},][]{1996A&AS..117..393B,
2016JOSS....1...58B} packages.
The procedure to measure the photometry encompassed the following steps.

First, we estimated the zero-point value of each \sdss frame.
We identified non-saturated bright stars and measured their instrumental magnitude 
with aperture photometry.
We then derived the slope and zero-point of individual frames
by comparing these values with the photometry from the \sdss \texttt{PhotoPrimary} catalog
\citep{2000AJ....120.1579Y},
which contains only stationary sources.

Using the \texttt{sep} package, we identified all sources in cut-out images
centered on the predicted location of the asteroid.
The \sdss images in different filters were obtained sequentially,
with a delay of \SI{17.7}{\second} between each of the \SI{54}{\second} exposures.
The position of the cut-out image of the asteroid hence changes in each filter, 
with the largest shift occuring between filters \filtg and \filtr.
Therefore, we identified the NEOs in these two filters using \skybot 
\citep{2006-ASPC-351-Berthier, 2016MNRAS.458.3394B},
since it provides the best S/N and brackets the other observations.
We then predicted the NEOs positions in other filters based on these determinations.
We next checked the images visually to select only those NEOs not blended with stars.
Whenever a NEO was observed on multiple epochs,
we co-added the asteroid-centered cut-out images to increase the asteroid S/N
prior to measuring its photometry.

We finally measured the magnitude of each NEO in each filter 
using an elliptical aperture
to account for the PSF elongation due to the fast motion (\Autoref{fig:photometry}).
We illustrate the improvement on the photometry in \Autoref{fig:2008_QC}.
These updated magnitudes are the ones used in the creation of the \neorocks
data set.

\section{Estimation of color uncertainties\label{app:error}}

\begin{figure}
  \centering
  \includegraphics[width=\hsize]{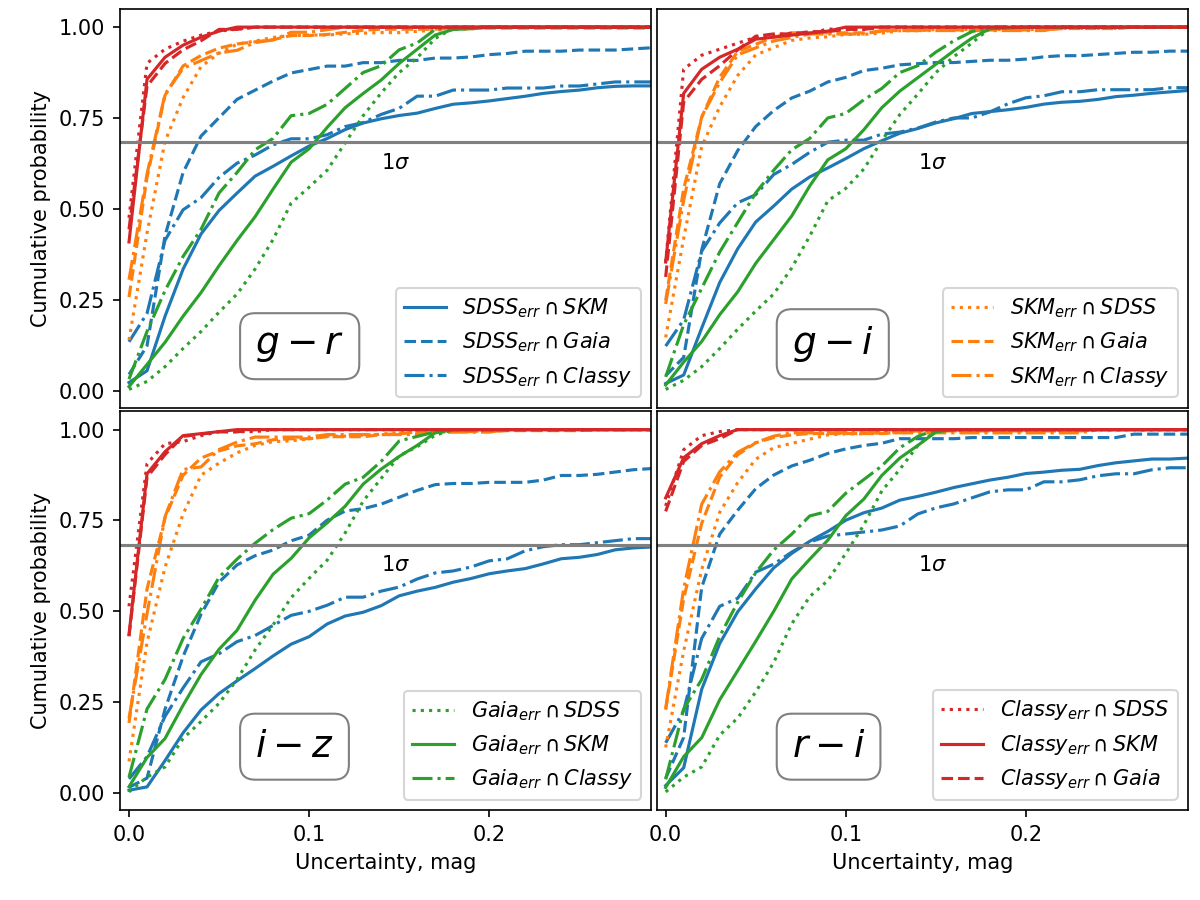}
  \caption{Cumulative distributions of 
   the difference of color for asteroids in \sdss and 
   the rest of the catalogs (blue), as well as 
   their color uncertainties obtained from photometry (orange).}
  \label{fig:neo_errors}
\end{figure}

In order to select the optimal color value amongst multiple catalogs, we had to take into account color value uncertainty. Nevertheless, there may be situations where the reported photometric errors, calculated via diverse methodologies, do not align. For instance, such discrepancies can arise when uncertainties are quantified as either standard deviations or standard errors, particularly when these uncertainties do not follow a normal distribution.

The availability of color estimates for the same asteroids in the different catalogs allowed us to compare the difference in color distribution with photometric uncertainties.
Color indexes, such as the \colorgr index, represent the difference in magnitude (brightness) between two different wavelength bands for a given object. Uncertainties in these indices can be calculated from the uncertainties in the photometric measurements for each band.

For example, in the g-r color index, the uncertainty can be calculated from the errors in the \filtg and \filtr magnitudes. For two different catalogs, we could represent these calculations as follows:

$gr1_{err} = \sqrt{g1_{err}^2 + r1_{err}^2}$

$gr2_{err} = \sqrt{g2_{err}^2 + r2_{err}^2}$.

Here, $g1_{err}^2$ and $r1_{err}^2$ are the uncertainties of the g and r photometry from the first catalog, and $g2_{err}$ and $r2_{err}$ are the uncertainties from the second catalog.
If we assume that the color of an asteroid does not change over time, we can calculate the difference in the color indices measured in two different catalogs. This can be done using the previously computed uncertainties:

$\Delta (gr1 - gr2) = \sqrt{gr1_{err}^2 + gr2_{err}^2}$,

where $\Delta (gr1 - gr2)$ is the difference in the g-r color index between the two catalogs and $gr1_{err}$ and $gr2_{err}$ are the uncertainties of this color index in the first and second catalogs, respectively.

Estimating the uncertainty of stellar objects is a complex task. While internal errors could provide a reasonable uncertainty estimate, systematic errors may distort these results. It is important to keep in mind that published uncertainties may potentially contain distortions that have not been accounted for. If we consider that the published uncertainties might not be accurate, and the true uncertainties are $gr1_{err}*k1$ and $gr2_{err}*k2$, where $k1$ and $k2$ are unknown factors, in this case, the difference in the color indices can be calculated as

$\Delta (gr1 - gr2) = \sqrt{gr1_{err}^2*k1^2 + gr2_{err}^2*k2^2}$.

In instances where there are more than two catalogs at our disposal, we can calculate the color difference between each pair of catalogs. For example, if we have three catalogs, we can formulate the following:

$\Delta (gr1 - gr2) = \sqrt{gr1_{err}^2*k1^2 + gr2_{err}^2*k2^2}$

$\Delta (gr1 - gr3) = \sqrt{gr1_{err}^2*k1^2 + gr3_{err}^2*k3^2}$

$\Delta (gr2 - gr3) = \sqrt{gr2_{err}^2*k2^2 + gr3_{err}^2*k3^2}$.

This formulation provides us with a system of three equations featuring three unknown variables ($k1$, $k2$, and $k3$). These equations can be resolved in order to estimate the authentic uncertainties inherent to each catalog.

In situations involving four catalogs (for instance, \sdss, \smss, \gaia, and \classy, in our example), we can compute the color differences between every pair, resulting in a system of six equations with four unknowns. This system is generally resolved using a least squares method. The solutions derived from this system would produce the estimated authentic uncertainties associated with each catalog.

We extracted the common asteroids from each of our four catalogs and obtained three samples for each of them. 
For example, for the \sdss catalog, we obtained \smss, \gaia, and \classy 
cross-match samples that contain \numb{54,283} \numb{27,158}, and \numb{1,807} 
of common asteroids, correspondingly.
Cumulative distributions of color errors for four colors are presented
in \Autoref{fig:neo_errors}, where we can see the typical photometry error distribution of the \sdss and \sm data that are limited by the magnitude. While the \gaia errors have a uniform distribution because the data have no dependence on the asteroid magnitude, the \classy data have no information about their errors, and therefore we generated random uniform errors in the range from 0 to 0.1 magnitudes.

\begin{table}[]
    \centering
    \begin{tabular}{lcccc}
        Color & \sdss & \sm & \gaia & \classy \\
        \hline
        $g-r$ &  0.779 &  2.594 &  0.539 &   0.629 \\
        $g-i$ &  0.716 &  2.695 &  0.542 &   0.906 \\
        $i-z$ &  0.321 &  2.240 &  0.513 &   0.367 \\
        $r-i$ &  0.822 &  1.633 &  0.361 &   0.285 \\
        \hline
        \\
    \end{tabular}
    \caption{Correction factors for catalog color uncertainty estimates.}
    \label{tab:individual_error}
\end{table}

The variation between the three distributions of the same catalog 
errors shows a different composition of the common samples.
The correction coefficients of color uncertainties for each catalog, calculated using the least squares method, are presented in \Autoref{tab:individual_error}.

\begin{figure}[ht]
  \centering
  \includegraphics[width=\hsize]{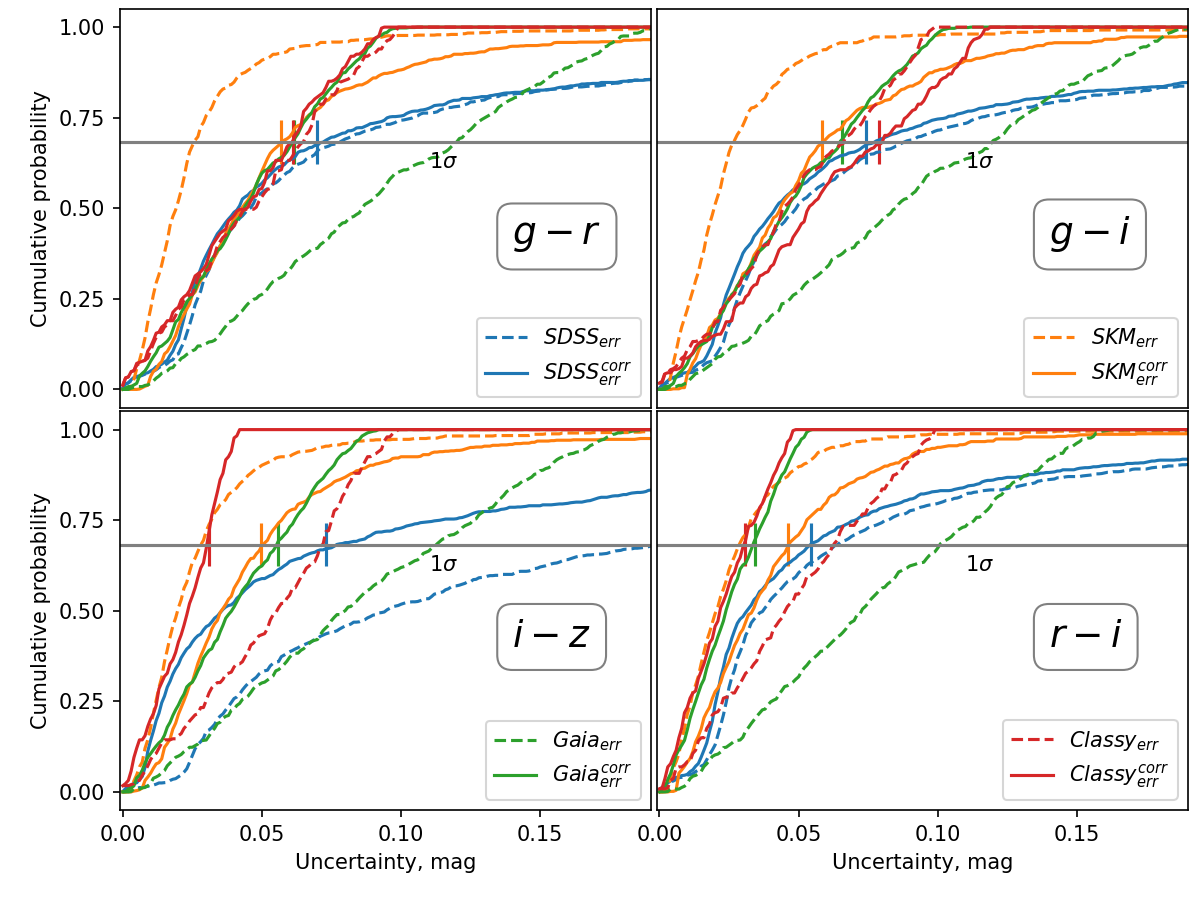}
    \caption{Cumulative distribution of the color uncertainty
    of asteroids present in \sdss \smss, \gaia, and \classy samples. 
    Vertical lines indicate the estimated values of color index difference errors.}
  \label{fig:neo_individual_errors}
\end{figure}

We subsequently calculated the cumulative distribution of color differences between asteroids found in varying catalogs. In \Autoref{fig:neo_individual_errors}, we depict the declared cumulative error distribution of each catalog. 
It is observable that the distribution of the declared \smss color uncertainties is overestimated compared with the computed distribution, especially within the $\smss \cap \sdss$ sample.
Conversely, the \gaia uncertainties seem to be underestimated, possibly owing to the manner in which we computed the uncertainties during the derivation of the color.

\section{Vikiri spectrum}\label{sec:virkki}

\begin{figure}[t]
    \centering
    \includegraphics[width=\hsize]{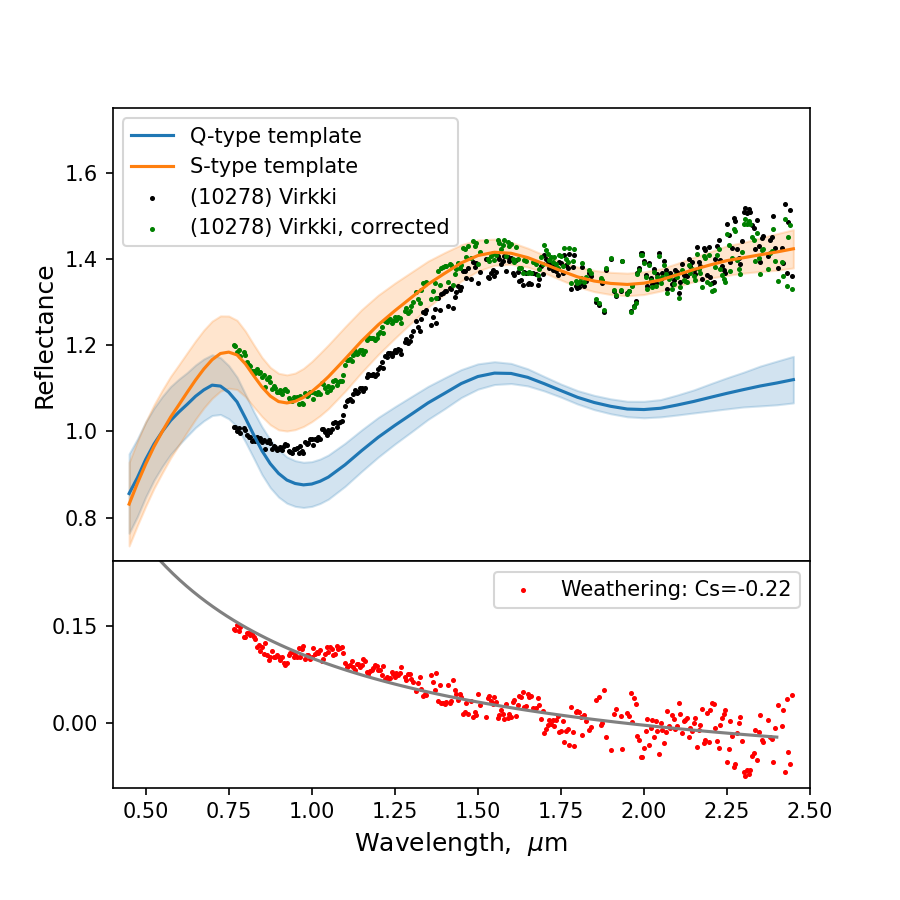}
    \caption{Near-infrared spectrum of (10278) Virkki
    (black dots),
    a former candidate for flyby by the ESA Hera mission.
    The orange line is the S-type asteroid spectrum template 
    from \citet{2022A&A...665A..26M}. Their ratio is typical
    of space weathering (red points in the bottom panel).
    Blue points present the reflectance corrected by the
    space weathering.}
    \label{fig:virkki}
\end{figure}

  We present a near-infrared spectrum
  of (10278) Virkki in \Autoref{fig:virkki}.
  This spectrum was collected with the 3-m IRTF located on Maunakea, 
  Hawaii, on October 14, 2020, through the MITHNEOS program
  \citep[][PI: DeMeo]{2019Icar..324...41B}.
  We used the SpeX NIR spectrograph \citep{2003-PASP-115-Rayner}
  combined with a 0.8x15\arcsec~slit in the low-resolution prism mode 
  to measure the spectra over the 0.7-2.5 $\mu$m wavelength range.
  Asteroid observations were bracketed with measurements of the following calibration stars, which are
  known to be very close spectral analogs to the Sun: Hyades 64 and
  \citet{1983AJ.....88..439L} stars 93-101 and 113-276.
   In-depth analysis of these calibration stars and additional stars used in MITHNEOS is provided in 
   \citet{2020ApJS..247...73M}.
   Data reduction and spectral extraction followed the procedure outlined in 
   \citet{2019Icar..324...41B}, with the Autospex software tool 
   \citep{2005Icar..175..175R}.

   These steps included trimming the images, creating a bad pixel map,
   flat-fielding the images, sky subtraction,
   tracing the spectra in both the wavelength and spatial dimensions, 
   co-adding the spectral images, extracting the spectra,
   performing wavelength calibration, and correcting for air-mass differences 
   between the asteroids and the corresponding solar analogs. 
   The resulting asteroid spectra were divided by the mean 
   stellar spectra to remove the solar gradient.

  Finally, we present the \gaia spectra for two candidates from the short list of seven candidates still considered 
  for a flyby by Hera in \Autoref{fig:hera}. We also present the
  \sdss/\sm colors of four other candidates.

\section{Catalog description}\label{sec:cat_desc}

We describe here the catalog of the NEAs we have released.
The catalog contains four colors (\colorgr, \colorgi, \colorri, and \coloriz), 
osculating elements, the most probable taxonomy, 
and the source region for each asteroid. 
The catalog presented here is available at the CDS via anonymous ftp to
\footnote{\url{http://cdsarc.u-strasbg.fr/}} or via
\footnote{\url{http://cdsarc.u-strasbg.fr/viz-bin/qcat?J/A+A/xxx/Axxx}}

\newcolumntype{P}[1]{>{\hspace{0pt}}p{#1}}
\begin{table}[h]
    \label{tab:cat_color_ind}
    \begin{tabular}{P{0.9cm}p{1.8cm}P{0.9cm}p{3.7cm}}
        \hline\hline
        \textbf{ID} &\textbf{Name} & \textbf{Unit} & \textbf{Description} \\ 
        \hline 
        1 & number & & SSO IAU number \\ 
        2 & name & & SSO IAU name \\ 
        3 & designation & & SSO IAU designation \\ 
        4 & dynclass & & NEOs dynamic class \\ 
        5-8 & m[$^*$color~list] & mag & Set of the color magnitude values \\
        9-13 & e[$^*$color~list] & mag & Set of the color uncertainties \\
        14-17 & d[$^*$color~list] & day & Set of time values between observations \\

        18 & complex & & The most probable complex \\
        19 & pcomplex & & Probability value of the complex \\
        20 & complex2 & & The second most probable complex \\
        21 & pcomplex2 & & Probability of the second most probable complex \\
        22 & origin & & Most probable asteroids' origin region \\
        23 & tmethod & & The method used for taxonomy classification \\
        24 & albedo & & Albedo value \\
        25 & a & au & Semi-major axis \\
        26 & e & & Eccentricity \\
        27 & i & deg & Inclination \\
        28 & H & mag & Absolute magnitude \\
        
        
            \hline
    \end{tabular}

    \text{$^*$ color list: g-r, g-i, r-i, i-z}
    \\
    \caption{Description of catalog that includes obtained colors, taxonomy, and orbital elements of NEOs.}
\end{table}

\newpage
\bibliographystyle{aa}
\bibliography{current}

\begin{thebibliography}{119}
\expandafter\ifx\csname natexlab\endcsname\relax\def\natexlab#1{#1}\fi

\bibitem[{Abell {et~al.}(2012)Abell, Barbee, Mink, Adamo, Alberding, Mazanek,
  Johnson, Yeomans, Chodas, Chamberlin, {et~al.}}]{2012-DPS-Abell}
Abell, P., Barbee, B.~W., Mink, R.~G., {et~al.} 2012, in AAS/Division for
  Planetary Sciences Meeting Abstracts, Vol.~44

\bibitem[{{Abell} {et~al.}(2015){Abell}, {Barbee}, {Chodas}, {Kawaguchi},
  {Landis}, {Mazanek}, \& {Michel}}]{2015aste.book..855A}
{Abell}, P.~A., {Barbee}, B.~W., {Chodas}, P.~W., {et~al.} 2015, in Asteroids
  IV, 855--880

\bibitem[{{Agostini} {et~al.}(2022){Agostini}, {Lucchetti}, {Pajola},
  {Epifani}, {Palumbo}, \& {Cremonese}}]{2022P&SS..21605476A}
{Agostini}, L., {Lucchetti}, A., {Pajola}, M., {et~al.} 2022, \planss, 216,
  105476

\bibitem[{{Alvarez-Candal} {et~al.}(2022{\natexlab{a}}){Alvarez-Candal},
  {Benavidez}, {Campo Bagatin}, \& {Santana-Ros}}]{2022A&A...657A..80A}
{Alvarez-Candal}, A., {Benavidez}, P.~G., {Campo Bagatin}, A., \&
  {Santana-Ros}, T. 2022{\natexlab{a}}, \aap, 657, A80

\bibitem[{{Alvarez-Candal} {et~al.}(2022{\natexlab{b}}){Alvarez-Candal},
  {Jimenez Corral}, \& {Colazo}}]{2022A&A...667A..81A}
{Alvarez-Candal}, A., {Jimenez Corral}, S., \& {Colazo}, M. 2022{\natexlab{b}},
  \aap, 667, A81

\bibitem[{{Astropy Collaboration} {et~al.}(2022){Astropy Collaboration},
  {Price-Whelan}, {Lim}, {Earl}, {Starkman}, {Bradley}, {Shupe}, {Patil},
  {Corrales}, {Brasseur}, {N{\"o}the}, {Donath}, {Tollerud}, {Morris},
  {Ginsburg}, {Vaher}, {Weaver}, {Tocknell}, {Jamieson}, {van Kerkwijk},
  {Robitaille}, {Merry}, {Bachetti}, {G{\"u}nther}, {Aldcroft},
  {Alvarado-Montes}, {Archibald}, {B{\'o}di}, {Bapat}, {Barentsen},
  {Baz{\'a}n}, {Biswas}, {Boquien}, {Burke}, {Cara}, {Cara}, {Conroy},
  {Conseil}, {Craig}, {Cross}, {Cruz}, {D'Eugenio}, {Dencheva}, {Devillepoix},
  {Dietrich}, {Eigenbrot}, {Erben}, {Ferreira}, {Foreman-Mackey}, {Fox},
  {Freij}, {Garg}, {Geda}, {Glattly}, {Gondhalekar}, {Gordon}, {Grant},
  {Greenfield}, {Groener}, {Guest}, {Gurovich}, {Handberg}, {Hart},
  {Hatfield-Dodds}, {Homeier}, {Hosseinzadeh}, {Jenness}, {Jones}, {Joseph},
  {Kalmbach}, {Karamehmetoglu}, {Ka{\l}uszy{\'n}ski}, {Kelley}, {Kern},
  {Kerzendorf}, {Koch}, {Kulumani}, {Lee}, {Ly}, {Ma}, {MacBride}, {Maljaars},
  {Muna}, {Murphy}, {Norman}, {O'Steen}, {Oman}, {Pacifici}, {Pascual},
  {Pascual-Granado}, {Patil}, {Perren}, {Pickering}, {Rastogi}, {Roulston},
  {Ryan}, {Rykoff}, {Sabater}, {Sakurikar}, {Salgado}, {Sanghi}, {Saunders},
  {Savchenko}, {Schwardt}, {Seifert-Eckert}, {Shih}, {Jain}, {Shukla}, {Sick},
  {Simpson}, {Singanamalla}, {Singer}, {Singhal}, {Sinha}, {Sip{\H{o}}cz},
  {Spitler}, {Stansby}, {Streicher}, {{\v{S}}umak}, {Swinbank}, {Taranu},
  {Tewary}, {Tremblay}, {Val-Borro}, {Van Kooten}, {Vasovi{\'c}}, {Verma}, {de
  Miranda Cardoso}, {Williams}, {Wilson}, {Winkel}, {Wood-Vasey}, {Xue},
  {Yoachim}, {Zhang}, {Zonca}, \& {Astropy Project
  Contributors}}]{astropy:2022}
{Astropy Collaboration}, {Price-Whelan}, A.~M., {Lim}, P.~L., {et~al.} 2022,
  \apj, 935, 167

\bibitem[{{Astropy Collaboration} {et~al.}(2018){Astropy Collaboration},
  {Price-Whelan}, {Sip{\H{o}}cz}, {G{\"u}nther}, {Lim}, {Crawford}, {Conseil},
  {Shupe}, {Craig}, {Dencheva}, {Ginsburg}, {Vand erPlas}, {Bradley},
  {P{\'e}rez-Su{\'a}rez}, {de Val-Borro}, {Aldcroft}, {Cruz}, {Robitaille},
  {Tollerud}, {Ardelean}, {Babej}, {Bach}, {Bachetti}, {Bakanov}, {Bamford},
  {Barentsen}, {Barmby}, {Baumbach}, {Berry}, {Biscani}, {Boquien}, {Bostroem},
  {Bouma}, {Brammer}, {Bray}, {Breytenbach}, {Buddelmeijer}, {Burke},
  {Calderone}, {Cano Rodr{\'\i}guez}, {Cara}, {Cardoso}, {Cheedella}, {Copin},
  {Corrales}, {Crichton}, {D'Avella}, {Deil}, {Depagne}, {Dietrich}, {Donath},
  {Droettboom}, {Earl}, {Erben}, {Fabbro}, {Ferreira}, {Finethy}, {Fox},
  {Garrison}, {Gibbons}, {Goldstein}, {Gommers}, {Greco}, {Greenfield},
  {Groener}, {Grollier}, {Hagen}, {Hirst}, {Homeier}, {Horton}, {Hosseinzadeh},
  {Hu}, {Hunkeler}, {Ivezi{\'c}}, {Jain}, {Jenness}, {Kanarek}, {Kendrew},
  {Kern}, {Kerzendorf}, {Khvalko}, {King}, {Kirkby}, {Kulkarni}, {Kumar},
  {Lee}, {Lenz}, {Littlefair}, {Ma}, {Macleod}, {Mastropietro}, {McCully},
  {Montagnac}, {Morris}, {Mueller}, {Mumford}, {Muna}, {Murphy}, {Nelson},
  {Nguyen}, {Ninan}, {N{\"o}the}, {Ogaz}, {Oh}, {Parejko}, {Parley}, {Pascual},
  {Patil}, {Patil}, {Plunkett}, {Prochaska}, {Rastogi}, {Reddy Janga},
  {Sabater}, {Sakurikar}, {Seifert}, {Sherbert}, {Sherwood-Taylor}, {Shih},
  {Sick}, {Silbiger}, {Singanamalla}, {Singer}, {Sladen}, {Sooley},
  {Sornarajah}, {Streicher}, {Teuben}, {Thomas}, {Tremblay}, {Turner},
  {Terr{\'o}n}, {van Kerkwijk}, {de la Vega}, {Watkins}, {Weaver}, {Whitmore},
  {Woillez}, {Zabalza}, \& {Astropy Contributors}}]{astropy:2018}
{Astropy Collaboration}, {Price-Whelan}, A.~M., {Sip{\H{o}}cz}, B.~M., {et~al.}
  2018, \aj, 156, 123

\bibitem[{{Astropy Collaboration} {et~al.}(2013){Astropy Collaboration},
  {Robitaille}, {Tollerud}, {Greenfield}, {Droettboom}, {Bray}, {Aldcroft},
  {Davis}, {Ginsburg}, {Price-Whelan}, {Kerzendorf}, {Conley}, {Crighton},
  {Barbary}, {Muna}, {Ferguson}, {Grollier}, {Parikh}, {Nair}, {Unther},
  {Deil}, {Woillez}, {Conseil}, {Kramer}, {Turner}, {Singer}, {Fox}, {Weaver},
  {Zabalza}, {Edwards}, {Azalee Bostroem}, {Burke}, {Casey}, {Crawford},
  {Dencheva}, {Ely}, {Jenness}, {Labrie}, {Lim}, {Pierfederici}, {Pontzen},
  {Ptak}, {Refsdal}, {Servillat}, \& {Streicher}}]{astropy:2013}
{Astropy Collaboration}, {Robitaille}, T.~P., {Tollerud}, E.~J., {et~al.} 2013,
  \aap, 558, A33

\bibitem[{{Barbary}(2016)}]{2016JOSS....1...58B}
{Barbary}, K. 2016, The Journal of Open Source Software, 1, 58

\bibitem[{{Belskaya} {et~al.}(2015){Belskaya}, {Cellino}, {Gil-Hutton},
  {Muinonen}, \& {Shkuratov}}]{2015aste.book..151B}
{Belskaya}, I., {Cellino}, A., {Gil-Hutton}, R., {Muinonen}, K., \&
  {Shkuratov}, Y. 2015, in Asteroids IV, 151--163

\bibitem[{{Belskaya} \& {Shevchenko}(2000)}]{2000Icar..147...94B}
{Belskaya}, I.~N. \& {Shevchenko}, V.~G. 2000, \icarus, 147, 94

\bibitem[{{Belton} {et~al.}(1995){Belton}, {Chapman}, {Thomas}, {Davies},
  {Greenberg}, {Klaasen}, {Byrnes}, {D'Amario}, {Synnott}, {Johnson}, {McEwen},
  {Merline}, {Davis}, {Petit}, {Storrs}, {Veverka}, \&
  {Zellner}}]{1995Natur.374..785B}
{Belton}, M.~J.~S., {Chapman}, C.~R., {Thomas}, P.~C., {et~al.} 1995, \nat,
  374, 785

\bibitem[{{Berthier} {et~al.}(2022){Berthier}, {Carry}, {Mahlke}, \&
  {Normand}}]{ssodnet}
{Berthier}, J., {Carry}, B., {Mahlke}, M., \& {Normand}, J. 2022, arXiv
  e-prints, arXiv:2209.10697

\bibitem[{{Berthier} {et~al.}(2023){Berthier}, {Carry}, {Mahlke}, \&
  {Normand}}]{2023A&A...671A.151B}
{Berthier}, J., {Carry}, B., {Mahlke}, M., \& {Normand}, J. 2023, \aap, 671,
  A151

\bibitem[{Berthier {et~al.}(2016)Berthier, Carry, Vachier, Eggl, \&
  Santerne}]{2016MNRAS.458.3394B}
Berthier, J., Carry, B., Vachier, F., Eggl, S., \& Santerne, A. 2016, \mnras,
  458, 3394

\bibitem[{Berthier {et~al.}(2006)Berthier, Vachier, Thuillot, Fernique,
  Ochsenbein, Genova, Lainey, \& Arlot}]{2006-ASPC-351-Berthier}
Berthier, J., Vachier, F., Thuillot, W., {et~al.} 2006, in Astronomical Society
  of the Pacific Conference Series, Vol. 351, Astronomical Data Analysis
  Software and Systems XV, ed. C.~{Gabriel}, C.~{Arviset}, D.~{Ponz}, \&
  S.~{Enrique}, 367

\bibitem[{{Bertin} \& {Arnouts}(1996)}]{1996A&AS..117..393B}
{Bertin}, E. \& {Arnouts}, S. 1996, \aaps, 117, 393

\bibitem[{{Binzel} {et~al.}(2019){Binzel}, {DeMeo}, {Turtelboom}, {Bus},
  {Tokunaga}, {Burbine}, {Lantz}, {Polishook}, {Carry}, {Morbidelli}, {Birlan},
  {Vernazza}, {Burt}, {Moskovitz}, {Slivan}, {Thomas}, {Rivkin}, {Hicks},
  {Dunn}, {Reddy}, {Sanchez}, {Granvik}, \& {Kohout}}]{2019Icar..324...41B}
{Binzel}, R.~P., {DeMeo}, F.~E., {Turtelboom}, E.~V., {et~al.} 2019, \icarus,
  324, 41

\bibitem[{Binzel {et~al.}(2010)Binzel, Morbidelli, Merouane, DeMeo, Birlan,
  Vernazza, Thomas, Rivkin, Bus, \& Tokunaga}]{2010Natur.463..331B}
Binzel, R.~P., Morbidelli, A., Merouane, S., {et~al.} 2010, \nat, 463, 331

\bibitem[{Binzel {et~al.}(2015)Binzel, Reddy, \&
  Dunn}]{2015-AsteroidsIV-Binzel}
Binzel, R.~P., Reddy, V., \& Dunn, T. 2015, Asteroids IV

\bibitem[{Binzel {et~al.}(2004)Binzel, Rivkin, Stuart, Harris, Bus, \&
  Burbine}]{2004Icar..170..259B}
Binzel, R.~P., Rivkin, A.~S., Stuart, J.~S., {et~al.} 2004, \icarus, 170, 259

\bibitem[{{Bohlin} {et~al.}(2014){Bohlin}, {Gordon}, \&
  {Tremblay}}]{2014PASP..126..711B}
{Bohlin}, R.~C., {Gordon}, K.~D., \& {Tremblay}, P.~E. 2014, \pasp, 126, 711

\bibitem[{Bradley {et~al.}(2020)Bradley, Sip{\H o}cz, Robitaille, Tollerud,
  Vin{\'{\i}}cius, Deil, Barbary, Wilson, Busko, G{\"u}nther, Cara, Conseil,
  Bostroem, Droettboom, Bray, Bratholm, Lim, Barentsen, Craig, Pascual, Perren,
  Greco, Donath, de~Val-Borro, Kerzendorf, Bach, Weaver, D'Eugenio, Souchereau,
  \& Ferreira}]{photutils}
Bradley, L., Sip{\H o}cz, B., Robitaille, T., {et~al.} 2020, astropy/photutils:
  1.0.0

\bibitem[{Brunetto {et~al.}(2015)Brunetto, Loeffler, Nesvorn{\'y}, Sasaki, \&
  Strazzulla}]{2015-AsteroidsIV-Brunetto}
Brunetto, R., Loeffler, M.~J., Nesvorn{\'y}, D., Sasaki, S., \& Strazzulla, G.
  2015, {Asteroid Surface Alteration by Space Weathering Processes}, ed.
  P.~Michel, F.~DeMeo, \& W.~F. Bottke, 597--616

\bibitem[{Brunetto {et~al.}(2006)Brunetto, Vernazza, Marchi, Birlan,
  Fulchignoni, Orofino, \& Strazzulla}]{2006Icar..184..327B}
Brunetto, R., Vernazza, P., Marchi, S., {et~al.} 2006, \icarus, 184, 327

\bibitem[{Burbine {et~al.}(1996)Burbine, Meibom, \&
  Binzel}]{1996MPS...31..607B}
Burbine, T.~H., Meibom, A., \& Binzel, R.~P. 1996, Meteoritics and Planetary
  Science, 31, 607

\bibitem[{Bus \& Binzel(2002)}]{2002-Icarus-158-BusI}
Bus, S.~J. \& Binzel, R.~P. 2002, \icarus, 158, 106

\bibitem[{Carry(2018)}]{2018AA...609A.113C}
Carry, B. 2018, \aap, 609, A113

\bibitem[{Carry {et~al.}(2010)Carry, Kaasalainen, Leyrat, Merline, Drummond,
  Conrad, Weaver, Tamblyn, Chapman, Dumas, Colas, Christou, Dotto, Perna,
  Fornasier, Bernasconi, Behrend, Vachier, Kryszczynska, Polinska, Fulchignoni,
  Roy, Naves, Poncy, \& Wiggins}]{2010-AA-523-Carry}
Carry, B., Kaasalainen, M., Leyrat, C., {et~al.} 2010, \aap, 523, A94

\bibitem[{Carry {et~al.}(2016)Carry, Solano, Eggl, \&
  DeMeo}]{2016Icar..268..340C}
Carry, B., Solano, E., Eggl, S., \& DeMeo, F. 2016, \icarus, 268, 340

\bibitem[{Carvano \& Davalos(2015)}]{2015-AA-580-Carvano}
Carvano, J.~M. \& Davalos, J. A.~G. 2015, \aap, 580, A98

\bibitem[{{Cellino} {et~al.}(2020){Cellino}, {Bendjoya}, {Delbo'}, {Galluccio},
  {Gayon-Markt}, {Tanga}, \& {Tedesco}}]{2020A&A...642A..80C}
{Cellino}, A., {Bendjoya}, P., {Delbo'}, M., {et~al.} 2020, \aap, 642, A80

\bibitem[{Chapman(2004)}]{2004AREPS..32..539C}
Chapman, C.~R. 2004, Annual Review of Earth and Planetary Sciences, 32, 539

\bibitem[{Chapman {et~al.}(1975)Chapman, Morrison, \&
  Zellner}]{1975-Icarus-25-Chapman}
Chapman, C.~R., Morrison, D., \& Zellner, B.~H. 1975, \icarus, 25, 104

\bibitem[{Chapman {et~al.}(1995)Chapman, Veverka, Thomas, Klaasen, Belton,
  Harch, McEwen, Johnson, Helfenstein, Davies, Merline, \&
  Denk}]{1995-Nature-374-Chapman}
Chapman, C.~R., Veverka, J., Thomas, P.~C., {et~al.} 1995, \nat, 374, 783

\bibitem[{{Christou} {et~al.}(2021){Christou}, {Borisov}, {Dell'Oro},
  {Cellino}, \& {Devog{\`e}le}}]{2021Icar..35413994C}
{Christou}, A.~A., {Borisov}, G., {Dell'Oro}, A., {Cellino}, A., \&
  {Devog{\`e}le}, M. 2021, \icarus, 354, 113994

\bibitem[{Cloutis {et~al.}(2015)Cloutis, Sanchez, Reddy, Gaffey, Binzel,
  Burbine, Hardersen, Hiroi, Lucey, Sunshine, \&
  Tait}]{2015-Icarus-252-Cloutis}
Cloutis, E.~A., Sanchez, J.~A., Reddy, V., {et~al.} 2015, \icarus, 252, 39

\bibitem[{{Colazo} {et~al.}(2022){Colazo}, {Alvarez-Candal}, \&
  {Duffard}}]{2022A&A...666A..77C}
{Colazo}, M., {Alvarez-Candal}, A., \& {Duffard}, R. 2022, \aap, 666, A77

\bibitem[{Collaboration {et~al.}(2016)Collaboration, Prusti, de~Bruijne, Brown,
  Vallenari, Babusiaux, Bailer-Jones, Bastian, Biermann, Evans, \&
  et~al.}]{2016AA...595A...1G}
Collaboration, G., Prusti, T., de~Bruijne, J. H.~J., {et~al.} 2016, \aap, 595,
  A1

\bibitem[{Consolmagno {et~al.}(2008)Consolmagno, Britt, \&
  Macke}]{2008-ChEG-68-Consolmagno}
Consolmagno, G., Britt, D., \& Macke, R. 2008, Chemie der Erde / Geochemistry,
  68, 1

\bibitem[{{Cruikshank} \& {Hartmann}(1984)}]{1984Sci...223..281C}
{Cruikshank}, D.~P. \& {Hartmann}, W.~K. 1984, Science, 223, 281

\bibitem[{{Delbo} {et~al.}(2014){Delbo}, {Libourel}, {Wilkerson}, {Murdoch},
  {Michel}, {Ramesh}, {Ganino}, {Verati}, \& {Marchi}}]{2014Natur.508..233D}
{Delbo}, M., {Libourel}, G., {Wilkerson}, J., {et~al.} 2014, \nat, 508, 233

\bibitem[{DeMeo {et~al.}(2009)DeMeo, Binzel, Slivan, \&
  Bus}]{2009Icar..202..160D}
DeMeo, F., Binzel, R.~P., Slivan, S.~M., \& Bus, S.~J. 2009, \icarus, 202, 160

\bibitem[{DeMeo \& Carry(2014)}]{2014Natur.505..629D}
DeMeo, F. \& Carry, B. 2014, \nat, 505, 629

\bibitem[{{DeMeo} \& {Carry}(2013)}]{2013Icar..226..723D}
{DeMeo}, F.~E. \& {Carry}, B. 2013, \icarus, 226, 723

\bibitem[{{DeMeo} {et~al.}(2023){DeMeo}, {Marsset}, {Polishook}, {Burt},
  {Binzel}, {Hasegawa}, {Granvik}, {Moskovitz}, {Earle}, {Bus}, {Thomas},
  {Rivkin}, \& {Slivan}}]{2023Icar..38915264D}
{DeMeo}, F.~E., {Marsset}, M., {Polishook}, D., {et~al.} 2023, \icarus, 389,
  115264

\bibitem[{{DeMeo} {et~al.}(2019){DeMeo}, {Polishook}, {Carry}, {Burt}, {Hsieh},
  {Binzel}, {Moskovitz}, \& {Burbine}}]{2019Icar..322...13D}
{DeMeo}, F.~E., {Polishook}, D., {Carry}, B., {et~al.} 2019, \icarus, 322, 13

\bibitem[{{Devog{\`e}le} {et~al.}(2019){Devog{\`e}le}, {Moskovitz}, {Thirouin},
  {Gustaffson}, {Magnuson}, {Thomas}, {Willman}, {Christensen}, {Person},
  {Binzel}, {Polishook}, {DeMeo}, {Hinkle}, {Trilling}, {Mommert}, {Burt}, \&
  {Skiff}}]{2019AJ....158..196D}
{Devog{\`e}le}, M., {Moskovitz}, N., {Thirouin}, A., {et~al.} 2019, \aj, 158,
  196

\bibitem[{{Di Martino} {et~al.}(1990){Di Martino}, {Ferreri}, {Fulchignoni},
  {De Angeles}, {Barucci}, {Lecacheux}, {Burchi}, \& {Di
  Paolantonio}}]{1990Icar...87..372D}
{Di Martino}, M., {Ferreri}, W., {Fulchignoni}, M., {et~al.} 1990, \icarus, 87,
  372

\bibitem[{{Doressoundiram} {et~al.}(1999){Doressoundiram}, {Weissman},
  {Fulchignoni}, {Barucci}, {Le Bras}, {Colas}, {Lecacheux}, {Birlan},
  {Lazzarin}, {Fornasier}, {Dotto}, {Barbieri}, {Sykes}, {Larson}, \&
  {Hergenrother}}]{1999A&A...352..697D}
{Doressoundiram}, A., {Weissman}, P.~R., {Fulchignoni}, M., {et~al.} 1999,
  \aap, 352, 697

\bibitem[{{Dotto} {et~al.}(2021){Dotto}, {Banaszkiewicz}, {Banchi}, {Barucci},
  {Bernardi}, {Birlan}, {Carry}, {Cellino}, {de Leon}, {Lazzarin}, {Mazzotta
  Epifani}, {Nomen Torres}, {Perna}, {Perozzi}, {Pravec}, {Sanchez Ortiz},
  {Snodgrass}, {Teodorescu}, \& {The Neorocks Team}}]{2021plde.confE.221D}
{Dotto}, E., {Banaszkiewicz}, M., {Banchi}, S., {et~al.} 2021, in 7th IAA
  Planetary Defense Conference, 221

\bibitem[{{Drube} {et~al.}(2015){Drube}, {Harris}, {Hoerth}, {Michel}, {Perna},
  \& {Sch{\"a}fer}}]{2015hchp.book..763D}
{Drube}, L., {Harris}, A.~W., {Hoerth}, T., {et~al.} 2015, in Handbook of
  Cosmic Hazards and Planetary Defense, 763--790

\bibitem[{{Erasmus} {et~al.}(2020){Erasmus}, {Navarro-Meza}, {McNeill},
  {Trilling}, {Sickafoose}, {Denneau}, {Flewelling}, {Heinze}, \&
  {Tonry}}]{2020ApJS..247...13E}
{Erasmus}, N., {Navarro-Meza}, S., {McNeill}, A., {et~al.} 2020, \apjs, 247, 13

\bibitem[{{Fitzsimmons} {et~al.}(2020){Fitzsimmons}, {Khan}, {K{\"u}ppers},
  {Michel}, \& {Pravec}}]{2020EPSC...14.1064F}
{Fitzsimmons}, A., {Khan}, M., {K{\"u}ppers}, M., {Michel}, P., \& {Pravec}, P.
  2020, in European Planetary Science Congress, EPSC2020--1064

\bibitem[{Fujiwara {et~al.}(2006)Fujiwara, Kawaguchi, Yeomans, Abe, Mukai,
  Okada, Saito, Yano, Yoshikawa, Scheeres, Barnouin-Jha, Cheng, Demura,
  Gaskell, Hirata, Ikeda, Kominato, Miyamoto, Nakamura, Sasaki, \&
  Uesugi}]{2006Sci...312.1330F}
Fujiwara, A., Kawaguchi, J., Yeomans, D.~K., {et~al.} 2006, Science, 312, 1330

\bibitem[{{Galluccio} {et~al.}(2022){Galluccio}, {Delbo}, {De Angeli},
  {Pauwels}, {Tanga}, {Mignard}, {Cellino}, {Brown}, {Muinonen}, {Penttila},
  {Jordan}, {Vallenari}, {Prusti}, {de Bruijne}, {Arenou}, {Babusiaux},
  {Biermann}, {Creevey}, {Ducourant}, {Evans}, {Eyer}, {Guerra}, {Hutton},
  {Jordi}, {Klioner}, {Lammers}, {Lindegren}, {Luri}, {Panem}, {Pourbaix},
  {Randich}, {Sartoretti}, {Soubiran}, {Walton}, {Bailer-Jones}, {Bastian},
  {Drimmel}, {Jansen}, {Katz}, {Lattanzi}, {van Leeuwen}, {Bakker}, {Cacciari},
  {Castaneda}, {Fabricius}, {Fouesneau}, {Fr{\'e}mat}, {Guerrier}, {Heiter},
  {Masana}, {Messineo}, {Mowlavi}, {Nicolas}, {Nienartowicz}, {Pailler},
  {Panuzzo}, {Riclet}, {Roux}, {Seabroke}, {Sordo}, {Th{\'e}venin},
  {Gracia-Abril}, {Portell}, {Teyssier}, {Altmann}, {Andrae}, {Audard},
  {Bellas-Velidis}, {Benson}, {Berthier}, {Blomme}, {Burgess}, {Busonero},
  {Busso}, {C{\'a}novas}, {Carry}, {Cheek}, {Clementini}, {Damerdji},
  {Davidson}, {de Teodoro}, {Nunez Campos}, {Delchambre}, {Dell Oro}, {Esquej},
  {Fern{\'a}ndez-Hern{\'a}ndez}, {Fraile}, {Garabato}, {Garc{\'\i}a-Lario},
  {Gosset}, {Haigron}, {Halbwachs}, {Hambly}, {Harrison}, {Hern{\'a}ndez},
  {Hestroffer}, {Hodgkin}, {Holl}, {Janssen}, {Jevardat de Fombelle},
  {Krone-Martins}, {Lanzafame}, {L{\"o}ffler}, {Marchal}, {Marrese},
  {Moitinho}, {Osborne}, {Pancino}, {Recio-Blanco}, {Reyl{\'e}}, {Riello},
  {Rimoldini}, {Roegiers}, {Rybizki}, {Sarro}, {Siopis}, {Smith}, {Sozzetti},
  {Utrilla}, {van Leeuwen}, {Abbas}, {{\'A}brah{\'a}m}, {Abreu Aramburu},
  {Aerts}, {Aguado}, {Ajaj}, {Aldea-Montero}, {Altavilla}, {{\'A}lvarez},
  {Alves}, {Anderson}, {Anglada Varela}, {Antoja}, {Baines}, {Baker},
  {Balaguer-N{\'u}nez}, {Balbinot}, {Balog}, {Barache}, {Barbato}, {Barros},
  {Barstow}, {Bartolom{\'e}}, {Bassilana}, {Bauchet}, {Becciani}, {Bellazzini},
  {Berihuete}, {Bernet}, {Bertone}, {Bianchi}, {Binnenfeld}, {Blanco-Cuaresma},
  {Boch}, {Bombrun}, {Bossini}, {Bouquillon}, {Bragaglia}, {Bramante},
  {Breedt}, {Bressan}, {Brouillet}, {Brugaletta}, {Bucciarelli}, {Burlacu},
  {Butkevich}, {Buzzi}, {Caffau}, {Cancelliere}, {Cantat-Gaudin}, {Carballo},
  {Carlucci}, {Carnerero}, {Carrasco}, {Casamiquela}, {Castellani},
  {Castro-Ginard}, {Chaoul}, {Charlot}, {Chemin}, {Chiaramida}, {Chiavassa},
  {Chornay}, {Comoretto}, {Contursi}, {Cooper}, {Cornez}, {Cowell}, {Crifo},
  {Cropper}, {Crosta}, {Crowley}, {Dafonte}, {Dapergolas}, {David}, {de
  Laverny}, {De Luise}, {De March}, {De Ridder}, {de Souza}, {de Torres}, {del
  Peloso}, {del Pozo}, {Delgado}, {Delisle}, {Demouchy}, {Dharmawardena},
  {Diakite}, {Diener}, {Distefano}, {Dolding}, {Enke}, {Fabre}, {Fabrizio},
  {Faigler}, {Fedorets}, {Fernique}, {Figueras}, {Fournier}, {Fouron},
  {Fragkoudi}, {Gai}, {Garcia-Gutierrez}, {Garcia-Reinaldos},
  {Garc{\'\i}a-Torres}, {Garofalo}, {Gavel}, {Gavras}, {Gerlach}, {Geyer},
  {Giacobbe}, {Gilmore}, {Girona}, {Giuffrida}, {Gomel}, {Gomez},
  {Gonz{\'a}lez-N{\'u}nez}, {Gonz{\'a}lez-Santamar{\'\i}a},
  {Gonz{\'a}lez-Vidal}, {Granvik}, {Guillout}, {Guiraud},
  {Guti{\'e}rrez-S{\'a}nchez}, {Guy}, {Hatzidimitriou}, {Hauser}, {Haywood},
  {Helmer}, {Helmi}, {Sarmiento}, {Hidalgo}, {Hadczuk}, {Hobbs}, {Holland},
  {Huckle}, {Jardine}, {Jasniewicz}, {Jean-Antoine Piccolo},
  {Jim{\'e}nez-Arranz}, {Juaristi Campillo}, {Julbe}, {Karbevska}, {Kervella},
  {Khanna}, {Kordopatis}, {Korn}, {Kosp{\'a}l}, {Kostrzewa-Rutkowska},
  {Kruszynska}, {Kun}, {Laizeau}, {Lambert}, {Lanza}, {Lasne}, {Le Campion},
  {Lebreton}, {Lebzelter}, {Leccia}, {Leclerc}, {Lecoeur-Taibi}, {Liao},
  {Licata}, {Lindstrom}, {Lister}, {Livanou}, {Lobel}, {Lorca}, {Loup},
  {Madrero Pardo}, {Magdaleno Romeo}, {Managau}, {Mann}, {Manteiga},
  {Marchant}, {Marconi}, {Marcos}, {Marcos Santos}, {Mar{\'\i}n Pina},
  {Marinoni}, {Marocco}, {Marshall}, {Polo}, {Mart{\'\i}n-Fleitas}, {Marton},
  {Mary}, {Masip}, {Massari}, {Mastrobuono-Battisti}, {Mazeh}, {McMillan},
  {Messina}, {Michalik}, {Millar}, {Mints}, {Molina}, {Molinaro}, {Moln{\'a}r},
  {Monari}, {Mongui{\'o}}, {Montegriffo}, {Montero}, {Mor}, {Mora},
  {Morbidelli}, {Morel}, {Morris}, {Muraveva}, {Murphy}, {Musella}, {Nagy},
  {Noval}, {Ocana}, {Ogden}, {Ordenovic}, {Osinde}, {Pagani}, {Pagano},
  {Palaversa}, {Palicio}, {Pallas-Quintela}, {Panahi}, {Payne-Wardenaar},
  {Penalosa Esteller}, {Petit}, {Pichon}, {Piersimoni}, {Pineau}, {Plachy},
  {Plum}, {Poggio}, {Prsa}, {Pulone}, {Racero}, {Ragaini}, {Rainer}, {Raiteri},
  {Ramos}, {Ramos-Lerate}, {Re Fiorentin}, {Regibo}, {Richards}, {Rios Diaz},
  {Ripepi}, {Riva}, {Rix}, {Rixon}, {Robichon}, {Robin}, {Robin}, {Roelens},
  {Rogues}, {Rohrbasser}, {Romero-G{\'o}mez}, {Rowell}, {Royer}, {Ruz Mieres},
  {Rybicki}, {Sadowski}, {S{\'a}ez N{\'u}nez}, {Sagrist{\`a} Sell{\'e}s},
  {Sahlmann}, {Salguero}, {Samaras}, {Sanchez Gimenez}, {Sanna}, {Santovena},
  {Sarasso}, {Schultheis}, {Sciacca}, {Segol}, {Segovia}, {S{\'e}gransan},
  {Semeux}, {Shahaf}, {Siddiqui}, {Siebert}, {Siltala}, {Silvelo}, {Slezak},
  {Slezak}, {Smart}, {Snaith}, {Solano}, {Solitro}, {Souami}, {Souchay},
  {Spagna}, {Spina}, {Spoto}, {Steele}, {Steidelm{\"u}ller}, {Stephenson},
  {S{\"u}veges}, {Surdej}, {Szabados}, {Szegedi-Elek}, {Taris}, {Taylor},
  {Teixeira}, {Tolomei}, {Tonello}, {Torra}, {Torra}, {Torralba Elipe},
  {Trabucchi}, {Tsounis}, {Turon}, {Ulla}, {Unger}, {Vaillant}, {van Dillen},
  {van Reeven}, {Vanel}, {Vecchiato}, {Viala}, {Vicente}, {Voutsinas},
  {Weiler}, {Wevers}, {Wyrzykowski}, {Yoldas}, {Yvard}, {Zhao}, {Zorec},
  {Zucker}, \& {Zwitter}}]{gaia3-spectra}
{Galluccio}, L., {Delbo}, M., {De Angeli}, F., {et~al.} 2022, arXiv e-prints,
  arXiv:2206.12174

\bibitem[{{Ginsburg} {et~al.}(2019){Ginsburg}, {Sip{\H o}cz}, {Brasseur},
  {Cowperthwaite}, {Craig}, {Deil}, {Guillochon}, {Guzman}, {Liedtke}, {Lian
  Lim}, {Lockhart}, {Mommert}, {Morris}, {Norman}, {Parikh}, {Persson},
  {Robitaille}, {Segovia}, {Singer}, {Tollerud}, {de Val-Borro}, {Valtchanov},
  {Woillez}, {The Astroquery collaboration}, \& {a subset of the astropy
  collaboration}}]{astroquery}
{Ginsburg}, A., {Sip{\H o}cz}, B.~M., {Brasseur}, C.~E., {et~al.} 2019, \aj,
  157, 98

\bibitem[{Gladman {et~al.}(1997)Gladman, Migliorini, Morbidelli, Zappala,
  Michel, Cellino, Froeschle, Levison, Bailey, \& Duncan}]{1997Sci...277..197G}
Gladman, B.~J., Migliorini, F., Morbidelli, A., {et~al.} 1997, Science, 277,
  197

\bibitem[{Gounelle {et~al.}(2006)Gounelle, Spurn{\'y}, \&
  Bland}]{2006-MPS-41-Gounelle}
Gounelle, M., Spurn{\'y}, P., \& Bland, P.~A. 2006, Meteoritics and Planetary
  Science, 41, 135

\bibitem[{Granvik \& Brown(2018)}]{2018-Icarus-311-Granvik}
Granvik, M. \& Brown, P. 2018, \icarus, 311, 271

\bibitem[{Granvik {et~al.}(2018)Granvik, Morbidelli, Jedicke, Bolin, Bottke,
  Beshore, Vokrouhlick{\'y}, Nesvorn{\'y}, \& Michel}]{2018-Icarus-312-Granvik}
Granvik, M., Morbidelli, A., Jedicke, R., {et~al.} 2018, \icarus, 312, 181

\bibitem[{Graves {et~al.}(2018)Graves, Minton, Hirabayashi, DeMeo, \&
  Carry}]{2018-Icarus-304-Graves}
Graves, K., Minton, D., Hirabayashi, M., DeMeo, F., \& Carry, B. 2018, \icarus,
  304, 162

\bibitem[{{Graves} {et~al.}(2019){Graves}, {Minton}, {Molaro}, \&
  {Hirabayashi}}]{2019Icar..322....1G}
{Graves}, K.~J., {Minton}, D.~A., {Molaro}, J.~L., \& {Hirabayashi}, M. 2019,
  \icarus, 322, 1

\bibitem[{{Harris} \& {Lagerros}(2002)}]{2002aste.book..205H}
{Harris}, A.~W. \& {Lagerros}, J.~S.~V. 2002, in Asteroids III, 205--218

\bibitem[{IMCCE(2021)}]{2021-riea}
IMCCE. 2021, Introduction aux éphémérides et phénomènes astronomiques, ed.
  J.~{Berthier}, P.~{Descamps}, \& F.~{Mignard} (edp sciences)

\bibitem[{K{\"u}ppers {et~al.}(2014)K{\"u}ppers, O'Rourke, Bockel{\'e}e-Morvan,
  Zakharov, Lee, von Allmen, Carry, Teyssier, Marston, M{\"u}ller, Crovisier,
  Barucci, \& Moreno}]{2014Natur.505..525K}
K{\"u}ppers, M., O'Rourke, L., Bockel{\'e}e-Morvan, D., {et~al.} 2014, \nat,
  505, 525

\bibitem[{{Landolt}(1983)}]{1983AJ.....88..439L}
{Landolt}, A.~U. 1983, \aj, 88, 439

\bibitem[{Lantz {et~al.}(2018)Lantz, Binzel, \& DeMeo}]{2018-Icarus-302-Lantz}
Lantz, C., Binzel, R.~P., \& DeMeo, F. 2018, \icarus, 302, 10

\bibitem[{{Lantz} {et~al.}(2017){Lantz}, {Brunetto}, {Barucci}, {Fornasier},
  {Baklouti}, {Bour{\c{c}}ois}, \& {Godard}}]{2017Icar..285...43L}
{Lantz}, C., {Brunetto}, R., {Barucci}, M.~A., {et~al.} 2017, \icarus, 285, 43

\bibitem[{{Lauretta} {et~al.}(2017){Lauretta}, {Balram-Knutson}, {Beshore},
  {Boynton}, {Drouet d'Aubigny}, {DellaGiustina}, {Enos}, {Golish},
  {Hergenrother}, {Howell}, {Bennett}, {Morton}, {Nolan}, {Rizk}, {Roper},
  {Bartels}, {Bos}, {Dworkin}, {Highsmith}, {Lorenz}, {Lim}, {Mink}, {Moreau},
  {Nuth}, {Reuter}, {Simon}, {Bierhaus}, {Bryan}, {Ballouz}, {Barnouin},
  {Binzel}, {Bottke}, {Hamilton}, {Walsh}, {Chesley}, {Christensen}, {Clark},
  {Connolly}, {Crombie}, {Daly}, {Emery}, {McCoy}, {McMahon}, {Scheeres},
  {Messenger}, {Nakamura-Messenger}, {Righter}, \&
  {Sandford}}]{2017SSRv..212..925L}
{Lauretta}, D.~S., {Balram-Knutson}, S.~S., {Beshore}, E., {et~al.} 2017, \ssr,
  212, 925

\bibitem[{Lazzarin {et~al.}(2004)Lazzarin, Marchi, Barucci, di~Martino, \&
  Barbieri}]{2004Icar..169..373L}
Lazzarin, M., Marchi, S., Barucci, M.~A., di~Martino, M., \& Barbieri, C. 2004,
  \icarus, 169, 373

\bibitem[{{Lazzarin} {et~al.}(2005){Lazzarin}, {Marchi}, {Magrin}, \&
  {Licandro}}]{2005MNRAS.359.1575L}
{Lazzarin}, M., {Marchi}, S., {Magrin}, S., \& {Licandro}, J. 2005, \mnras,
  359, 1575

\bibitem[{{Lucas} {et~al.}(2019){Lucas}, {Emery}, {MacLennan},
  {Pinilla-Alonso}, {Cartwright}, {Lindsay}, {Reddy}, {Sanchez}, {Thomas}, \&
  {Lorenzi}}]{2019Icar..322..227L}
{Lucas}, M.~P., {Emery}, J.~P., {MacLennan}, E.~M., {et~al.} 2019, \icarus,
  322, 227

\bibitem[{{Lupton} {et~al.}(1999){Lupton}, {Gunn}, \&
  {Szalay}}]{1999AJ....118.1406L}
{Lupton}, R.~H., {Gunn}, J.~E., \& {Szalay}, A.~S. 1999, \aj, 118, 1406

\bibitem[{{Mahlke} {et~al.}(2022){Mahlke}, {Carry}, \&
  {Mattei}}]{2022A&A...665A..26M}
{Mahlke}, M., {Carry}, B., \& {Mattei}, P.~A. 2022, \aap, 665, A26

\bibitem[{Marchi {et~al.}(2012)Marchi, Paolicchi, \&
  Richardson}]{2012-MNRAS-421-Marchi}
Marchi, S., Paolicchi, P., \& Richardson, D.~C. 2012, \mnras, 421, 2

\bibitem[{{Marsset} {et~al.}(2020){Marsset}, {DeMeo}, {Binzel}, {Bus},
  {Burbine}, {Burt}, {Moskovitz}, {Polishook}, {Rivkin}, {Slivan}, \&
  {Thomas}}]{2020ApJS..247...73M}
{Marsset}, M., {DeMeo}, F.~E., {Binzel}, R.~P., {et~al.} 2020, \apjs, 247, 73

\bibitem[{{Marsset} {et~al.}(2022){Marsset}, {DeMeo}, {Burt}, {Polishook},
  {Binzel}, {Granvik}, {Vernazza}, {Carry}, {Bus}, {Slivan}, {Thomas},
  {Moskovitz}, \& {Rivkin}}]{2022AJ....163..165M}
{Marsset}, M., {DeMeo}, F.~E., {Burt}, B., {et~al.} 2022, \aj, 163, 165

\bibitem[{{Masiero} {et~al.}(2021){Masiero}, {Wright}, \&
  {Mainzer}}]{2021PSJ.....2...32M}
{Masiero}, J.~R., {Wright}, E.~L., \& {Mainzer}, A.~K. 2021, \psj, 2, 32

\bibitem[{McSween {et~al.}(2006)McSween, Lauretta, \&
  Leshin}]{2006-MESS2-McSween}
McSween, H.~Y., Lauretta, D.~S., \& Leshin, L.~A. 2006, Meteorites and the
  Early Solar System II, 53

\bibitem[{{Michel} {et~al.}(2022){Michel}, {K{\"u}ppers}, {Bagatin}, {Carry},
  {Charnoz}, {Leon}, {Fitzsimmons}, {Gordo}, {Green}, {H{\'e}rique}, {Juzi},
  {Karatekin}, {Kohout}, {Lazzarin}, {Murdoch}, {Okada}, {Palomba}, {Pravec},
  {Snodgrass}, {Tortora}, {Tsiganis}, {Ulamec}, {Vincent}, {W{\"u}nnemann},
  {Zhang}, {Raducan}, {Dotto}, {Chabot}, {Cheng}, {Rivkin}, {Barnouin},
  {Ernst}, {Stickle}, {Richardson}, {Thomas}, {Arakawa}, {Miyamoto},
  {Nakamura}, {Sugita}, {Yoshikawa}, {Abell}, {Asphaug}, {Ballouz}, {Bottke},
  {Lauretta}, {Walsh}, {Martino}, \& {Carnelli}}]{2022PSJ.....3..160M}
{Michel}, P., {K{\"u}ppers}, M., {Bagatin}, A.~C., {et~al.} 2022, \psj, 3, 160

\bibitem[{Nesvorn{\'y} {et~al.}(2010)Nesvorn{\'y}, Bottke, Vokrouhlick{\'y},
  Chapman, \& Rafkin}]{2010Icar..209..510N}
Nesvorn{\'y}, D., Bottke, W.~F., Vokrouhlick{\'y}, D., Chapman, C.~R., \&
  Rafkin, S. 2010, \icarus, 209, 510

\bibitem[{{Nesvorn{\'y}} {et~al.}(2005){Nesvorn{\'y}}, {Jedicke}, {Whiteley},
  \& {Ivezi{\'c}}}]{2005Icar..173..132N}
{Nesvorn{\'y}}, D., {Jedicke}, R., {Whiteley}, R.~J., \& {Ivezi{\'c}}, {\v{Z}}.
  2005, \icarus, 173, 132

\bibitem[{{Noguchi} {et~al.}(2011){Noguchi}, {Nakamura}, {Kimura}, {Zolensky},
  {Tanaka}, {Hashimoto}, {Konno}, {Nakato}, {Ogami}, {Fujimura}, {Abe}, {Yada},
  {Mukai}, {Ueno}, {Okada}, {Shirai}, {Ishibashi}, \&
  {Okazaki}}]{2011Sci...333.1121N}
{Noguchi}, T., {Nakamura}, T., {Kimura}, M., {et~al.} 2011, Science, 333, 1121

\bibitem[{Parker {et~al.}(2008)Parker, Ivezi{\'c}, Juri{\'c}, Lupton, Sekora,
  \& Kowalski}]{2008-Icarus-198-Parker}
Parker, A., Ivezi{\'c}, {\v Z}., Juri{\'c}, M., {et~al.} 2008, \icarus, 198,
  138

\bibitem[{{Perna} {et~al.}(2018){Perna}, {Barucci}, {Fulchignoni}, {Popescu},
  {Belskaya}, {Fornasier}, {Doressoundiram}, {Lantz}, \&
  {Merlin}}]{2018PSS..157...82P}
{Perna}, D., {Barucci}, M.~A., {Fulchignoni}, M., {et~al.} 2018, \planss, 157,
  82

\bibitem[{Polishook {et~al.}(2017)Polishook, Jacobson, Morbidelli, \&
  Aharonson}]{2017NatAs...1E.179P}
Polishook, D., Jacobson, S.~A., Morbidelli, A., \& Aharonson, O. 2017, Nature
  Astronomy, 1, 0179

\bibitem[{Popescu {et~al.}(2018)Popescu, Licandro, Carvano, Stoicescu,
  de~Le{\'o}n, Morate, Boac{\u{a}}, \& Cristescu}]{2018-AA-617-Popescu}
Popescu, M., Licandro, J., Carvano, J.~M., {et~al.} 2018, \aap, 617, A12

\bibitem[{{Popescu} {et~al.}(2018){Popescu}, {Perna}, {Barucci}, {Fornasier},
  {Doressoundiram}, {Lantz}, {Merlin}, {Belskaya}, \&
  {Fulchignoni}}]{2018MNRAS.477.2786P}
{Popescu}, M., {Perna}, D., {Barucci}, M.~A., {et~al.} 2018, \mnras, 477, 2786

\bibitem[{Rayner {et~al.}(2003)Rayner, Toomey, Onaka, Denault, Stahlberger,
  Vacca, Cushing, \& Wang}]{2003-PASP-115-Rayner}
Rayner, J.~T., Toomey, D.~W., Onaka, P.~M., {et~al.} 2003, Publications of the
  Astronomical Society of the Pacific, 115, 362

\bibitem[{{Reddy} {et~al.}(2015){Reddy}, {Dunn}, {Thomas}, {Moskovitz}, \&
  {Burbine}}]{2015aste.book...43R}
{Reddy}, V., {Dunn}, T.~L., {Thomas}, C.~A., {Moskovitz}, N.~A., \& {Burbine},
  T.~H. 2015, in Asteroids IV, 43--63

\bibitem[{{Rivkin} {et~al.}(2005){Rivkin}, {Binzel}, \&
  {Bus}}]{2005Icar..175..175R}
{Rivkin}, A.~S., {Binzel}, R.~P., \& {Bus}, S.~J. 2005, \icarus, 175, 175

\bibitem[{{Rivkin} {et~al.}(2021){Rivkin}, {Chabot}, {Stickle}, {Thomas},
  {Richardson}, {Barnouin}, {Fahnestock}, {Ernst}, {Cheng}, {Chesley}, {Naidu},
  {Statler}, {Barbee}, {Agrusa}, {Moskovitz}, {Terik Daly}, {Pravec},
  {Scheirich}, {Dotto}, {Della Corte}, {Michel}, {K{\"u}ppers}, {Atchison}, \&
  {Hirabayashi}}]{2021PSJ.....2..173R}
{Rivkin}, A.~S., {Chabot}, N.~L., {Stickle}, A.~M., {et~al.} 2021, \psj, 2, 173

\bibitem[{{Rivkin} {et~al.}(2007){Rivkin}, {Trilling}, {Thomas}, {DeMeo},
  {Spahr}, \& {Binzel}}]{2007Icar..192..434R}
{Rivkin}, A.~S., {Trilling}, D.~E., {Thomas}, C.~A., {et~al.} 2007, \icarus,
  192, 434

\bibitem[{{Rodrigo} \& {Solano}(2020)}]{2020sea..confE.182R}
{Rodrigo}, C. \& {Solano}, E. 2020, in XIV.0 Scientific Meeting (virtual) of
  the Spanish Astronomical Society, 182

\bibitem[{{Rodrigo} {et~al.}(2012){Rodrigo}, {Solano}, \&
  {Bayo}}]{2012ivoa.rept.1015R}
{Rodrigo}, C., {Solano}, E., \& {Bayo}, A. 2012, {SVO Filter Profile Service
  Version 1.0}, IVOA Working Draft 15 October 2012

\bibitem[{Rodrigo {et~al.}(2012)Rodrigo, Solano, Bayo, \&
  Rodrigo}]{2012-IVOA-Rodrigo}
Rodrigo, C., Solano, E., Bayo, A., \& Rodrigo, C. 2012, {SVO Filter Profile
  Service Version 1.0}, Tech. rep.

\bibitem[{{Ruesch} {et~al.}(2016){Ruesch}, {Platz}, {Schenk}, {McFadden},
  {Castillo-Rogez}, {Quick}, {Byrne}, {Preusker}, {O'Brien}, {Schmedemann},
  {Williams}, {Li}, {Bland}, {Hiesinger}, {Kneissl}, {Neesemann}, {Schaefer},
  {Pasckert}, {Schmidt}, {Buczkowski}, {Sykes}, {Nathues}, {Roatsch},
  {Hoffmann}, {Raymond}, \& {Russell}}]{2016Sci...353.4286R}
{Ruesch}, O., {Platz}, T., {Schenk}, P., {et~al.} 2016, Science, 353, aaf4286

\bibitem[{{Sanchez} {et~al.}(2014){Sanchez}, {Reddy}, {Kelley}, {Cloutis},
  {Bottke}, {Nesvorn{\'y}}, {Lucas}, {Hardersen}, {Gaffey}, {Abell}, \& {Le
  Corre}}]{2014Icar..228..288S}
{Sanchez}, J.~A., {Reddy}, V., {Kelley}, M.~S., {et~al.} 2014, \icarus, 228,
  288

\bibitem[{Sanchez {et~al.}(2012)Sanchez, Reddy, Nathues, Cloutis, Mann, \&
  Hiesinger}]{2012Icar..220...36S}
Sanchez, J.~A., Reddy, V., Nathues, A., {et~al.} 2012, \icarus, 220, 36

\bibitem[{Sasaki {et~al.}(2001)Sasaki, Nakamura, Hamabe, Kurahashi, \&
  Hiroi}]{2001Natur.410..555S}
Sasaki, S., Nakamura, K., Hamabe, Y., Kurahashi, E., \& Hiroi, T. 2001, \nat,
  410, 555

\bibitem[{{Scheeres} {et~al.}(2020){Scheeres}, {McMahon}, {Wood}, {Bierhaus},
  {Benner}, {Hartzell}, {Hayne}, {Hopkins}, {Jedicke}, {LeCorre}, {Naidu},
  {Pravec}, \& {Ravine}}]{2020LPI....51.1965S}
{Scheeres}, D.~J., {McMahon}, J.~W., {Wood}, J., {et~al.} 2020, in 51st Annual
  Lunar and Planetary Science Conference, Lunar and Planetary Science
  Conference, 1965

\bibitem[{{Sergeyev} \& {Carry}(2021)}]{SergeyevCarry2021}
{Sergeyev}, A.~V. \& {Carry}, B. 2021, \aap, 652, A59

\bibitem[{{Sergeyev} {et~al.}(2022){Sergeyev}, {Carry}, {Onken}, {Devillepoix},
  {Wolf}, \& {Chang}}]{Sergeyev2022}
{Sergeyev}, A.~V., {Carry}, B., {Onken}, C.~A., {et~al.} 2022, \aap, 658, A109

\bibitem[{Solano {et~al.}(2014)Solano, Rodrigo, Pulido, \&
  Carry}]{2014AN....335..142S}
Solano, E., Rodrigo, C., Pulido, R., \& Carry, B. 2014, Astronomische
  Nachrichten, 335, 142

\bibitem[{Strazzulla {et~al.}(2005)Strazzulla, Dotto, Binzel, Brunetto,
  Barucci, Blanco, \& Orofino}]{2005Icar..174...31S}
Strazzulla, G., Dotto, E., Binzel, R.~P., {et~al.} 2005, \icarus, 174, 31

\bibitem[{{Tachibana} {et~al.}(2022){Tachibana}, {Sawada}, {Okazaki}, {Takano},
  {Sakamoto}, {Miura}, {Okamoto}, {Yano}, {Yamanouchi}, {Michel}, {Zhang},
  {Schwartz}, {Thuillet}, {Yurimoto}, {Nakamura}, {Noguchi}, {Yabuta},
  {Naraoka}, {Tsuchiyama}, {Imae}, {Kurosawa}, {Nakamura}, {Ogawa}, {Sugita},
  {Morota}, {Honda}, {Kameda}, {Tatsumi}, {Cho}, {Yoshioka}, {Yokota},
  {Hayakawa}, {Matsuoka}, {Sakatani}, {Yamada}, {Kouyama}, {Suzuki}, {Honda},
  {Yoshimitsu}, {Kubota}, {Demura}, {Yada}, {Nishimura}, {Yogata}, {Nakato},
  {Yoshitake}, {Suzuki}, {Furuya}, {Hatakeda}, {Miyazaki}, {Kumagai}, {Okada},
  {Abe}, {Usui}, {Ireland}, {Fujimoto}, {Yamada}, {Arakawa}, {Connolly},
  {Fujii}, {Hasegawa}, {Hirata}, {Hirata}, {Hirose}, {Hosoda}, {Iijima},
  {Ikeda}, {Ishiguro}, {Ishihara}, {Iwata}, {Kikuchi}, {Kitazato}, {Lauretta},
  {Libourel}, {Marty}, {Matsumoto}, {Michikami}, {Mimasu}, {Miura}, {Mori},
  {Nakamura-Messenger}, {Namiki}, {Nguyen}, {Nittler}, {Noda}, {Noguchi},
  {Ogawa}, {Ono}, {Ozaki}, {Senshu}, {Shimada}, {Shimaki}, {Shirai}, {Soldini},
  {Takahashi}, {Takei}, {Takeuchi}, {Tsukizaki}, {Wada}, {Yamamoto},
  {Yoshikawa}, {Yumoto}, {Zolensky}, {Nakazawa}, {Terui}, {Tanaka}, {Saiki},
  {Yoshikawa}, {Watanabe}, \& {Tsuda}}]{2022Sci...375.1011T}
{Tachibana}, S., {Sawada}, H., {Okazaki}, R., {et~al.} 2022, Science, 375, 1011

\bibitem[{Taylor(2005)}]{2005ASPC..347...29T}
Taylor, M.~B. 2005, in Astronomical Society of the Pacific Conference Series,
  Vol. 347, Astronomical Data Analysis Software and Systems XIV, ed.
  P.~{Shopbell}, M.~{Britton}, \& R.~{Ebert}, 29

\bibitem[{Tholen(1984)}]{1984-PhD-Tholen}
Tholen, D.~J. 1984, PhD thesis, Arizona Univ., Tucson.

\bibitem[{{Tholen}(1989)}]{1989aste.conf.1139T}
{Tholen}, D.~J. 1989, in Asteroids II, ed. R.~P. {Binzel}, T.~{Gehrels}, \&
  M.~S. {Matthews}, 1139--1150

\bibitem[{Thomas {et~al.}(2012)Thomas, Trilling, \&
  Rivkin}]{2012-Icarus-219-Thomas}
Thomas, C.~A., Trilling, D.~E., \& Rivkin, A.~S. 2012, \icarus, 219, 505

\bibitem[{Usui {et~al.}(2013)Usui, Kasuga, Hasegawa, Ishiguro, Kuroda,
  M{\"u}ller, Ootsubo, \& Matsuhara}]{2013-ApJ-762-Usui}
Usui, F., Kasuga, T., Hasegawa, S., {et~al.} 2013, \apj, 762, 56

\bibitem[{{Vallenari} {et~al.}(2022){Vallenari}, {Brown}, {Prusti}, {de
  Bruijne}, {Arenou}, {Babusiaux}, {Biermann}, {Creevey}, {Ducourant}, {Evans},
  {Eyer}, {Guerra}, {Hutton}, {Jordi}, {Klioner}, {Lammers}, {Lindegren},
  {Luri}, {Mignard}, {Panem}, {Pourbaix}, {Randich}, {Sartoretti}, {Soubiran},
  {Tanga}, {Walton}, {Bailer-Jones}, {Bastian}, {Drimmel}, {Jansen}, {Katz},
  {Lattanzi}, {van Leeuwen}, {Bakker}, {Cacciari}, {Casta{\~n}eda}, {De
  Angeli}, {Fabricius}, {Fouesneau}, {Fr{\'e}mat}, {Galluccio}, {Guerrier},
  {Heiter}, {Masana}, {Messineo}, {Mowlavi}, {Nicolas}, {Nienartowicz},
  {Pailler}, {Panuzzo}, {Riclet}, {Roux}, {Seabroke}, {Sordo{\o}rcit},
  {Th{\'e}venin}, {Gracia-Abril}, {Portell}, {Teyssier}, {Altmann}, {Andrae},
  {Audard}, {Bellas-Velidis}, {Benson}, {Berthier}, {Blomme}, {Burgess},
  {Busonero}, {Busso}, {C{\'a}novas}, {Carry}, {Cellino}, {Cheek},
  {Clementini}, {Damerdji}, {Davidson}, {de Teodoro}, {Nu{\~n}ez Campos},
  {Delchambre}, {Dell'Oro}, {Esquej}, {Fern{\'a}ndez-Hern{\'a}ndez}, {Fraile},
  {Garabato}, {Garc{\'\i}a-Lario}, {Gosset}, {Haigron}, {Halbwachs}, {Hambly},
  {Harrison}, {Hern{\'a}ndez}, {Hestroffer}, {Hodgkin}, {Holl}, {Jan{\ss}en},
  {Jevardat de Fombelle}, {Jordan}, {Krone-Martins}, {Lanzafame},
  {L{\"o}ffler}, {Marchal}, {Marrese}, {Moitinho}, {Muinonen}, {Osborne},
  {Pancino}, {Pauwels}, {Recio-Blanco}, {Reyl{\'e}}, {Riello}, {Rimoldini},
  {Roegiers}, {Rybizki}, {Sarro}, {Siopis}, {Smith}, {Sozzetti}, {Utrilla},
  {van Leeuwen}, {Abbas}, {{\'A}brah{\'a}m}, {Abreu Aramburu}, {Aerts},
  {Aguado}, {Ajaj}, {Aldea-Montero}, {Altavilla}, {{\'A}lvarez}, {Alves},
  {Anders}, {Anderson}, {Anglada Varela}, {Antoja}, {Baines}, {Baker},
  {Balaguer-N{\'u}{\~n}ez}, {Balbinot}, {Balog}, {Barache}, {Barbato},
  {Barros}, {Barstow}, {Bartolom{\'e}}, {Bassilana}, {Bauchet}, {Becciani},
  {Bellazzini}, {Berihuete}, {Bernet}, {Bertone}, {Bianchi}, {Binnenfeld},
  {Blanco-Cuaresma}, {Blazere}, {Boch}, {Bombrun}, {Bossini}, {Bouquillon},
  {Bragaglia}, {Bramante}, {Breedt}, {Bressan}, {Brouillet}, {Brugaletta},
  {Bucciarelli}, {Burlacu}, {Butkevich}, {Buzzi}, {Caffau}, {Cancelliere},
  {Cantat-Gaudin}, {Carballo}, {Carlucci}, {Carnerero}, {Carrasco},
  {Casamiquela}, {Castellani}, {Castro-Ginard}, {Chaoul}, {Charlot}, {Chemin},
  {Chiaramida}, {Chiavassa}, {Chornay}, {Comoretto}, {Contursi}, {Cooper},
  {Cornez}, {Cowell}, {Crifo}, {Cropper}, {Crosta}, {Crowley}, {Dafonte},
  {Dapergolas}, {David}, {David}, {de Laverny}, {De Luise}, {De March}, {De
  Ridder}, {de Souza}, {de Torres}, {del Peloso}, {del Pozo}, {Delbo},
  {Delgado}, {Delisle}, {Demouchy}, {Dharmawardena}, {Di Matteo}, {Diakite},
  {Diener}, {Distefano}, {Dolding}, {Edvardsson}, {Enke}, {Fabre}, {Fabrizio},
  {Faigler}, {Fedorets}, {Fernique}, {Fienga}, {Figueras}, {Fournier},
  {Fouron}, {Fragkoudi}, {Gai}, {Garcia-Gutierrez}, {Garcia-Reinaldos},
  {Garc{\'\i}a-Torres}, {Garofalo}, {Gavel}, {Gavras}, {Gerlach}, {Geyer},
  {Giacobbe}, {Gilmore}, {Girona}, {Giuffrida}, {Gomel}, {Gomez},
  {Gonz{\'a}lez-N{\'u}{\~n}ez}, {Gonz{\'a}lez-Santamar{\'\i}a},
  {Gonz{\'a}lez-Vidal}, {Granvik}, {Guillout}, {Guiraud},
  {Guti{\'e}rrez-S{\'a}nchez}, {Guy}, {Hatzidimitriou}, {Hauser}, {Haywood},
  {Helmer}, {Helmi}, {Sarmiento}, {Hidalgo}, {Hilger}, {H{\l}adczuk}, {Hobbs},
  {Holland}, {Huckle}, {Jardine}, {Jasniewicz}, {Jean-Antoine Piccolo},
  {Jim{\'e}nez-Arranz}, {Jorissen}, {Juaristi Campillo}, {Julbe}, {Karbevska},
  {Kervella}, {Khanna}, {Kontizas}, {Kordopatis}, {Korn}, {K{\'o}sp{\'a}l},
  {Kostrzewa-Rutkowska}, {Kruszy{\'n}ska}, {Kun}, {Laizeau}, {Lambert},
  {Lanza}, {Lasne}, {Le Campion}, {Lebreton}, {Lebzelter}, {Leccia}, {Leclerc},
  {Lecoeur-Taibi}, {Liao}, {Licata}, {Lindstr{\o}m}, {Lister}, {Livanou},
  {Lobel}, {Lorca}, {Loup}, {Madrero Pardo}, {Magdaleno Romeo}, {Managau},
  {Mann}, {Manteiga}, {Marchant}, {Marconi}, {Marcos}, {Marcos Santos},
  {Mar{\'\i}n Pina}, {Marinoni}, {Marocco}, {Marshall}, {Polo},
  {Mart{\'\i}n-Fleitas}, {Marton}, {Mary}, {Masip}, {Massari},
  {Mastrobuono-Battisti}, {Mazeh}, {McMillan}, {Messina}, {Michalik}, {Millar},
  {Mints}, {Molina}, {Molinaro}, {Moln{\'a}r}, {Monari}, {Mongui{\'o}},
  {Montegriffo}, {Montero}, {Mor}, {Mora}, {Morbidelli}, {Morel}, {Morris},
  {Muraveva}, {Murphy}, {Musella}, {Nagy}, {Noval}, {Oca{\~n}a}, {Ogden},
  {Ordenovic}, {Osinde}, {Pagani}, {Pagano}, {Palaversa}, {Palicio},
  {Pallas-Quintela}, {Panahi}, {Payne-Wardenaar}, {Pe{\~n}alosa Esteller},
  {Penttil{\"a}}, {Pichon}, {Piersimoni}, {Pineau}, {Plachy}, {Plum}, {Poggio},
  {Pr{\v{s}}a}, {Pulone}, {Racero}, {Ragaini}, {Rainer}, {Raiteri}, {Rambaux},
  {Ramos}, {Ramos-Lerate}, {Re Fiorentin}, {Regibo}, {Richards}, {Rios Diaz},
  {Ripepi}, {Riva}, {Rix}, {Rixon}, {Robichon}, {Robin}, {Robin}, {Roelens},
  {Rogues}, {Rohrbasser}, {Romero-G{\'o}mez}, {Rowell}, {Royer}, {Ruz Mieres},
  {Rybicki}, {Sadowski}, {S{\'a}ez N{\'u}{\~n}ez}, {Sagrist{\`a} Sell{\'e}s},
  {Sahlmann}, {Salguero}, {Samaras}, {Sanchez Gimenez}, {Sanna},
  {Santove{\~n}a}, {Sarasso}, {Schultheis}, {Sciacca}, {Segol}, {Segovia},
  {S{\'e}gransan}, {Semeux}, {Shahaf}, {Siddiqui}, {Siebert}, {Siltala},
  {Silvelo}, {Slezak}, {Slezak}, {Smart}, {Snaith}, {Solano}, {Solitro},
  {Souami}, {Souchay}, {Spagna}, {Spina}, {Spoto}, {Steele},
  {Steidelm{\"u}ller}, {Stephenson}, {S{\"u}veges}, {Surdej}, {Szabados},
  {Szegedi-Elek}, {Taris}, {Taylo}, {Teixeira}, {Tolomei}, {Tonello}, {Torra},
  {Torra}, {Torralba Elipe}, {Trabucchi}, {Tsounis}, {Turon}, {Ulla}, {Unger},
  {Vaillant}, {van Dillen}, {van Reeven}, {Vanel}, {Vecchiato}, {Viala},
  {Vicente}, {Voutsinas}, {Weiler}, {Wevers}, {Wyrzykowski}, {Yoldas}, {Yvard},
  {Zhao}, {Zorec}, {Zucker}, \& {Zwitter}}]{gaia3-survey}
{Vallenari}, A., {Brown}, A.~G.~A., {Prusti}, T., {et~al.} 2022, arXiv
  e-prints, arXiv:2208.00211

\bibitem[{Vernazza {et~al.}(2009)Vernazza, Binzel, Rossi, Fulchignoni, \&
  Birlan}]{2009-Nature-458-Vernazza}
Vernazza, P., Binzel, R.~P., Rossi, A., Fulchignoni, M., \& Birlan, M. 2009,
  \nat, 458, 993

\bibitem[{{Viles} {et~al.}(2010){Viles}, {Ehlmann}, {Wilson}, {Cebula}, {Page},
  \& {Bourke}}]{2010GeoRL..3718201V}
{Viles}, H., {Ehlmann}, B., {Wilson}, C.~F., {et~al.} 2010, \grl, 37, L18201

\bibitem[{Waszczak {et~al.}(2015)Waszczak, Chang, Ofek, Laher, Masci, Levitan,
  Surace, Cheng, Ip, Kinoshita, Helou, Prince, \&
  Kulkarni}]{2015-AJ-150-Waszczak}
Waszczak, A., Chang, C.-K., Ofek, E.~O., {et~al.} 2015, \aj, 150, 75

\bibitem[{{Yano} {et~al.}(2022){Yano}, {Fujita}, {Kusumoto}, {Mimasu},
  {Yoshikawa}, \& {Tsuda}}]{2022cosp...44.3264Y}
{Yano}, H., {Fujita}, M., {Kusumoto}, T., {et~al.} 2022, in 44th COSPAR
  Scientific Assembly. Held 16-24 July, Vol.~44, 3264

\bibitem[{{York} {et~al.}(2000){York}, {Adelman}, {Anderson}, {Anderson},
  {Annis}, {Bahcall}, {Bakken}, {Barkhouser}, {Bastian}, {Berman}, {Boroski},
  {Bracker}, {Briegel}, {Briggs}, {Brinkmann}, {Brunner}, {Burles}, {Carey},
  {Carr}, {Castander}, {Chen}, {Colestock}, {Connolly}, {Crocker}, {Csabai},
  {Czarapata}, {Davis}, {Doi}, {Dombeck}, {Eisenstein}, {Ellman}, {Elms},
  {Evans}, {Fan}, {Federwitz}, {Fiscelli}, {Friedman}, {Frieman}, {Fukugita},
  {Gillespie}, {Gunn}, {Gurbani}, {de Haas}, {Haldeman}, {Harris}, {Hayes},
  {Heckman}, {Hennessy}, {Hindsley}, {Holm}, {Holmgren}, {Huang}, {Hull},
  {Husby}, {Ichikawa}, {Ichikawa}, {Ivezi{\'c}}, {Kent}, {Kim}, {Kinney},
  {Klaene}, {Kleinman}, {Kleinman}, {Knapp}, {Korienek}, {Kron}, {Kunszt},
  {Lamb}, {Lee}, {Leger}, {Limmongkol}, {Lindenmeyer}, {Long}, {Loomis},
  {Loveday}, {Lucinio}, {Lupton}, {MacKinnon}, {Mannery}, {Mantsch}, {Margon},
  {McGehee}, {McKay}, {Meiksin}, {Merelli}, {Monet}, {Munn}, {Narayanan},
  {Nash}, {Neilsen}, {Neswold}, {Newberg}, {Nichol}, {Nicinski}, {Nonino},
  {Okada}, {Okamura}, {Ostriker}, {Owen}, {Pauls}, {Peoples}, {Peterson},
  {Petravick}, {Pier}, {Pope}, {Pordes}, {Prosapio}, {Rechenmacher}, {Quinn},
  {Richards}, {Richmond}, {Rivetta}, {Rockosi}, {Ruthmansdorfer}, {Sandford},
  {Schlegel}, {Schneider}, {Sekiguchi}, {Sergey}, {Shimasaku}, {Siegmund},
  {Smee}, {Smith}, {Snedden}, {Stone}, {Stoughton}, {Strauss}, {Stubbs},
  {SubbaRao}, {Szalay}, {Szapudi}, {Szokoly}, {Thakar}, {Tremonti}, {Tucker},
  {Uomoto}, {Vanden Berk}, {Vogeley}, {Waddell}, {Wang}, {Watanabe},
  {Weinberg}, {Yanny}, {Yasuda}, \& {SDSS Collaboration}}]{2000AJ....120.1579Y}
{York}, D.~G., {Adelman}, J., {Anderson}, John~E., J., {et~al.} 2000, \aj, 120,
  1579

\bibitem[{Yurimoto {et~al.}(2011)Yurimoto, Abe, Abe, Ebihara, Fujimura,
  Hashiguchi, Hashizume, Ireland, Itoh, Katayama, Kato, Kawaguchi, Kawasaki,
  Kitajima, Kobayashi, Meike, Mukai, Nagao, Nakamura, Naraoka, Noguchi,
  Okazaki, Park, Sakamoto, Seto, Takei, Tsuchiyama, Uesugi, Wakaki, Yada,
  Yamamoto, Yoshikawa, \& Zolensky}]{2011-Science-333-Yurimoto}
Yurimoto, H., Abe, K.-i., Abe, M., {et~al.} 2011, Science, 333, 1116

\end{thebibliography}

\end{document}